\documentclass{IAU-TrB}

\usepackage{graphicx}
\usepackage{natbib}
\usepackage{hyperref}

\usepackage{lineno}
\title[SOLAR ACTIVITY]     %% header right hand page %%
{}

\author[DIVISION II COMMISSION 10]   %% header left hand page %%
{}

\pubyear{2015}
\volume{Volume XXIXA}
\pagerange{1--?}
\date{\today}
\jname{Transactions IAU, Volume XXIXA}
\editors{Thierry Montmerle, ed.}
\begin{document}

\maketitle
\maketitle

{\bf

\large
\begin{tabbing}
\hspace*{65mm}       \=                                              \kill
DIVISION II \\COMMISSION 10         \> SOLAR ACTIVITY                                     \\
                     \> {\it (ACTIVITE SOLAIRE)}                             \\
\end{tabbing}

\normalsize

\begin{tabbing}
\hspace*{65mm}       \=                                              \kill
PRESIDENT            \> Carolus J. Schrijver      \\
VICE-PRESIDENT       \> Lyndsay Fletcher    \\
PAST PRESIDENT       \> Lidia {van Driel-Gesztelyi}                    \\
ORGANIZING COMMITTEE \> Ayumi Asai, Paul S.\ Cally,\\ 
                     \> Paul Charbonneau, Sarah E.\ Gibson, \\
                     \>  Daniel Gomez, Siraj S.\ Hasan,\\
                     \>  Astrid M.\ Veronig, Yihua Yan\\
\end{tabbing}

\bigskip
\noindent
%TITLE
}

\small

%\title[Commission 10. Solar Activity]{COMMISSION 10. Solar Activity}
%\author[Commission 10]{PRESIDENT: C.J. Schrijver$^1$ \\
% VICE-PRESIDENT: L. Fletcher$^2$ \\
%ORGANIZING COMMITTEE: A. Asai$^3$, P. Cally$^4$, \hbox{P. Charbonneau$^5$}, S. Gibson$^6$, D. G{\'o}mez$^7$, S.S. Hasan$^8$, \hbox{L. {van Driel-Gesztelyi}$^9$}, A. Veronig$^{10}$, Y. Yan$^{11}$}
%\affiliation{$^1$ Lockheed Martin Advanced Technology Center, Palo Alto, California, USA;
%$^2$ SUPA School of Physics and Astronomy, University of Glasgow, Glasgow, Scotland;
%$^3$ Unit of Synergetic Studies for Space, Kyoto University, Yamashina, Kyoto, Japan;
%$^4$ School of Mathematical Sciences, Monash University, Clayton, Victoria, Australia;
%$^5$ D{\'e}partement de Physique et Calcul Qu{\'e}bec, Universit{\'e} de Montr{\'e}al, Montr{\'e}al, Canada;
%$^6$ High Altitude Observatory, National Center for Atmospheric Research, Boulder, Colorado, USA;
%$^7$ Instituto de Astronomía y F{\'i}sica del Espacio, Buenos Aires, Argentina;
%$^8$ Indian Institute of Astrophysics, Koramangala, Bangalore, India;
%$^9$ Observatoire de Paris, LESIA, CNRS, UPMC Universit{\'e} Paris-Diderot; Mullard Space Science Laboratory, University College London;
%$^{10}$ Institute of Physics/IGAM, University of Graz, Graz, Austria;
%$^{11}$ National Astronomical Observatories, Chinese Academy of Sciences, Beijing, China}

%\pubyear{2015}
%\volume{\ldots}  %% insert here IAU Transactions No.
%\pagerange{\ldots}
%\date{?? and in revised form ??}
%\setcounter{page}{1}
%\jname{REPORTS ON ASTRONOMY \ldots}
%\editors{\ldots}

\begin{abstract}
  After more than half a century of community support related to the
  science of ``solar activity'', IAU's Commission 10 was formally
  discontinued in 2015, to be succeeded by C.E2 with the same area of
  responsibility. On this occasion, we look back at the growth of the
  scientific disciplines involved around the world over almost a full
  century. Solar activity and fields of research looking into the
  related physics of the heliosphere continue to be vibrant and
  growing, with currently over 2,000 refereed publications appearing
  per year from over 4,000 unique authors, publishing in dozens of
  distinct journals and meeting in dozens of workshops and conferences
  each year. The size of the rapidly growing community and of the
  observational and computational data volumes, along with the
  multitude of connections into other branches of astrophysics, pose
  significant challenges; aspects of these challenges are beginning to
  be addressed through, among others, the development of new systems
  of literature reviews, machine-searchable archives for data and
  publications, and virtual observatories. As customary in these
  reports, we highlight some of the research topics that have seen
  particular interest over the most recent triennium, specifically
  active-region magnetic fields, coronal thermal structure, coronal
  seismology, flares and eruptions, and the variability of solar
  activity on long time scales. We close with a collection of
  developments, discoveries, and surprises that illustrate the
  range and dynamics of the discipline.
\end{abstract}

\firstsection % if your document starts with a section,


\begin{table}[t]
\caption{Overview of Commission 10 leadership and triennial reports
  (as available in ADS) from 1961 onward, i.e. for the period that C10
  operated under the banner of ``Solar Activity'' or ``Activit{\'e}
  Solaire''. Reports flagged with an asterisk appear not to be
  available on line.}\label{tab:oc10}
\begin{center}
\begin{tabular}{lll}
\hline
Years & President and Vice President & ADS bibcode\\
\hline
%% 1919: Solar radiation
%1932-1935  & Brunner, W.               & \\
%1935-1938  & Brunner, W.    & \\
%1938-1948  & Brunner , W.     & \\
%1948-1952  &	Waldmeier, M.      & \\
%% Ph{\'e}nom{\`e}nes Photospherique
%1952-1955  & d’Azambuja, L.     & \\
%1955-1958  & d’Azambuja, L.   & \\
%1958-1961  & Severny, A. B. & \\
1961-1964  & Severny, A. B. , Ellison, M. A. & *\\
1964-1967  & Svestka, Z. , Jefferies, J. T. & *\\
1967-1970  &	Svestka, Z. , Jefferies, J. T.	&  1970IAUTA..14...71S\\
1970-1973  &	Jefferies, J. T., Kiepenheuer, K. O. & 1973IAUTA..15...75J \\
1973-1976  &	Kiepenheuer, K. O., Newkirk, G. A.   & *\\
1976-1979  & Newkirk, G., Bumba, V.  & 1979IAUTA..17b..11N  \\
1979-1982  &	Bumba, V., Tandberg-Hanssen, E.& 1982IAUTA..18...55B  \\
1982-1985  & Tandberg-Hanssen, E., Pick, M. & 1985IAUTA..19...57T \\
1985-1988  & Pick, M., Priest, E. R. & 1988IAUTA..20...55P  \\
1988-1991  & Priest, E. R., Gaizauskas, V. & 1991IAUTA..21...53P  \\
1991-1994  &	Gaizauskas, V., Engvold, O. & *1994IAUTA..22...53G  \\
1994-1997  &	Engvold, O., 	Ai, G.	& *1997IAUTA..23..121E   \\
1997-2000  &	Ai, G., Benz, A. O. &* \\
2000-2003  &	Benz, A. O., Melrose, D. B.& *\\
2003-2006  & Melrose, D. B., Klimchuk, J. A. & 2007IAUTA..26...75M\\
2006-2009  & Klimchuk, J. A., van Driel-Gesztelyi, L.  & 2009IAUTA..27...79K   \\
2009-2012  & van Driel-Gesztelyi, L. , Schrijver, C. J.& 2012IAUTA..28...69V   \\
2012-2015  & Schrijver, C. J., Fletcher, L.  & {\em (this report)}\\
\hline
\end{tabular}
\end{center}
\end{table}
\section{Historical context}
The IAU was founded in 1919, with the first General Assembly occurring
in 1922 in Rome, Italy. The standing commissions that focused on what
we nowadays capture under the term ``solar physics'' evolved over the
following decades. A ``Commission 10'' was instituted by 1922, but
under the title of ``Solar Radiation''. By 1935 that had been changed
to ``Sunspots and sunspot numbers'', and by 1952 it had transitioned
to ``Photospheric phenomena''. In parallel to Commission 10 there were
the other solar-oriented Commissions 11 through 15 dealing with
subjects such as the solar atmosphere, eclipses, rotation, and
spectroscopy (with, as noted in the 1961 report by Commission 10, a
``lack of clear demarcation lines among the [three] solar
commissions''). It was not until 1961 that ``Solar Activity'' became
the title of Commission 10, lasting until its transition into
Commission C.E2 in the overall reorganization of the IAU's structure
in 2015.

The scientific discipline focusing on solar activity continued to grow
after Commission 10 settled on its final name, seeing among the many
activities of its community the launch of a dedicated journal ``Solar
Physics'' in 1967 and a marked advance in access to solar corona and
inner heliosphere with the space-based Apollo Telescope Mount on
Skylab in 1973.  At present, we have ground- and space-based
observatories looking at the Sun and innermost heliosphere from
different perspectives, in a range of wavelengths, and probing the
Sun's internal dynamics using helioseismology. But although these
observatories provide a wealth of information and insight into the
workings of our neighboring star, we struggle to provide the research
community with a comprehensive view of the phenomena captured under
the term ``Solar Activity''. That is certainly not a new challenge:
for example, in the 1970-1973 report, then President Jefferies of C10
notes that ``We believe that the major direction in which priority
should be placed to facilitate the understanding of solar activity
lies in the provision of space and ground based observatories
specifically designed to complement each other's capabilities.''

The early reports of Commission 10 published in the ``Reports on
astronomy'' could highlight many of the key developments in the field
overall. With the rapid growth of the discipline from the 1940s
through the 1970s that quickly became impossible. From the middle of
the 20th Century onward, severe selections had to be made in order for
the task of the writing of a progress report to remain feasible. The
need for such major down-selects on what to cover clearly concerned the
members of the Organizing Committees of the Commission (the presidents
and vice-presidents are listed in Table~\ref{tab:oc10}). For example,
in the report on the period 1967-1970, the Commission President
Zde{\~n}ek Svestka apologizes for the ``fairly severe'' selection made
in the Commission's overview with 32 pages of text as ``only about one
third of all published papers could have been mentioned in the
references''. By 2015, with some 6,000 papers appearing per triennium
(as discussed in the next section), simply reading all published
papers is too challenging a task. Even with severe selections applied
by the Organizing Committee of Commission 10, the 1967-1970 report was
still deemed too long: John Jefferies notes in the subsequent report
for 1970-1972 that the General Secretary requested that the reports
should ``concentrate on the more important developments'' (which led
to 34 pages of text in that cycle). The most recent reports condense
the increased number of publications into 10 to 20 pages of text,
selecting highlights only in the view of the members of the Organizing
Committee.

For this report on 2012-2015 we would face a similarly daunting
selection task. Instead, we chose to focus on three aspects of our
community: the health of the community itself, new developments in the areas of
instrumentation and IT infrastructure, and a discussion of scientific developments 
based on citations within the community and impressions of the OC members of C10.

\section{Trends in the research community and its publications}
As C10 transitions into C.E2 we take the opportunity to review the
community's size and publication activity. For this, we use the tools
provided by the Astrophysics Data System (ADS\footnote{URL:
  http://adsabs.harvard.edu/abstract$\_$service.html}), which enables
searches over all the major trade publications in astrophysics in
general. We reviewed the number of refereed publications per year
going back over a century, and quantified the population of active
researchers and their publication productivity.

\begin{figure}[t]
\centering
\includegraphics[width=0.65\textwidth]{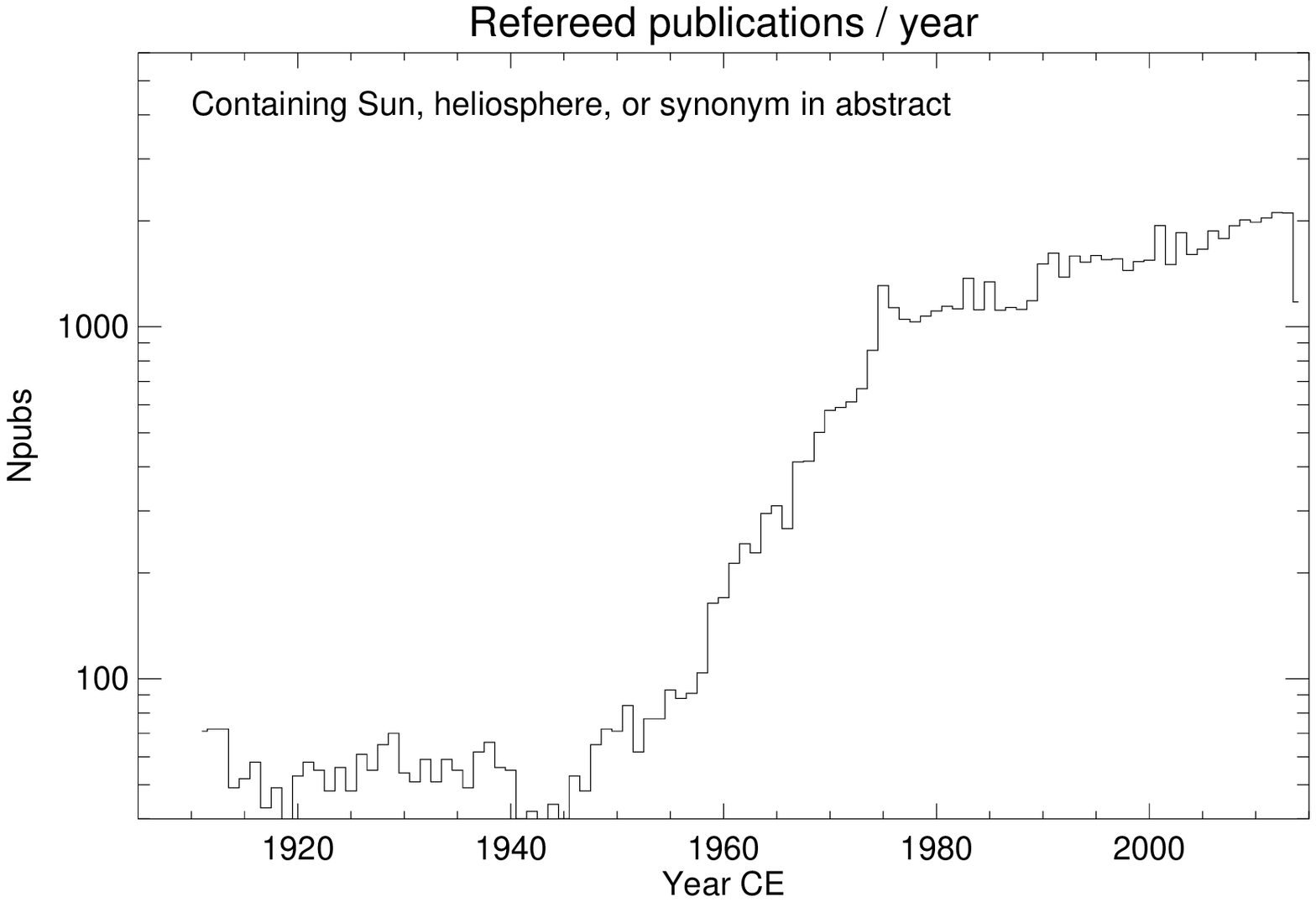}
\caption{Number of refereed publications per year with abstracts focusing on Sun or heliosphere (as returned by ADS) from 1911 through 2014.}
\label{fig:yearlystats}
%\end{figure}
%\begin{figure}[t]
\centering
\includegraphics[width=0.65\textwidth]{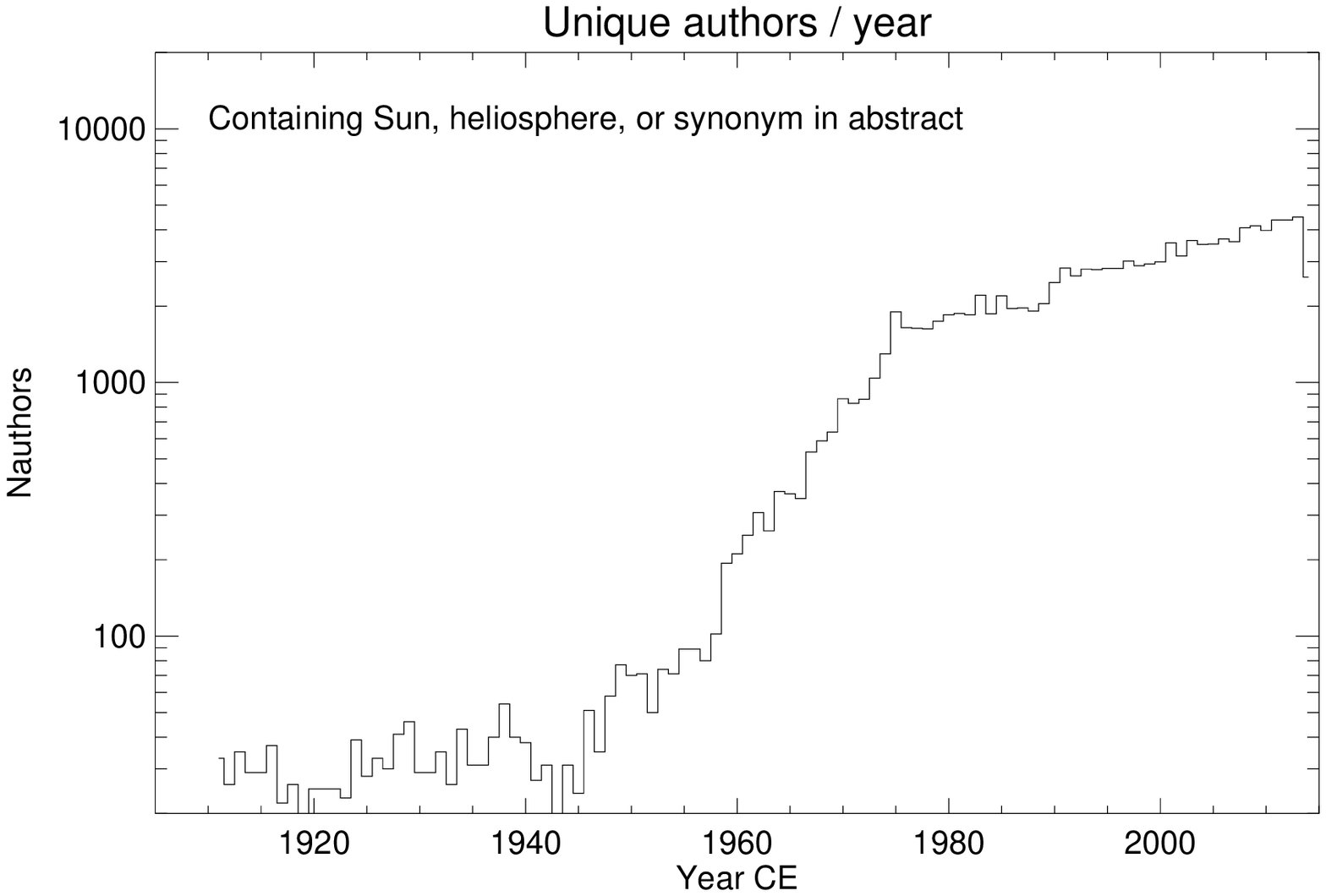}
\caption{Number of unique authors for each year in publications per year with abstracts focusing on Sun or heliosphere (as returned by ADS) from 1911 through 2014.}
\label{fig:nauthors}
\end{figure}
The study of phenomena related to ``solar activity'' often involves
other aspects of the Sun (such as internal dynamics, dynamo, or
surface field patterns) and they are obviously not limited to the Sun
but drive phenomena throughout the heliosphere. We therefore do not
attempt a separation by research disciplines along the somewhat
arbitrary dividing lines between the IAU Commissions in what was
Division II and is now Division E\footnote{See
  \hbox{http://www.iau.org/science/scientific$\_$bodies/divisions/}},
all the more so because of the shifts in focus of the Commissions
related to solar physics since their inception after 1919 as noted
above. Consequently, we searched ADS for abstracts of refereed
publications in the ``Astronomy'' database, either mentioning the Sun
or heliosphere or their synonyms. We filter out at least many of the
papers that do not deal with Sun/heliosphere that come into the search
results because their abstracts include, for example, a unit like
``solar mass''; to do so, we exclude abstracts that contain one or
more of the following words or word groups: cluster, dwarf,
extrasolar, galaxy, gravitational, ice, kpc, solar system, stellar,
binary, sunset, sunrise, eclipse, solar cell, solar occultation,
interstellar medium, and supernova.  Sampling the returned titles and
abstracts suggests that the sample we study is dominated by far by
papers that do indeed focus on Sun and heliosphere, while of course
also including topics such as climate forcing, the physics of the
upper atmospheres of planets, comets in the solar wind, cosmic-ray
modulation, and weathering of lunar surface materials.

\begin{figure}[t]
\centering
\includegraphics[width=0.65\textwidth]{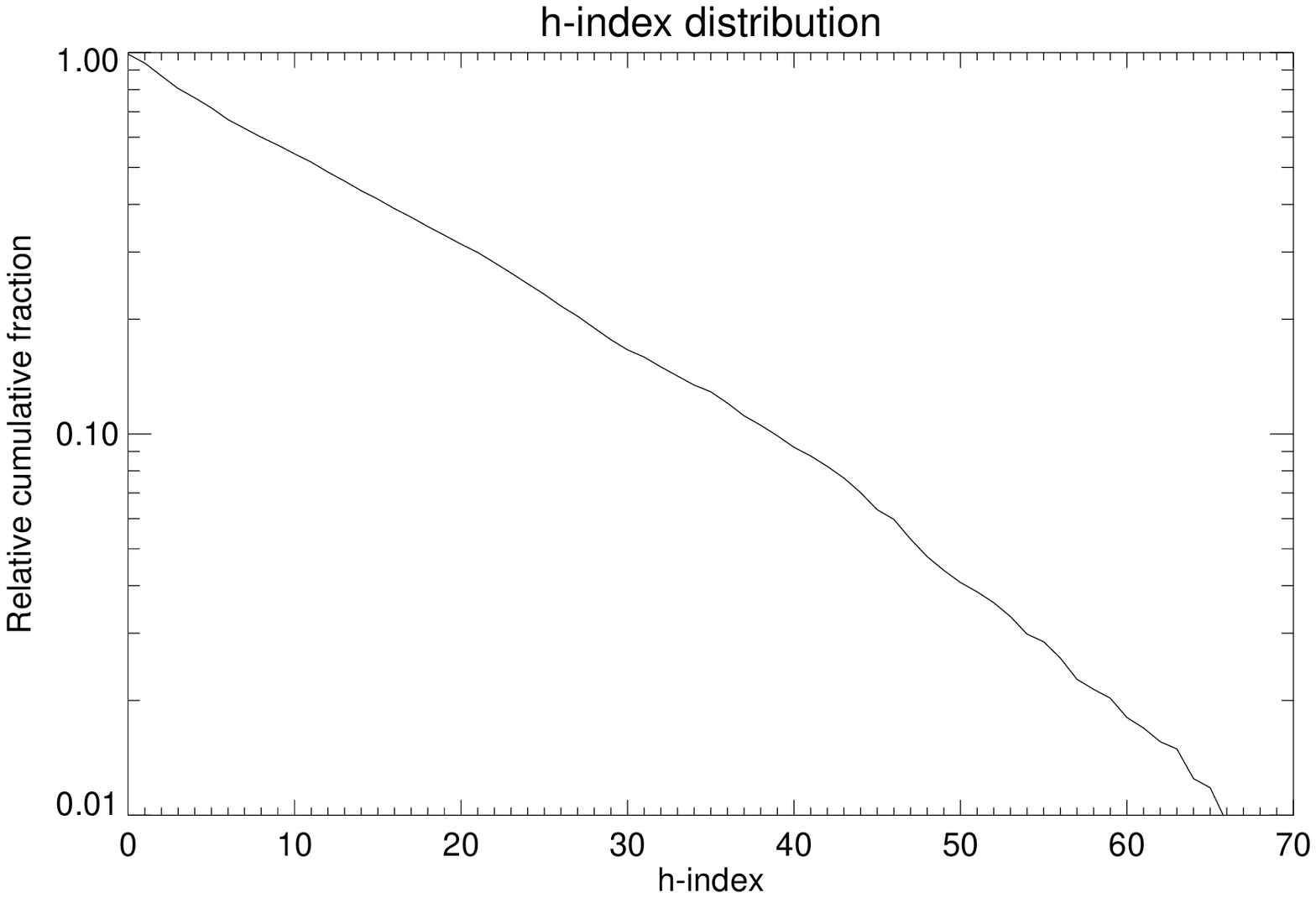}
\caption{Distribution of $H$ indices for all authors publishing on Sun and heliosphere in 2014, based on ADS citation counts.}
\label{fig:hindex}
%\end{figure}
%\begin{figure}[t]
\centering
\includegraphics[width=0.65\textwidth]{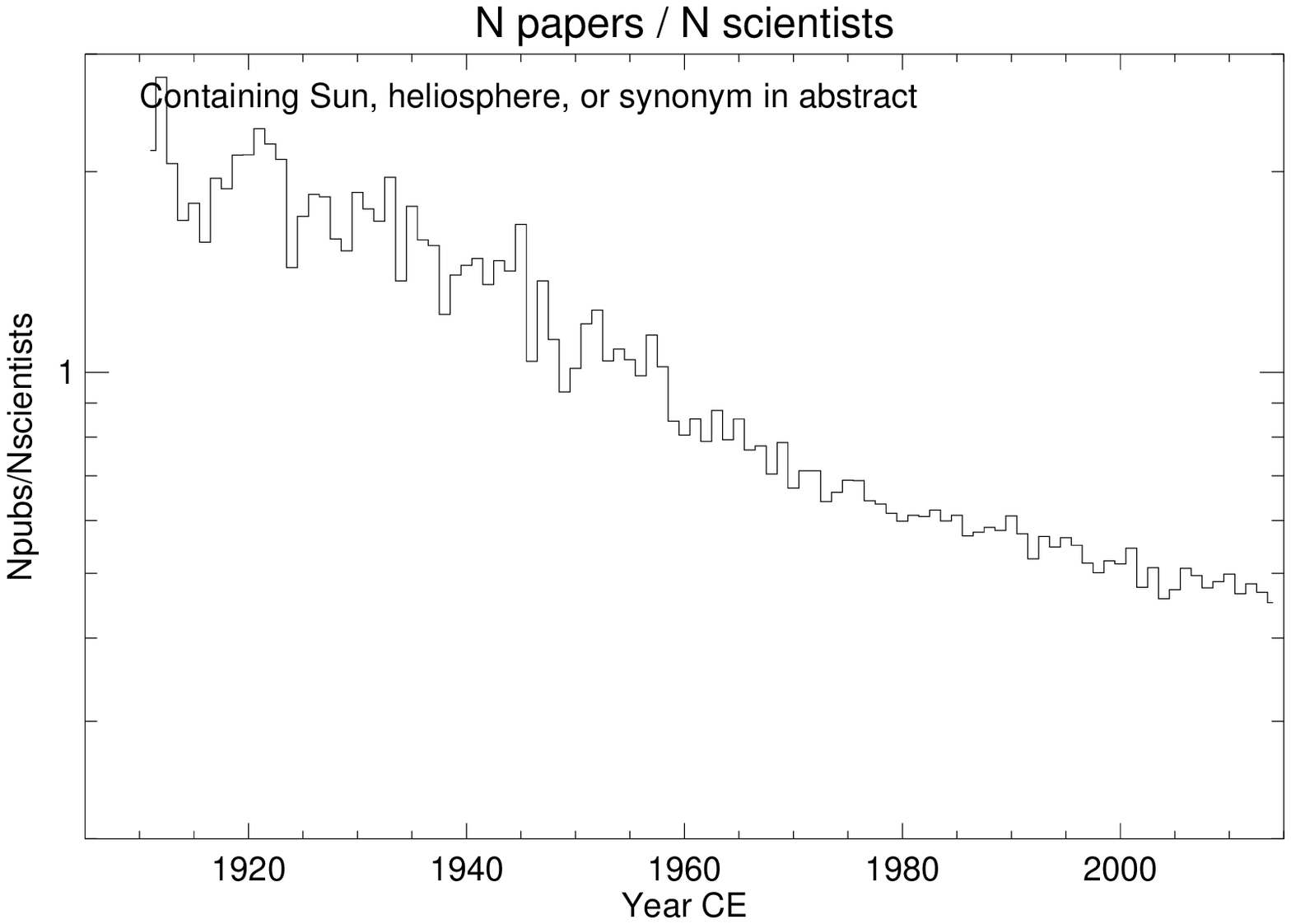}
\caption{Number of refereed publications per author with abstracts focusing on Sun or heliosphere (as returned by ADS) from 1911 through 2014.}
\label{fig:nauthorsperpaper}
\end{figure}
The ADS searches suggest that the productivity of the world-wide
community researching the Sun and heliosphere continues to grow
steadily if measured through its publications
(Fig.~\ref{fig:yearlystats}). A rapid growth in the number of refereed
publications that started after the Second World War continued up to
about 1975. After that, the growth slowed drastically, transitioning
to a sustained increase that doubles the number of refereed
publications on a time scale of approximately 40 years, reaching a
total of some 2200 refereed publications by 2014.

\begin{figure}[t]
\centering
\includegraphics[width=0.65\textwidth]{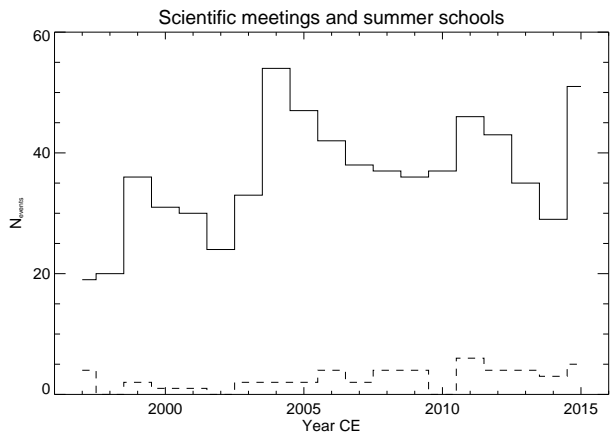}
\caption{Number international scientific meetings (solid) and summer schools (dashed) from 1997 through 2014 (from the SOHO web site: http://sohowww.nascom.nasa.gov/community/).}
\label{fig:meetings}
\end{figure}
Such automated searches enable us to process a lot of
information, but it is not readily possible to avoid a distortion of the
statistics associated with the author names. For one thing, authors
with identical family names and initials for their given names are not
differentiated. Also, authors who publish with different spellings or
composites of their family names ({\it e.g.}  married and maiden names) or
their initials will be counted as separate individuals. These effects
bias the total number of publications and researchers involved, but we
expect their impacts to be limited when we review fractional trends over the
years, as we do not expect very substantial changes of the relative
impacts of these biases over the years.

The number of unique author names (subject to the above-noted caveat
of the way we used the ADS system) contributing to refereed
publications shows a trend that roughly mimics that of the number of
publications: flat in the first half of the 20th Century with a total population of merely some 40 active researchers globally, then rapidly
growing after WW~II, and eventually growing exponentially from about 1975 onward
up to a total approaching 5,000. This growth rate of $\approx
2.5$\%/yr is about twice the growth rate of the world's overall
population (which averaged at $\approx 1.3$\%/yr over the same period;
\citealp{prb2013}), suggestive of a
rather healthy growth in the research discipline of solar and
heliospheric sciences and related fields over that of the general population.

Many of the authors in the population publishing in the refereed
literature are in the field only for a short time, or publish only
infrequently as team members on, for example, instrument and facility
papers. If we look for the population of authors who have worked in
the field of solar-heliospheric sciences long enough to have
contributed several papers that have been cited a few times, we can
use the so-called $H$ index. That index is computed by ranking all
refereed publications by an author in decreasing order of the number
of citations, and then looking for the rank in that list where rank
and number of citations equal.
The downward-cumulative $H$ index distribution for authors publishing
on Sun and heliosphere in 2014 (based on citations in ADS, and using
refereed as well as unrefereed publications) is shown in
Fig.~\ref{fig:hindex}.

For the present purpose of selecting researchers with a few years of
activity in the field, we take a rather arbitrary $H$-index threshold
of 5, i.e., looking for authors with at least 5 refereed publications
on their record that have each been cited at least 5
times. Fig.~\ref{fig:hindex} shows that this includes about two thirds
of all publishing authors in 2014.

The number of publications per year grows somewhat slower than the
population of contributing authors, so that the number of publications
per author shows a steady decrease over the past century, trending to
just under half a refereed publication per year per author (see
Fig.~\ref{fig:nauthorsperpaper}). That average value is subject to a
large range for the population: some authors may not publish for years
in a row, while others may exceed a dozen in particularly productive
years, for example as members of sizable teams working with new
instrumentation.

\begin{table}[t]
\caption{IAU meetings related to solar activity in the triennium 2012-2014.}\label{tab:meetings}
\begin{center}
\begin{tabular}{ll}
\hline
2014& IAUS305 Polarimetry: the Sun to stars and stellar environments\\
2013& IAUS302 Magnetic fields throughout stellar evolution\\
2013& IAUS300 Nature of prominences and their role in space weather\\
2012& IAUS294 Solar and astrophysical dynamos and magnetic activity\\
\hline
\end{tabular}
\end{center}
\end{table}
The community exchanges information efficiently at scientific
meetings.  Fig.~\ref{fig:meetings} shows that the number of such
events tends to increase over the years, with marked fluctuations from
year to year, averaging around 40 meetings per year over the past
decade. The Symposia supported by the IAU through Commission 10 in the
last three years are listed in Table~\ref{tab:meetings}.  Summer
schools in which new generations of researchers are given broader or
deeper views into the community's activities appear to grow slowly in
frequency, trending towards about five events per year (dashed line in
Fig.~\ref{fig:meetings}).

Although the scientific community working on ``solar activity'' appears
healthy and growing, there is a clear need to improve how we
communicate the excitement about our science to the general public:
for example, only 8 in the most recent 400 press releases and news
articles listed by the
AAS\footnote{http://aas.org/astronomy-in-the-news, accessed on
  2015/09/03.} were related to some aspect of solar activity.

\section{Trends in observational capabilities}
The observational infrastructure enabling the study of solar physics
has seen dramatic advances over the past five years. Among these, we
highlight a few:

In 2009, the two STEREO spacecraft (launched into Earth-trailing and
Earth-leading orbits in 2006; \citealp{2008SSRv..136....5K}) passed the quadrature points relative to
the Sun-Earth line. When combined with Earth-perspective viewing, the
following years enabled a view of the entire solar surface, for the
first time in history showing us the evolution of an entire stellar
atmosphere.

The Atmospheric Imaging Assembly (AIA; \citealp{aiainstrument}) on
board the Solar Dynamics Observatory (SDO - launched in 2010;
\citealp{2012SoPh..275....3P}) provides uninterrupted observing of the
outer atmosphere of the entire Earth-facing side of the Sun at a
cadence of 12\,s and a resolution of close to an arcsecond. Combined
with magnetography and helioseismology with the Helioseismic and
Magnetic Imager (HMI; \citealp{2012SoPh..275..207S}), as well as
Sun-as-a-star spectroscopy in the EUV with the EUV Variability
Experiment (EVE; \citealp{2012SoPh..275..115W}), this powerful SDO
spacecraft sends down well over 1\,TB/day. The primary data and
higher-level derivatives fill a data archive that now exceeds 7\,PB
and holds over 96\%\ of all data ever taken from space in the domain
that focuses on the Sun and the physics of the Sun-Earth connections.

Instrumentation with high spatial or spectral resolution is flown on
the JAXA-led Hinode/Solar-B mission (launched in 2006; \citealp{kosugi+etl2007}) and the NASA Small Explorer IRIS
(Interface Region Imaging Spectrograph, launched in 2013; \citealp{2014SoPh..289.2733D}) that offer
images with resolutions between 0.1 and 0.3\,arcsec, combined with
spectroscopy in the visible and ultraviolet. RHESSI \citep{2002SoPh..210....3L}, launched in 2002,
is continuing to provide unique spectroscopic images of the Sun at
high energy.  

These space-based instruments provide important
access to the domain from photosphere to corona, critically complemented
by ground-based telescopes and their instrumentation, as well as by
rocket experiments such as Hi-C \citep{2014ApJ...787L..10W}.

In the optical domain, the 1.6\,m New Solar Telescope (NST,
\citealp{2012SPIE.8444E..03G}) at Big Bear in the U.S.A.\ and the
1.5-meter German GREGOR solar telescope at Tenerife in Spain
\citep{2012AN....333..796S} have been in regular operation since
2012. The 1-m New Vacuum Solar Telescope (NVST,
\citealp{2014RAA....14..705L}) at Fuxian Lake in southwest China is
also in regular operation recently. CRISP (Crisp Imaging
Spectro-polarimeter) at the Swedish 1-m Solar Telescope (SST;
\citealp{2008A&A...489..429V}) reached 0.13\,arcsec spatial resolution
and high polarimetric sensitivity aided by post-processing. All these
telescopes have capacities close to their diffraction limit due to
advanced designs and excellent seeing conditions.  We have seen
glimpses into very high-resolution ground-based coronal observing as
well with, for the first time, synoptic observations of coronal Stokes
polarimetry in the near infrared by the Coronal Multichannel
Polarimeter (CoMP, \citealp{2008SoPh..247..411T}) telescope.

In the radio domain, the recently commissioned Chinese Spectral
Radioheliograph (CSRH;
\citealp{2009EM&P..104...97Y,2013PASJ...65S..18W}) at Mingantu in
Inner Mongolia of China (renamed as Mingantu Ultrawide Spectral
Radioheliograph-MUSER) is a radioheliograph operating with the widest
frequency range ever reached from 400\,MHz to 15\,GHz, with a high
temporal, spatial, and spectral resolution. Recent non-solar dedicated
radio arrays such as the Murchison Wide-field Array (MWA,
\citealp{2011ApJ...728L..27O}) and the Low-Frequency Array for Radio
astronomy (LOFAR, \citealp{2011pre7.conf..507M}) have obtained
spectroscopic solar imaging at metric and lower frequencies.  The
recently upgraded Karl G.\ Jansky Very Large Array (EVLA) has provided
solar radio dynamic imaging spectroscopy of type III bursts at
decimeter wavelengths \citep{2013ApJ...763L..21C}.  Even the
millimeter domain and beyond is seeing major advances with the Atacama
Millimeter/sub-millimeter Array (ALMA; \citealp{2015IAUGA..2257295B})
observatory coming on line for solar observing as it continues to
complete its construction phase.

The observations of tens of thousands of Sun-like stars by the NASA
Kepler satellite is offering yet more insights into the physics of the Sun,
ranging from an improved understanding of internal structure and
dynamics (with asteroseismic techniques) to a view of rare, extremely
energetic flares. Kepler data, with ground-based follow-up studies,
suggest that solar flares may occur with energies up to several
hundred times higher than observed directly in the past half century
with space-based instrumentation \citep{2014EOSTr..95Q.201S}.

\section{Trends in IT infrastructure for research and data}
The volume of information that needs to be processed by solar
researchers is increasing rapidly. In terms of data to be analyzed, we
have definitely reached the petabyte era. This is true for
observational data in the archives of SDO, but also in computer
experiments in which single snapshot data dumps of the advanced codes
can exceed a TB.

A tendency towards open data policies means that we have ever more
access to large volumes and a daunting diversity of data. That
complicates finding, processing, and integrating data. Infrastructure
support for, for example, the Virtual Solar Observatory (VSO),
SolarSoft IDL (SSWIDL), and the Heliophysics Events Knowledgebase and
Registry (HEK, HER) are critically important to enable efficient
utilization of the growing data diversity and volume. The community
lags in strategic thinking about these meta-infrastructural elements,
both where current support and future expansion or replacement are
concerned. There has also been a recent movement towards open-source
data analysis software, with the development of the SunPy solar
analysis environment in Python \citep{mumford-proc-scipy-2013}.

A similar flood of information is found with scientific
publications, which exceed 2,000 refereed publications per year (see
above). Here, the support infrastructure of ADS is of critical
value. The wide diversity of journals in which the works of colleagues
are published requires subscription access to many publications, at
costs that are increasingly hard to bear for relatively small research
groups; here, preprint services such as arXiv and MaxMillennium play
significant roles in making the community aware of what is going
on. ``Living reviews'' as offered by the free on-line journal Living
Reviews in Solar Physics enable new researchers to understand the
context of their work and established researchers in one sub-specialty
a quick introduction into adjacent areas by their peers.

Among the difficulties the solar activity community also faces is that
many solar and inner-heliospheric events are studied by different
groups and published in different journals. Finding studies on a
particular solar region of interest is hampered by inconsistent use of
the characterizing spatio-temporal coordinates of events which may be
found in abstract, main text, tables, appendices, and sometimes only
marked in figures that are not machine-readable. The IAU adopted a
standard convention for this in 2009 (\citealp{2010SoPh..263....1.})
in its SOL standard (short for Solar Object Locator). Its use is
encouraged by journal editors including those of Solar Physics, the
Astrophysical Journal, and the Journal of Geophysical Research. Broad
usage of the SOL standard would enable computer searches of related
publications, enabling researchers to put new (meta-)studies in
broader contexts.

\section{Trends in research directions and key findings}
For the highlights touched upon in this report, we opted for two
criteria to identify topics of interest. One is to mention specific
areas of note in the opinion of the Organizing Committee that may be
new developments, are highly specialized yet significant, may concern
new instrumentation or methods, or are otherwise deemed to be
developments that may grow to see more activity in terms of
publications in the near future.

The other criterion we applied is to be guided towards topics of
frequent activity by the community itself by looking at the most cited
works. Such a selection does introduce a bias toward the papers
published early in the period reviewed, of course, but our purpose is
not to identify the most-cited works per se, but rather to find the
dominant themes within the set of these works that apparently resonate
strongly within the community already within the 3-y window from which
they are selected.

We searched for the most-cited refereed publications from ADS with the
terms ``solar.activity'', ``coronal.mass.ejection'', ``solar flare'',
``solar prominence'', or synonym(s) in their abstracts in the period
2012-2014. Within this set, we identify the following
themes (sorted alphabetically): active-region magnetic fields; coronal
thermal structure; coronal seismology; solar flares and eruptions; and the
Sun-in-time related aspects of long-term solar variability
including cosmic-ray modulation. We close with a collection of developments, discoveries, and surprises.

\subsection{Active-region magnetic fields} 
The photospheric magnetic field of active regions forms the foundation
of the overlying atmosphere. Its evolution --~through emergence,
displacement, and submergence of flux~-- is key to driving eruptive
and explosive events in the corona and into the heliosphere. Until
recently, generally only line-of-sight magnetic field maps were
available for this work \citep[with its routine observations enabling
the study of increasingly large samples; one example of a large study is the work
by][who analyse the magnetic properties of $\sim$2,000 active regions
looking for signatures likely involved in
flaring]{2015A&A...579A..64A}. Nowadays, regular vector-magnetic
determinations from the observed polarization signals are available
from sources that include Hinode, SOLIS, and SDO/HMI. The sensitivity
of such vector field maps allows the detection of lasting changes in
the photospheric field when comparing pre- to post-flare observations
\citep[{\it e.g.} ][]{2012ApJ...745L..17W}. Temporal resolution is so good
that coronal events can be tightly bracketed to try to understand the
causes of flares and eruptions, as well as the changes in energy and
helicity involved; for example, \cite{2012ApJ...748...77S} analyze a
series of nonlinear force-free field models for the evolution in the
energy of an active region around the time of a major eruption.

The increasing availability of vector-magnetic data enables
statistical studies on the properties of active regions and their
activity heretofore possible only on line-of-sight magnetograms. For
example, \cite{2015ApJ...798..135B} use a data base of vector field
maps of over 2,000 active regions to train a machine-learning
algorithm to attempt forecasting of large
flares. \cite{2014ApJ...788..150S} review 3,000 vector field maps of
some 60 regions to compare estimates of free energy with flare
rates. \cite{2015GeoRL..42.5702T} analyze a sample of nearly 200
coronal mass ejections to reveal that flux, twist, and proxies for
free energy tend to set upper limits to the speed of CMEs emanating
from the active regions studied. \cite{2015PASJ...67....6O} review
hundreds of vector magnetograms of a sample of 80 active regions to
study helicity and twist parameters to test for the influence of
Coriolis forcing.

One long sought-after goal of flare and CME physics is to use models
of the solar atmospheric field to understand why field configurations
destabilize and under what conditions destabilization begins and
proceeds. In a ``meta-analysis'' review, \cite{2015SoPh..tmp...64S}
discuss recent developments, including the use of 3D field
extrapolations that suggest that topological structures (notably null
points and hyperbolic flux tubes) may be involved in the
triggering and generally as tracers of reconnection processes early
on.

But even as the availability of vector-magnetic field maps becomes
routine, the realization is growing that by themselves they appear
insufficient to provide generally adequate lower-boundary conditions
to understand the solar atmospheric activity. For example, a group of
modelers using a variety of non-linear force-free field algorithms
concludes in a series of studies \citep[see, {\it e.g.} ][and references
therein]{derosa+etal2008,2015SoPh..290.1159T,2015arXiv150805455D} that
snapshot vector-magnetic maps do not suffice to obtain a reliable
coronal field model with accurate energy or helicity measurements,
with effects of instrumental resolution and field of view, as well as
the model geometry (cartesian vs.\ spherical) all compounding the
problems. New developments include the use of coronal loop
observations to guide non-potential field models either from a single
perspective as is possible currently
\citep[{\it e.g.} ][]{2014ApJ...783..102M} or by using stereoscopic data
from existing or future space- based instrumentation such as STEREO,
Solar Orbiter, and SDO
\citep[{\it e.g.} ][]{derosa+etal2008,2015arXiv150604713A}. Methods to
follow the evolution of active-region fields based on data driving are
also being developed, using the uninterrupted stream of (vector)
magnetograms now available from space-based platforms
\citep[{\it e.g.} ][based on the MHD-like magnetofrictional
approximation]{2012ApJ...757..147C,2015SpWea..13..369F}, even reaching
up to global MHD field models from near the solar surface into the
heliosphere \citep{2015SoPh..290.1507H}. Others are developing MHD
methods to study CME initiation based on observed surface field
evolution \citep[{\it e.g.} ][]{2014Natur.514..465A}.  Fundamental
difficulties with these emerging methods include the difficulty in
measuring the transverse field in areas of relatively low flux
densities, the intrinsic 180-degree ambiguity in the field direction
given a magnitude for the transverse component, the need to
constrain the electric field to drive the model's evolution, and ultimately
the quantitative comparison with solar observables to determine the
verisimilitude of the model fields. 

Even as our ability to observe and process rapidly growing data volumes on
active region fields increases, we remain puzzled by the Sun's
atmospheric magnetic field: we have yet to understand how large
amounts of energy can sometimes be stored to eventually be explosively
converted into flares and CMEs, while in other cases the energy
is either not stored or is not explosively released. For a recent review of our
understanding of the magnetic field in the solar atmosphere, and the
variety of methods used to observe and study it, we refer to
\cite{2014A&ARv..22...78W}, and references therein.

\subsection{Coronal thermal structure and heating} 
The X-ray solar corona is made of complex arrays of magnetic flux
tubes anchored on both sides to the photosphere, confining a
relatively dense and hot plasma. This optically thin plasma is almost
fully ionized, with temperatures above 1\,MK, and emitting mostly
in the extreme UV to X-rays with intensities proportional to the square of
its density. Significant progress was made in studying the physical
and morphological features of coronal loops in a series of very
successful EUV and X-ray solar missions, since the first observational
evidence of the presence of coronal loops provided by rocket missions
in the mid-1960s \citep{Giacconi1965}.

In recent years, several solar missions were launched and started
producing spectacular images and data in different spectral
ranges. The high spatial and temporal resolution of these instruments
and the complementarity between these data sets, poses new challenges
to understand the heating and dynamics of coronal loops. The launch in
2010 of the Solar Dynamics Observatory
\citep[SDO,][]{2012SoPh..275....3P} allows continuous observation of
the whole Sun with high temporal and spatial resolution with AIA, EVE,
and HMI. In particular, AIA images span at least 1.3 solar diameters
in multiple wavelengths, at about 1\,arcsec in spatial resolution and
at a cadence of about 10\,seconds. Coronal loops observed by AIA have
been found to be highly variable and highly structured in space, time,
and temperature, challenging the traditional view of these loops as
isothermal structures and favoring the case for multi-thermal
cross-field temperature distributions.  Because the thermal
conductivity is severely reduced in the directions perpendicular to
the magnetic field, a spatially intermittent heating mechanism might
give rise to a multi-threaded structure in the internal structure of
loops. One possible example of such an intermittent heating process is
MHD turbulence, which is expected to produce fine scale structuring
within loops all the way beyond the resolution of current observations
\citep{Gomez1993,2014ApJ...787...87V}. A recent study by
\cite{Brooks2012} combining spectroscopic data from the EUV Imaging
Spectrometer \citep[EIS,][]{Culhane2007} aboard Hinode and SDO/AIA
images, shows that most of their loops must be composed of a number of
spatially unresolved threads.

More recently, the sounding rocket mission High-resolution Coronal
Imager \citep[Hi-C,][]{Cirtain2013}, achieved an unprecedented spatial
resolution 0.2\,arcsec in EUV images. \cite{2014ApJ...787L..10W} find that
the finely structured corona, down to the 0.2'' resolution, is
concentrated in the moss and in areas of sheared field, where the
heating is intense. This result suggests that heating is on smaller
spatial scales than AIA and that it could be sporadic. These results
are consistent with differential emission measure (DEM) analysis that
study the distribution of temperature across loops. \cite{Warren2012}
present a systematic study of the differential emission measure
distribution in active region cores, using data from EIS and the X-Ray
Telescope (XRT) aboard Hinode. Their results suggest that while the
hot active region emission might be close to equilibrium, warm active
regions may be dominated by evolving million degree loops in the
core. More recently, \cite{Schmelz2014} used XRT and EIS data as well
as images from SDO/AIA, and found that cooler loops tend to have
comparatively narrower DEM widths. While the DEM distribution of warm
loops could be explained through bundles of threads with different
temperatures, cooler loops are consistent with narrow DEMs and perhaps
even isothermal plasma. The authors then speculate that warm,
multi-thermal, multi-threaded loops might correspond to plasma being
heated, while cool loops are composed of threads which have had time
to cool to temperatures of about a million degrees, thus resembling a
single isothermal loop.

The Interface Region Imaging Spectrograph \citep[IRIS,][]{DePontieu2014}
was launched in 2013, and provides crucial information to understand
coronal heating, by tracing the flow of energy and plasma from the
chromosphere and transition region up to the corona. IRIS obtains high
resolution UV spectra and images with high spatial (0.33\,arcsec) and
temporal (1\,s) resolution. Recent IRIS observations show rather fast
variations (20-60\,s) of intensity and velocity on spatial scales
smaller than 500\,km at the foot points of hot coronal loops
\citep{Testa2014}. These observations were interpreted as the result
of heating by electron beams generated in small and impulsive heating
events (the so-called coronal nanoflares).

Theoretical models of coronal heating have been traditionally
classified into AC or DC, depending on the time scales involved in the
driving at the loop foot points: (a) AC or wave models, for which the
energy is provided by waves at the Sun's photosphere, with timescales
much faster than the time it takes an Alfv{\'e}n wave to cross the
loop; (b) DC or stress models, which assume that energy dissipation
takes place by magnetic stresses driven by slow foot point motions
(compared to the Alfv{\'e}n wave crossing time) at the Sun’s
photosphere. Although these scenarios seem mutually exclusive, two
common factors prevail: (i) the ultimate energy source is the kinetic
energy of the sub-photospheric velocity field, (ii) the existence of
fine scale structure is essential to speed up the dissipation
mechanisms invoked \citep{Gomez2011}. For a coronal heating mechanism
to be considered viable, the input energy must be compatible with
observed energy losses in active regions, estimated by
\cite{Withbroe1977} to be $\approx 1 \times 10^7 {\rm erg}\ {\rm
  cm}^{-2}\ {\rm s}^{-1}$. \cite{Welsch2015} used high-resolution
observations of plage magnetic fields made with the Solar Optical
Telescope aboard Hinode to estimate the vertical Poynting flux being
injected into the corona, obtaining values of about $\approx 5 \times
10^7 {\rm erg}\ {\rm cm}^{-2}\ {\rm s}^{-1}$, which suffices to heat
the plasma.

\subsection{Coronal seismology} 
Coronal seismology, like terrestrial seismology and traditional
helioseismology, provides a means of probing the background state of
the medium through which the waves propagate. Although the corona is
not hidden from view like the interiors of Earth and Sun, it is very
difficult to directly measure plasma and magnetic properties such as
density $\rho$, magnetic field $\mathbf{B}$, or transport
coefficients. Observations of oscillations in the coronal plasma can
potentially provide powerful constraints on these quantities, for
example through the Alfv{\'e}n velocity $\mathbf{B}/\sqrt{\mu\rho}$
\citep[\emph{e.g.}, see the \emph{Living Review} of][]{NakVer05aa}.

But unlike the solar interior and terrestrial examples where waves are
but perturbations on the background state, the coronal seismic waves
are crucially important to the energy balance of their host
medium. Coronal heating and solar wind acceleration are widely thought
to result at least in part from waves \citep{2010LRSP....7....4O}. An
overview of coronal seismology as of 2012 is provided by
\citet{De-Nak12aa}.

The last 5--8 years have seen an explosion in coronal wave studies due
to the advent of new instrumentation, such as the ground-based Coronal
Multichannel Polarimeter \citep[CoMP,][]{2008SoPh..247..411T} and the
space-based Atmospheric Imaging Assembly (AIA) on the Solar Dynamics
Observatory (SDO). Both have revealed ubiquitous Alfv{\'e}n-like (i.e.,
transverse to the magnetic field) coronal oscillations
\citep[][respectively]{TomMcIKei07aa,McIde-Car11aa}, though the
interpretation of their exact natures -- Alfv{\'e}n or kink -- is
controversial \citep{VanNakVer08aa}. The term ``Alfv{\'e}nic'' is
commonly used to encompass both wave types. In either case, the
increased resolution of the AIA observations \citep{WhiVer12aa} has
allowed the detection of sufficient oscillatory power to potentially
power both the corona and solar wind \citep{McIde-Car11aa}, though
precise mechanisms are not currently known with certainty. The
concentration of oscillatory power in the few-minute period band, and
in particular at around 5-minutes, strongly suggests a link with the
Sun's internal $p$-mode global oscillations. Consistency in the
estimation of Alfv{\'e}n speed between seismic techniques and magnetic
field extrapolation is confirmed by \citet{VerVanFou13aa} using AIA
data from a flare-induced coronal loop oscillation.

Time-Distance techniques applied to CoMP observations indicate a
preponderance of outward propagating waves over inward propagation,
even in closed loop structures, suggesting \emph{in situ} dissipation
(or mode conversion) on a timescale comparable to the Alfv{\'e}n
crossing time \citep{TomMcI09aa}.

Disentangling Alfv{\'e}n and kink waves is addressed in depth by
\citet{MatJesErd13aa}. In the magnetically structured solar
atmosphere, the only true Alfv{\'e}n wave is torsional, and there is
considerable interest in identifying these in observations because of
the amount of energy they could potentially contribute to the outer
atmosphere. However, being incompressive, Alfv{\'e}n waves are not seen
in intensity, and torsional Alfv{\'e}n waves are also difficult to
detect in Doppler \citep{De-Pas12aa,McIDe-12aa}. Recently though,
0.33-arcsec high-resolution observations of the chromosphere
and transition region (TR) with NASA's Interface Region Imaging
Spectrograph (IRIS), coordinated with the Swedish Solar Telescope,
have revealed widespread twisting motions across quiet Sun, coronal
holes, and active regions alike that often seem to be associated with
heating \citep{De-RouMcI14aa}. This must presumably extend to the
corona as well.

Dissipation of Alfv{\'e}n waves in the solar atmosphere has long been
thought to rely largely on the generation of Alfv{\'e}n turbulence
through nonlinear interaction between counter-propagating waves
\citep{Cravan05aa,vanAsgCra11aa}. Observations with CoMP showing
enhanced high-frequency power near the apex of coronal loops
\citep{LiuMcIDe-14aa,De-McIThr14aa} possibly supports this view.

The presumed acceleration of the solar wind by Alfv{\'e}n turbulent
energy deposition poses the challenge of identifying and explaining
counter-propagating Alfv{\'e}n waves in open magnetic field regions
required to produce that turbulence. \cite{MorTomPin15aa} confirm the
presence of these counter-propagating waves using CoMP. These
observations also provide evidence of a link to the $p$-mode spectrum,
which presumably relies on magneto\-acoustic-to-Alfv{\'e}n mode
conversion occurring in the lower atmosphere
\citep{CalHan11aa,HanCal12aa}.

\subsection{Flares}
\subsubsection{The flare's impact on the lower solar atmosphere}
The solar chromosphere is where most of the energy of a solar flare is
dissipated and radiated in bright linear structures called flare
ribbons, and recent work on this topic has been dominated by
observations from IRIS. The high spectral resolution available with
IRIS shows complex spectral line profiles in the Si IV transition
region line at 1394\AA\ and 1403\AA\, sometimes with two or three
time-varying gaussian components within one IRIS spatial pixel
\citep{2015ApJ...810....4B}.  The hot ``coronal'' line of Fe XXI
(1354\AA) on the other hand shows only a single strongly blue-shifted
component \citep{2015ApJ...810...45B,2015ApJ...807L..22G} originating
in compact ribbon sources at the beginning of a flare, with no hot
``stationary component'' present. The presence of this line also
demonstrates the high temperatures reached by the chromosphere in
flares as deduced from Hinode/EIS observations
\citep{2013ApJ...762..133B,2013ApJ...767...83G}. With EIS, we see
1.5-3\,MK redshifts \citep{2013ApJ...766..127Y} confirming earlier
reports, as well as significant non-thermal broadening. This means
that the momentum-conserving condensation front that is produced by
flare heating and paired with the evaporation flow contains hot
plasma. In turn this implies that the condensation front originates
relatively high in the chromosphere, otherwise such high temperatures
would not be possible in the standard (electron-beam-driven) model of
flares. This is somewhat at odds with recent measurements of element
abundance \citep{2014ApJ...786L...2W} which look more photospheric
than coronal, suggesting that up-flowing evaporated material comes
from low down in the chromosphere, below where the normal
fractionation by first ionization potential sets in.

Flare optical (or white-light -- WL) emission continues to be difficult to observe and
difficult to explain. Optical foot points characterized using 3-filter
observations with Hinode/SOT
\citep{2013ApJ...776..123W,2014ApJ...783...98K} could be explained by modest
temperature increases of the photospheric black body. The other main
proposed radiation mechanism is recombination emission, and flare
continuum in the near UV (beyond the Balmer edge) observed with IRIS
has an intensity consistent with this \citep{2014ApJ...794L..23H}, but
the tell-tale Balmer jump has not been seen. Co-spatial hard X-ray (HXR) and
white-light flare sources have been observed using RHESSI and SDO/HMI,
and require that both emissions are produced a few hundred kilometres
above the photosphere \citep{2012ApJ...753L..26M,2015ApJ...802...19K},
at a height corresponding to the temperature minimum region, and
beyond the range expected for HXR-emitting electrons arriving from the
corona (unless the chromosphere is under-dense compared to
expectations). It may be possible to generate optical emission from
the temperature minimum region, for example by modest heating due to
ion-neutral damping \citep{2013ApJ...765...81R}, but the presence of
non-thermal electrons in this plasma is harder to explain. Also
indicating the flare's impact on the dense lower atmosphere, there
have been many more reports of ``sun quakes'' ---~flare seismic
emission \citep[e.g.][]{2012SoPh..280..335A,2013SoPh..284..315Z}~---
but the mechanical driver remains uncertain; there are correspondences
with either HXR sources (pressure pulses from electron-beam driven
shocks) or magnetic transients (Lorentz forces) in some but not all
cases.  High-resolution ground-based flare observations using the New
Solar Telescope at BBSO show extraordinary fine structure, on a
sub-arcsecond scale, in flare ribbons and footprints
\citep{2013ApJ...769..112D,2014ApJ...788L..18S}, setting the scene for
future observations with DKIST.

\subsubsection{Magnetic-field evolution and energetics}
Examination of the changes in photospheric vector field occurring at
the flare impulsive phase by
\cite{2012ApJ...759...50P,2013SoPh..287..415P} using the HMI on SDO
suggests strong, permanent and abrupt variations in the vertical
component of the Lorentz force at the photosphere consistent with a
downward `collapse' of magnetic loops, and changes in the horizontal
component mostly parallel to the neutral line in opposite directions
on each side, indicating a decrease of shear near the neutral
line. The brighter the flare (as expressed in the GOES class), the larger are both the total
(area-integrated) change in the magnetic field and the change of
Lorentz force \citep{2012ApJ...757L...5W}, although
\cite{2014ApJ...788..150S} found that indicators of magnetic
non-potentiality (e.g. rotation, shear and helicity changes) are more
closely associated with flares ---~see for example the study of
flare-associated rotating sunspots by \cite{2012ApJ...761...60V}.  In
the corona, imaging and spectroscopic observations show compelling
evidence for plasma flows suggestive of those expected around a
coronal reconnection region
\citep[e.g.][]{2014ApJ...797L..14T,2013NatPh...9..489S}.

The growing number of observed flares in archives and the increased
coverage of flares in wavelength space and in domains from surface to
heliosphere is enabling an improved assessment of energy budgets. For
example, \cite{2012ApJ...759...71E} quantify energies of an ensemble
of large, eruptive flares to find, among others that it appears that
the energy in accelerated particles during the initial phases of the
flare suffice to supply the energy eventually radiated in the flare
across the spectrum, and that that total energy is statistically just
under the bulk kinetic energy in associated coronal mass ejections.

\subsubsection{Particle acceleration and transport}
The central problem in solar flare theory remains the acceleration of
the non-thermal electrons required to explain observed chromospheric
HXR sources. Observations with RHESSI and SDO show that coronal
electron acceleration can be very efficient;
\cite{2014ApJ...780..107K} find that essentially all electrons in a
coronal source of density a few times $10^9$\,cm$^{-3}$ are energized
to above around 10\,keV.  Kappa distributions, which are found to be a
better fit than a standard thermal plus non-thermal distribution in
coronal HXR sources \citep{2015ApJ...799..129O}, are shown to arise
naturally in an acceleration region when there is a balance between
diffusive acceleration and collisions, in the absence of significant
escape from the acceleration region \citep{2014ApJ...796..142B}. In the
electron-beam model of a flare, electrons must of course escape the
corona to produce the chromospheric HXR sources, and the number flux
requirements have always been somewhat problematic. This may be
alleviated if electrons are boosted by wave-particle interactions in
the corona; a quasi-linear simulation of coronal electron propagation
shows that wave-particle interaction with the high phase-velocity
Langmuir waves generated by density inhomogeneities can accelerate
beam electrons to higher energies, reducing the requirement on
electron flux at energies of a few tens of keV by up to a factor ten
\citep{2013A&A...550A..51H}. \cite{2014A&A...563A..51V} study a model
in which electrons are re-accelerated in the chromosphere, concluding
that this also reduces demands on putative electron beam fluxes
(however requirements on the flare chromosphere energy source, not
directly addressed in this model, remain the same). Chromospheric
(re-) acceleration models may also produce electron angular distributions
which are more isotropic, consistent with the angular distributions inferred from inversion of RHESSI mean electron flux spectra, accounting for photospheric X-ray albedo, by 
\cite{2013SoPh..284..405D}. A completely different view by
\cite{2014SoPh..289..881M} is that the electron energization in flares
takes place in a parallel electric field that develops close to or in
the chromosphere, in a region of anomalous resistivity, if energy is
transported Alfv\'enically ---~specifically by inertial Alfv\'en
waves. However, the orthodoxy remains that energy transport is by
electron beams, and this is now being tested against observations
using beam-driven radiation hydrodynamics codes, the output of which
can be compared with, for example, IRIS \citep{2015ApJ...804...56R}
and EVE and AIA data \citep{2015A&A...578A..72K}, so far with mixed
success.

\subsection{Coronal mass ejections}
The understanding of the initiation and evolution of coronal mass
ejections (CMEs) has tremendously profited from the combination of
coronagraphic observations with high cadence imaging in the EUV
together with the multi-perspective view provided by the STEREO
mission, as well as from increasingly sophisticated MHD and
thermodynamic modeling \cite[for reviews
see][]{2012LRSP....9....3W,2014IAUS..300..184A}.  Magnetic flux ropes
play a key role in the physics of CMEs. But there is a long debate
whether flux ropes are pre-existing or formed during the
eruption. \cite{2013ApJ...764..125P} observed the formation of a flux
rope during a confined flare in high-cadence SDO/AIA EUV
imagery. Within hours after its formation, the flux rope became
unstable and erupted resulting in a CME/flare event.
For other CMEs, for example those associated with quiescent prominence-cavity systems, a variety of observations indicate a pre-existing flux rope which may erupt bodily as a CME (see e.g. Figure 12 of \citealp{2015ASSL..415..323G}).
\cite{2013SoPh..284..179V} synthesized 16 years of coronagraphic and
EUV observations with MHD simulations, and found that flux ropes are a
common structure in CMEs; in at least 40\% a clear flux rope structure
could be identified.  In addition, they established a new
``two-front'' morphology consisting of a faint front followed by
diffuse emission and the bright CME leading edge. The faint front is
suggestive of a wave or shock front driven by the CME.

The high-cadence six-passband SDO/AIA EUV imagery allows to perform
differential emission measure (DEM) analysis on solar flares and CMEs
to study their multi-thermal dynamics \citep{2012A&A...539A.146H}. It
was shown that the CME core region, typically identified as the
embedded flux rope, is hot (8 -- 10 MK) indicative of magnetic
reconnection being involved. In contrast, the CME leading front has
temperatures similar to the pre-eruptive corona but of higher
densities suggesting that the front is a result of compression of the
ambient coronal plasma
\citep{2012ApJ...761...62C,2013A&A...553A..10H}. The hot flux rope is a key indicator to the physical processes involved in the early acceleration phase of the CME \citep{2012ApJ...758...60F,2013ApJ...769L..25C}.

The environment of CMEs is important for the development of non-radial
propagation. \cite{2012ApJ...744...66Z} report that during solar
minimum conditions CMEs originating from high latitudes can be easily
deflected toward the heliospheric current sheet, thus eventually
becoming geo-effective. \cite{2013SoPh..287..391P} showed that coronal
holes nearby the CME initiation site can cause strong deflections of
CMEs.

Modeling of the initiation of CMEs continues to provide insights
into the various forces and mechanisms that may be involved:
initiation may involve the kink instability
\citep{2012ApJ...746...67K}, sunspot rotation, reduction of tension of
the overlying field \citep{2013SoPh..286..453T}, torus instability
\citep{2014ApJ...789...46K}, and the breakout process
\citep{2012ApJ...760...81K,2013ApJ...764...87L}. Which dominates
under which conditions and how commonly these occur remain topics of
future work.

\subsection{Large-scale EUV waves}
Since their discovery by the SOHO/EIT instrument about 15 years ago, there
has been a vivid debate about the physical nature of large-scale EUV
waves, i.e.\ whether they
are true wave phenomena or propagating disturbances related to the
magnetic restructuring due to the erupting CME. In the recent years,
there has been tremendous progress in the understanding of these
intriguing phenomena thanks to the unprecedented observations
available, in particular the high-cadence EUV imagery in six
wavelengths bands by SDO/AIA combined with the STEREO multi-point view
which allowed for the first time to follow EUV waves in full-Sun maps
\citep{2012ApJ...756..143O}. There seems now relatively broad
consensus that large-scale EUV waves are often fast-mode magnetosonic
waves (of large amplitude or shocks), driven by the strong lateral
expansion of the CME \cite[see reviews
by][]{2012SoPh..281..187P,2014SoPh..289.3233L}. A number of detailed
case studies revealed that the CME lateral front and the EUV wave
appear originally co-spatial. But when the lateral CME expansion slows
down, the EUV wave decouples from the driver and then propagates
freely, adjusting to the local fast-mode speed of the medium
\citep[e.g.][]{2012ApJ...745L...5C,2012ApJ...756..143O}. Three-dimensional
thermodynamic MHD modeling of well observed EUV waves also supports
these findings, showing the outer fast-mode EUV wave front followed by
another bright front indicating the CME component
\citep{2012ApJ...750..134D}. Statistical studies of EUV waves based on
SDO/AIA \citep{2013ApJ...776...58N} and STEREO/EUVI data
\citep{2014SoPh..289.4563M} revealed EUV wave speeds that range from
close to the fast magnetosonic speed in the quiet corona to values
well above, the fastest ones exceeding
1000~km~s$^{-1}$. \cite{2014SoPh..289.4563M} showed that at least half
of the EUV waves under study show significant deceleration, and a
distinct anti-correlation between the starting speed and the
deceleration, providing further evidence for a freely propagating
fast-mode wave. The association rate of EUV waves with type II bursts,
which are indicative of shock waves in the solar corona, may be as
high as 50\% \citep{2013ApJ...776...58N}.
Detailed case studies provided a number of further characteristics
suggestive of the wave nature, such as reflection and refraction of
EUV waves at coronal holes and active regions, transmission into
coronal holes as well as the initiation of secondary waves by the
arrival of the wave at structures of high Alfv\'en speed
\citep{2012ApJ...746...13L,2012ApJ...756..143O,2012ApJ...754....7S,2013SoPh..286..201K} and for one case, \cite{2015ApJ...799..224L} have evaluated the EUV wave's initial energy using a blast-wave approximation, to be around 10\%\ of that of the associated CME. \cite{2012ApJ...753...52L}
discovered quasi-periodic fast-mode wave trains within a large-scale
EUV Wave with a periodicity of 2~min, running ahead of the laterally
expanding CME flanks. \cite{2012ApJ...745L..18A} presented the first
simultaneous observations of the propagation of a large-scale EUV wave
and an H$\alpha$ Moreton wave, showing that the wave fronts evolve
co-spatially indicating that they are both signatures of a fast
magnetosonic wave
pulse.

\subsection{CME evolution in the heliosphere}
The STEREO mission, often in combination with SOHO/LASCO, offers
observations of CMEs all the way from their origin on the Sun, and of
their propagation in interplanetary space to beyond 1\,AU from outside
the Sun-Earth line. These data combined with a multitude of other
in-situ space missions have been vividly used to connect remote
sensing CME observations to the field and plasma data observed by
in-situ spacecraft, to constrain models of interplanetary CME
propagation, to study CME-CME interaction, and to forecast CME arrival
times and speeds with the ultimate aim improving the prediction of
their geo-effectiveness.

\cite{2012ApJ...746...64H} and \cite {2013ApJ...769...43D} tracked a
flux rope all the way from its solar origin to its in-situ signatures
at 1\,AU using the STEREO SECCHI EUV imagers, coronagraphs and
wide-angle heliospheric imagers. They establish that the cavity in the
classic three-part CME is the feature that becomes the magnetic cloud,
implying material ahead of the cavity is piled-up material from the
corona or the solar wind.

Modeling of the interplanetary propagation of CMEs makes use of
empirical, analytic and numerical approaches.  The analytical
``drag-based model" (DBM) is based on the hypothesis that the Lorentz
forces driving a CME eruption ceases in the upper corona and that
beyond a certain distance the interplanetary CME (ICME) dynamics is
governed solely by the interaction of the ICME and the ambient solar
wind plasma \citep{2013SoPh..285..295V}. From the observational side,
a variety of reconstruction methods have been developed and applied to
the Heliospheric Imager data including one- as well as two-spacecraft
(stereoscopic) observations and inclusion of in-situ data and radio
type II bursts to better constrain the propagation direction, distance
and speed profile of CMEs in interplanetary space
\citep{2012SoPh..276..293R,2013SoPh..285..411M,2013JGRA..118.6866C,2013ApJ...769...45L}. These
efforts result in comparable typical uncertainties in the CME arrival
time of about half a day
\citep{2014ApJS..213...21V,2013SoPh..285..349L,2014ApJ...787..119M}.

Studies using STEREO Heliospheric Imager data and in-situ plasma and
field measurements established that the interaction of CMEs in the
inner heliosphere, due to a faster CME launched after a slower one,
seems to be a common and important phenomenon
\citep{2012ApJ...746L..15L,2012ApJ...750...45H,2012ApJ...759...68L,2012ApJ...749...57T,2012ApJ...758...10M,2013ApJ...769...45L,2014ApJ...785...85T}. The
interaction process may cause deflection or merging of CMEs, and
either deceleration or acceleration of merged CME fronts (including
heating and compression). \cite{2014NatCo...5E3481L}, reporting in a
fast CME causing an extreme storm, speculate that the interaction
between two successively launched CMEs resulted in the extreme
enhancement of the magnetic field of the ejecta that was observed
in-situ near 1\,AU.

%\subsubsection{Solar Energetic Particles} [heliospheric phenomenon?]

\nocite{2015ApJ...809..149A}
\nocite{2013SoPh..282..335B}
\nocite{2015ApJ...806..169B}
\nocite{2014JGRA..119..680C}
\nocite{2014SSRv..186...35C}
\nocite{2013ApJ...777L..29C}
\nocite{2014ApJ...789...35F}
\nocite{2015ApJ...808L..28J}
\nocite{2014SSRv..186..561K}
\nocite{2014ApJ...782...93H}
\nocite{2012ApJ...755L..22K}
\nocite{2013ApJ...778...11M}
\nocite{2012ApJ...753..146M}
\nocite{2013ApJ...762...73N}
\nocite{2014SoPh..289..441N}
\nocite{2014A&A...568A.113P}
\nocite{pavai_etal_2015}
\nocite{2013ApJ...778L..38S}
\nocite{2014ApJ...792..142U}
\nocite{2013ApJ...774L..29Z}
\subsection{Sun-in-time}
The unusually deep and temporally-extended activity minimum between
sunspot cycles 23 and 24, followed by a slowly rising and low
amplitude cycle 24, has led to renewed interest in the underlying
causes of solar cycle fluctuations, including Grand Minima. Much
attention has focused on the so-called Babcock-Leighton solar cycle
models, in which the regeneration of the solar surface dipole takes
place via the decay of active regions. Most extant versions of these
dynamo models are geometrically (axisymmetric) and dynamically
(kinematic) simplified, yet they do remarkably well at reproducing
many observed solar cycle characteristics (see, {\em e.g.}, Karak {\em et
al.} 2014, and references therein).  Explanations for the extended
cycle 23-24 minimum and low amplitude cycle 24 have been sought in
terms of variations in the meridional flow expected to thread the
solar convection zone (Upton and Hathaway 2014), and patterns of
active region emergence and associated feedback on surface flows
(Cameron {\em et al.} 2014; Jiang {\em et al.} 2015).  These successes of the
Babcock-Leighton modelling framework have however been challenged by
helioseismic measurements (Zhao {\em et al.} 2013; Schad {\em et al.} 2013)
indicating that the meridional flow within the convection zone has a
far more complex cellular structure than assumed in the majority of
these mean-field-like solar cycle models. Possible avenues out of this
conundrum are being explored (see, {\em e.g.}, Hazra {\em et al.} 2014; Belucz {\em et
al.} 2015).

Much effort has also been invested in implementing various form of
data assimilation schemes in dynamo models, with the aim of achieving
improved forecasting of the amplitude and timing of future sunspot
cycles.  At this point in time no existing dynamo model-based
forecasting scheme has done significantly better than the known
precursor skill of the solar surface magnetic dipole moment at times
of cycle minima, nonetheless progress is likely forthcoming in this
area.

Global magnetohydrodynamical simulations of solar convection have also
progressed rapidly in recent years, with many research groups
worldwide now running simulations producing large-scale magnetic
fields undergoing polarity reversals ({\em e.g.}  Masada {\em et al.} 2013; Nelson
{\em et al.} 2013; Fan and Fang 2014; Passos and Charbonneau 2014; Warnecke
{\em et al.} 2014).  Due to computing limitations all these simulations run
in parameter regimes still far removed from solar interior
conditions. Nonetheless, many are producing tantalizingly solar-like
features, including rotational torsional oscillations (Beaudoin {\em et
al.} 2013) equatorward propagation of activity ``belts'' (K\"apyl\"a {\em et
al.} 2012; Warnecke {\em et al.} 2014; Augustson {\em et al.} 2015) cyclic in-phase
magnetic modulation of convective energy transport (Cossette {\em et
al.} 2013) and Grand Minima-like interruptions of cyclic behavior
(Augustson {\em et al.} 2015).  One particularly interesting feature is the
spontaneous production of magnetic flux tube-like structures within
the convection zone, as reported in Nelson {\em et al.} (2013). These were
found to rise to the top of the simulation domain, partly through
magnetic buoyancy, while maintaining their orientation in a manner
compatible with Hale' polarity laws (Nelson {\em et al.} 2014). This has
revived the idea that dynamo action could be wholly contained within
the solar convective envelope, rather than relying on the tachocline
for the formation and storage of the magnetic flux ropes eventually
giving rise to sunspots.

Major efforts have also taken place in reinterpreting and reanalyzing
historical observations of magnetic activity. Noteworthy in this
respect are the analyses of tilt angle patterns for bipolar magnetic
regions (see Pavai {\em et al.} 2015, and references therein), and
reanalysis of polar faculae data by Munoz-Jaramillo {\em et al.} (2012). Of
particular importance is the recently completed revision of the
international sunspot number (SSN) time series.  SSN values for the
period 1947-present now account for a discontinuity in the manner of
counting spot groups having occurred at the Locarno reference station
(Clette {\em et al.} 2014).  Correcting for this leads to a significant
($\simeq 20\,$\%) decrease in SSN values during the space
era. Consequently, reconstructions of solar activity into the distant
past using the SSN as a backbone to extrapolate space-borne
measurements will need to be reassessed.

%\subsection{Long-term solar activity from cosmic-ray modulation}
Radionuclides generated by the atmospheric impact of galactic cosmic
rays (GCRs) provide a crucial proxy for the evolution in the Sun's
activity \citep{usoskin_2008,2013LRSP...10....3P} on time scales
longer than a few years or a decade, depending on the radionuclide and
its deposition in terrestrial natural archives. New ice core data on
$^{10}$Be and tree ring data on $^{14}$C have been combined to provide
better understanding of climate impacts on these records: a joint
analysis of composite tree ring data with ice cores from Greenland and
Antarctica have enabled the separation of the common signal (assumed
to be dominated by solar and heliospheric variability) from
terrestrial variability \citep{2012PNAS..109.5967S}. From this, we now
have 94 centuries of data on a proxy for solar activity. But
translating that proxy into details of solar activity that may drive
space weather and terrestrial climate remains a challenge, as reviewed
by, {\it e.g.}  \cite{2013ARA&A..51..311S}.

\subsection{Developments, discoveries, and surprises}
And then, of course, there were numerous surprising realizations and
discoveries, in the real world as much as in the rapidly growing
virtual world). We mention merely a small sampling in no particular
order: a weak solar cycle following an uncommonly low and long solar
minimum \citep{2013ApJ...779....2M}; a series of X-class flares from
AR\,12192 none of which were associated with a CME, contrasting with
statistics to date
\citep{2015ApJ...808L..24C,2015ApJ...804L..28S,2015ApJ...801L..23T};
use of a Sun-grazing comet to probe the high corona and its connection
to the innermost heliosphere
\citep{lovejoymhd2013,2014ApJ...788..152R}; an extremely large amount
of dense, cool plasma falling back onto the Sun following a massive
filament eruption providing a close-up example of distant accretion
processes \citep{2013Sci...341..251R}; reports of enormously energetic
flares from what would appear to be Sun-like stars
\citep{2014PASJ...66L...4N} and potential evidence for strong SEP
events associated with very energetic solar flaring from $^{14}$C
records albeit without obvious auroral counterparts
\citep{2012Natur.486..240M,2013A&A...552L...3U,2015AN....336..225N};
the successful creation of realistic looking sunspots in the computer
\citep{2014ApJ...785...90R}; the revision of sunspot numbers that
suggests no long-term increase in solar activity occurred over the
past few hundred years \citep{2014SSRv..186...35C}; radiative
magneto-convective simulations have reached resolution scales of a few
kilometers, and suggest comparable energy densities in magnetic and
kinetic reservoirs \citep{2014ApJ...789..132R}; IRIS observations
uncovering rapidly-evolving low-lying loops at transition region
temperatures, heretofore inferred from emission measure studies but
never yet observed \citep{2014Sci...346E.315H}; a new model was
proposed for coronal heating based on magnetic gradient pumping
\citep{2014ApJ...795..140T}; non-potential field models for a
continuously-driven corona over a 16-year period was achieved
\citep{2014SoPh..289..631Y}; a solar eruption in July of 2012 that
would likely have powered a century-level extreme geomagnetic storm,
as for the Carrington-Hodgson flare of 1859, had it enveloped Earth
\citep{2013SpWea..11..585B}; the realization that Stokes' theorem
combined with the induction equation could explain why polar fields
should be a good indicator for the strength of the next sunspot cycle
\citep{2015Sci...347.1333C}; the simulation of a sequence of
homologous CMEs and demonstration of so-called "canniballistic"
behavior \citep{2013ApJ...778L...8C}; the discovery of nested
toroidal line-of-sight flows and “lagomorphic” coronal polarimetric
signatures within coronal cavities indicating the presence of magnetic
flux ropes \citep{2013ApJ...770L..28B}; rapidly rotating magnetic
structures ("magnetic tornadoes") have been identified, which provide
a channel of energy and twist from the solar surface to the corona
\citep{2011ApJ...741L...7Z,2012Natur.486..505W}; and the X-class flare
SOL2014-03-29T made history by becoming ``the best-observed flare of
all time''
(\href{http://www.nasa.gov/content/goddard/nasa-telescopes-coordinate-best-ever-flare-observations/#.VhJ4TrQ9Yow}{according
  to NASA}) as the ground-based Dunn Solar Telescope and the
space-based IRIS, RHESSI, and SDO all observed it in detail.

\acknowledgement We gratefully acknowledge making use of NASA's
Astrophysics Data System in the writing of this report.

%\bibliographystyle{spr-mp-sola}
%\bibliography{ref_karel.bib,CoronalSeismology.bib,IAU_C10_gomez.bib,iaureport2015_fromPaulChar.bib,cm10.bib,yan.bib,cm10_flares.bib}

\begin{thebibliography}{201}
% BibTex style file: spr-mp-sola.bst (nameyear), 2015-03-09
\ifx\bisbn     \undefined \def\bisbn  #1{ISBN #1}\fi
\ifx\binits    \undefined \def\binits#1{#1}\fi
\ifx\bauthor   \undefined \def\bauthor#1{#1}\fi
\ifx\batitle   \undefined \def\batitle#1{#1}\fi
\ifx\bjtitle   \undefined \def\bjtitle#1{\textit{#1}}\fi
\ifx\bvolume   \undefined \def\bvolume#1{\textbf{#1}}\fi
\ifx\byear     \undefined \def\byear#1{#1}\fi
\ifx\bissue    \undefined \def\bissue#1{#1}\fi
\ifx\bfpage    \undefined \def\bfpage#1{#1}\fi
\ifx\blpage    \undefined \def\blpage #1{#1}\fi
\ifx\burl      \undefined \def\burl#1{\textsf{#1}}\fi
\ifx\href      \undefined \def\href#1#2{\textsf{#2}}\fi
\ifx\betal     \undefined \def\betal{\textit{et al.}}\fi
\ifx\bctitle   \undefined \def\bctitle#1{#1}\fi
\ifx\beditor   \undefined \def\beditor#1{#1}\fi
\ifx\bbtitle   \undefined \def\bbtitle#1{\textit{#1}}\fi
\ifx\bedition  \undefined \def\bedition#1{#1}\fi
\ifx\bseriesno \undefined \def\bseriesno#1{\textbf{#1}}\fi
\ifx\blocation \undefined \def\blocation#1{#1}\fi
\ifx\bsertitle \undefined \def\bsertitle#1{\textit{#1}}\fi
\ifx\bsnm      \undefined \def\bsnm#1{#1}\fi
\ifx\bsuffix   \undefined \def\bsuffix#1{#1}\fi
\ifx\bparticle \undefined \def\bparticle#1{#1}\fi
\ifx\barticle  \undefined \def\barticle#1{}\fi
\ifx\binstitute  \undefined \def\binstitute#1{#1}\fi
\ifx\bpublisher  \undefined \def\bpublisher#1{#1}\fi
\ifx\doiurl    \undefined
  \def\doiurl#1{\href{http://dx.doi.org/#1}{\textsf{DOI}}:#1}\fi
\ifx\arxivurl  \undefined
  \def\arxivurl#1{\href{http://arxiv.org/abs/#1}{\textsf{arXiv}}:#1}\fi
\ifx\adsurl    \undefined
  \def\adsurl#1{\href{http://adsabs.harvard.edu/abs/#1}{\textsf{ADS}}:#1}\fi
\ifx\botherref \undefined \def\botherref#1{}\fi
\ifx\url       \undefined \def\url#1{\textsf{#1}}\fi
\ifx\bchapter  \undefined \def\bchapter#1{}\fi
\ifx\bbook     \undefined \def\bbook#1{}\fi
\ifx\bcomment  \undefined \def\bcomment#1{#1}\fi
\ifx\oauthor   \undefined \def\oauthor#1{#1}\fi
\ifx\citeauthoryear \undefined\def \citeauthoryear#1{#1}\fi
\ifx\endbibitem\undefined \def\endbibitem{}\fi
\ifx\bconflocation  \undefined \def\bconflocation#1{#1} \fi

\bibitem[\protect\citeauthoryear{{Al-Ghraibah}, {Boucheron}, and
  {McAteer}}{2015}]{2015A&A...579A..64A}
\begin{barticle}
\bauthor{\bsnm{{Al-Ghraibah}}, \binits{A.}},
\bauthor{\bsnm{{Boucheron}}, \binits{L.E.}},
\bauthor{\bsnm{{McAteer}}, \binits{R.T.J.}}:
\byear{2015},
\batitle{{An automated classification approach to ranking photospheric proxies
  of magnetic energy build-up}}.
\bjtitle{A{\&}A}
\bvolume{579},
\bfpage{A64}.
\doiurl{10.1051/0004-6361/201525978}.
\end{barticle}
\endbibitem

\bibitem[\protect\citeauthoryear{{Alvarado-G{\'o}mez}
  \textit{et~al.}}{2012}]{2012SoPh..280..335A}
\begin{barticle}
\bauthor{\bsnm{{Alvarado-G{\'o}mez}}, \binits{J.D.}},
\bauthor{\bsnm{{Buitrago-Casas}}, \binits{J.C.}},
\bauthor{\bsnm{{Mart{\'{\i}}nez-Oliveros}}, \binits{J.C.}},
\bauthor{\bsnm{{Lindsey}}, \binits{C.}},
\bauthor{\bsnm{{Hudson}}, \binits{H.}},
\bauthor{\bsnm{{Calvo-Mozo}}, \binits{B.}}:
\byear{2012},
\batitle{{Magneto-Acoustic Energetics Study of the Seismically Active Flare of
  15 February 2011}}.
\bjtitle{Solar Phys.}
\bvolume{280},
\bfpage{335}.
\doiurl{10.1007/s11207-012-0009-6}.
\end{barticle}
\endbibitem

\bibitem[\protect\citeauthoryear{{Amari}, {Canou}, and
  {Aly}}{2014}]{2014Natur.514..465A}
\begin{barticle}
\bauthor{\bsnm{{Amari}}, \binits{T.}},
\bauthor{\bsnm{{Canou}}, \binits{A.}},
\bauthor{\bsnm{{Aly}}, \binits{J.-J.}}:
\byear{2014},
\batitle{{Characterizing and predicting the magnetic environment leading to
  solar eruptions}}.
\bjtitle{Nature}
\bvolume{514},
\bfpage{465}.
\doiurl{10.1038/nature13815}.
\end{barticle}
\endbibitem

\bibitem[\protect\citeauthoryear{{Asai}
  \textit{et~al.}}{2012}]{2012ApJ...745L..18A}
\begin{barticle}
\bauthor{\bsnm{{Asai}}, \binits{A.}},
\bauthor{\bsnm{{Ishii}}, \binits{T.T.}},
\bauthor{\bsnm{{Isobe}}, \binits{H.}},
\bauthor{\bsnm{{Kitai}}, \binits{R.}},
\bauthor{\bsnm{{Ichimoto}}, \binits{K.}},
\bauthor{\bsnm{{UeNo}}, \binits{S.}},
\bauthor{\bsnm{{Nagata}}, \binits{S.}},
\bauthor{\bsnm{{Morita}}, \binits{S.}},
\bauthor{\bsnm{{Nishida}}, \binits{K.}},
\bauthor{\bsnm{{Shiota}}, \binits{D.}},
\bauthor{\bsnm{{Oi}}, \binits{A.}},
\bauthor{\bsnm{{Akioka}}, \binits{M.}},
\bauthor{\bsnm{{Shibata}}, \binits{K.}}:
\byear{2012},
\batitle{{First Simultaneous Observation of an H{$\alpha$} Moreton Wave, EUV
  Wave, and Filament/Prominence Oscillations}}.
\bjtitle{Astrophys. J., Lett.}
\bvolume{745},
\bfpage{L18}.
\doiurl{10.1088/2041-8205/745/2/L18}.
\end{barticle}
\endbibitem

\bibitem[\protect\citeauthoryear{{Aschwanden}, {Schrijver}, and
  {Malanushenko}}{2015}]{2015arXiv150604713A}
\begin{botherref}
\oauthor{\bsnm{{Aschwanden}}, \binits{M.J.}},
\oauthor{\bsnm{{Schrijver}}, \binits{C.J.}},
\oauthor{\bsnm{{Malanushenko}}, \binits{A.}}:
2015,
{Blind Stereoscopy of the Coronal Magnetic Field}.
\textit{ArXiv e-prints}.
\end{botherref}
\endbibitem

\bibitem[\protect\citeauthoryear{{Augustson}
  \textit{et~al.}}{2015}]{2015ApJ...809..149A}
\begin{barticle}
\bauthor{\bsnm{{Augustson}}, \binits{K.}},
\bauthor{\bsnm{{Brun}}, \binits{A.S.}},
\bauthor{\bsnm{{Miesch}}, \binits{M.}},
\bauthor{\bsnm{{Toomre}}, \binits{J.}}:
\byear{2015},
\batitle{{Grand Minima and Equatorward Propagation in a Cycling Stellar
  Convective Dynamo}}.
\bjtitle{Astrophys. J.}
\bvolume{809},
\bfpage{149}.
\doiurl{10.1088/0004-637X/809/2/149}.
\end{barticle}
\endbibitem

\bibitem[\protect\citeauthoryear{{Aulanier}}{2014}]{2014IAUS..300..184A}
\begin{bchapter}
\bauthor{\bsnm{{Aulanier}}, \binits{G.}}:
\byear{2014},
\bctitle{{The physical mechanisms that initiate and drive solar eruptions}}.
In: \beditor{\bsnm{{Schmieder}}, \binits{B.}},
\beditor{\bsnm{{Malherbe}}, \binits{J.-M.}},
\beditor{\bsnm{{Wu}}, \binits{S.T.}} (eds.)
\bbtitle{IAU Symposium},
\bsertitle{IAU Symposium}
\bseriesno{300},
\bfpage{184}.
\doiurl{10.1017/S1743921313010958}.
\end{bchapter}
\endbibitem

\bibitem[\protect\citeauthoryear{{Baker}
  \textit{et~al.}}{2013}]{2013SpWea..11..585B}
\begin{barticle}
\bauthor{\bsnm{{Baker}}, \binits{D.N.}},
\bauthor{\bsnm{{Li}}, \binits{X.}},
\bauthor{\bsnm{{Pulkkinen}}, \binits{A.}},
\bauthor{\bsnm{{Ngwira}}, \binits{C.M.}},
\bauthor{\bsnm{{Mays}}, \binits{M.L.}},
\bauthor{\bsnm{{Galvin}}, \binits{A.B.}},
\bauthor{\bsnm{{Simunac}}, \binits{K.D.C.}}:
\byear{2013},
\batitle{{A major solar eruptive event in July 2012: Defining extreme space
  weather scenarios}}.
\bjtitle{Space Weather}
\bvolume{11},
\bfpage{585}.
\doiurl{10.1002/swe.20097}.
\end{barticle}
\endbibitem

\bibitem[\protect\citeauthoryear{{Bastian}
  \textit{et~al.}}{2015}]{2015IAUGA..2257295B}
\begin{barticle}
\bauthor{\bsnm{{Bastian}}, \binits{T.S.}},
\bauthor{\bsnm{{Barta}}, \binits{M.}},
\bauthor{\bsnm{{Brajsa}}, \binits{R.}},
\bauthor{\bsnm{{Chen}}, \binits{B.}},
\bauthor{\bsnm{{De Pontieu}}, \binits{B.}},
\bauthor{\bsnm{{Fleishman}}, \binits{G.}},
\bauthor{\bsnm{{Gary}}, \binits{D.}},
\bauthor{\bsnm{{Hales}}, \binits{A.}},
\bauthor{\bsnm{{Hills}}, \binits{R.}},
\bauthor{\bsnm{{Hudson}}, \binits{H.}},
\bauthor{\bsnm{{Iwai}}, \binits{K.}},
\bauthor{\bsnm{{Shimojo}}, \binits{M.}},
\bauthor{\bsnm{{White}}, \binits{S.}},
\bauthor{\bsnm{{Wedemeyer}}, \binits{S.}},
\bauthor{\bsnm{{Yan}}, \binits{Y.}}:
\byear{2015},
\batitle{{The Atacama Large Millimeter/Submillimeter Array: a New Asset for
  Solar and Heliospheric Physics}}.
\bjtitle{IAU General Assembly}
\bvolume{22},
\bfpage{57295}.
\end{barticle}
\endbibitem

\bibitem[\protect\citeauthoryear{{B{\c a}k-St{\c e}{\'s}licka}
  \textit{et~al.}}{2013}]{2013ApJ...770L..28B}
\begin{barticle}
\bauthor{\bsnm{{B{\c a}k-St{\c e}{\'s}licka}}, \binits{U.}},
\bauthor{\bsnm{{Gibson}}, \binits{S.E.}},
\bauthor{\bsnm{{Fan}}, \binits{Y.}},
\bauthor{\bsnm{{Bethge}}, \binits{C.}},
\bauthor{\bsnm{{Forland}}, \binits{B.}},
\bauthor{\bsnm{{Rachmeler}}, \binits{L.A.}}:
\byear{2013},
\batitle{{The Magnetic Structure of Solar Prominence Cavities: New
  Observational Signature Revealed by Coronal Magnetometry}}.
\bjtitle{Astrophys. J., Lett.}
\bvolume{770},
\bfpage{L28}.
\doiurl{10.1088/2041-8205/770/2/L28}.
\end{barticle}
\endbibitem

\bibitem[\protect\citeauthoryear{{Beaudoin}
  \textit{et~al.}}{2013}]{2013SoPh..282..335B}
\begin{barticle}
\bauthor{\bsnm{{Beaudoin}}, \binits{P.}},
\bauthor{\bsnm{{Charbonneau}}, \binits{P.}},
\bauthor{\bsnm{{Racine}}, \binits{E.}},
\bauthor{\bsnm{{Smolarkiewicz}}, \binits{P.K.}}:
\byear{2013},
\batitle{{Torsional Oscillations in a Global Solar Dynamo}}.
\bjtitle{Solar Phys.}
\bvolume{282},
\bfpage{335}.
\doiurl{10.1007/s11207-012-0150-2}.
\end{barticle}
\endbibitem

\bibitem[\protect\citeauthoryear{{Belucz}, {Dikpati}, and
  {Forg{\'a}cs-Dajka}}{2015}]{2015ApJ...806..169B}
\begin{barticle}
\bauthor{\bsnm{{Belucz}}, \binits{B.}},
\bauthor{\bsnm{{Dikpati}}, \binits{M.}},
\bauthor{\bsnm{{Forg{\'a}cs-Dajka}}, \binits{E.}}:
\byear{2015},
\batitle{{A Babcock-Leighton Solar Dynamo Model with Multi-cellular Meridional
  Circulation in Advection- and Diffusion-dominated Regimes}}.
\bjtitle{Astrophys. J.}
\bvolume{806},
\bfpage{169}.
\doiurl{10.1088/0004-637X/806/2/169}.
\end{barticle}
\endbibitem

\bibitem[\protect\citeauthoryear{{Bian}
  \textit{et~al.}}{2014}]{2014ApJ...796..142B}
\begin{barticle}
\bauthor{\bsnm{{Bian}}, \binits{N.H.}},
\bauthor{\bsnm{{Emslie}}, \binits{A.G.}},
\bauthor{\bsnm{{Stackhouse}}, \binits{D.J.}},
\bauthor{\bsnm{{Kontar}}, \binits{E.P.}}:
\byear{2014},
\batitle{{The Formation of Kappa-distribution Accelerated Electron Populations
  in Solar Flares}}.
\bjtitle{Astrophys. J.}
\bvolume{796},
\bfpage{142}.
\doiurl{10.1088/0004-637X/796/2/142}.
\end{barticle}
\endbibitem

\bibitem[\protect\citeauthoryear{{Bobra} and
  {Couvidat}}{2015}]{2015ApJ...798..135B}
\begin{barticle}
\bauthor{\bsnm{{Bobra}}, \binits{M.G.}},
\bauthor{\bsnm{{Couvidat}}, \binits{S.}}:
\byear{2015},
\batitle{{Solar Flare Prediction Using SDO/HMI Vector Magnetic Field Data with
  a Machine-learning Algorithm}}.
\bjtitle{Astrophys. J.}
\bvolume{798},
\bfpage{135}.
\doiurl{10.1088/0004-637X/798/2/135}.
\end{barticle}
\endbibitem

\bibitem[\protect\citeauthoryear{{Brannon}, {Longcope}, and
  {Qiu}}{2015}]{2015ApJ...810....4B}
\begin{barticle}
\bauthor{\bsnm{{Brannon}}, \binits{S.R.}},
\bauthor{\bsnm{{Longcope}}, \binits{D.W.}},
\bauthor{\bsnm{{Qiu}}, \binits{J.}}:
\byear{2015},
\batitle{{Spectroscopic Observations of an Evolving Flare Ribbon Substructure
  Suggesting Origin in Current Sheet Waves}}.
\bjtitle{Astrophys. J.}
\bvolume{810},
\bfpage{4}.
\doiurl{10.1088/0004-637X/810/1/4}.
\end{barticle}
\endbibitem

\bibitem[\protect\citeauthoryear{Brooks, Warren, and
  Ugarte-Urra}{2012}]{Brooks2012}
\begin{barticle}
\bauthor{\bsnm{Brooks}, \binits{D.H.}},
\bauthor{\bsnm{Warren}, \binits{H.P.}},
\bauthor{\bsnm{Ugarte-Urra}, \binits{I.}}:
\byear{2012},
\batitle{Solar coronal loops resolved by hinode and the solar dynamics
  observatory}.
\bjtitle{The Astrophysical Journal Letters}
\bvolume{755}(\bissue{2}),
\bfpage{L33}.
\bisbn{2041-8205}.
\burl{http://stacks.iop.org/2041-8205/755/i=2/a=L33}.
\end{barticle}
\endbibitem

\bibitem[\protect\citeauthoryear{{Brosius}}{2013}]{2013ApJ...762..133B}
\begin{barticle}
\bauthor{\bsnm{{Brosius}}, \binits{J.W.}}:
\byear{2013},
\batitle{{Chromospheric Evaporation in Solar Flare Loop Strands Observed with
  the Extreme-ultraviolet Imaging Spectrometer on Board Hinode}}.
\bjtitle{Astrophys. J.}
\bvolume{762},
\bfpage{133}.
\doiurl{10.1088/0004-637X/762/2/133}.
\end{barticle}
\endbibitem

\bibitem[\protect\citeauthoryear{{Brosius} and
  {Daw}}{2015}]{2015ApJ...810...45B}
\begin{barticle}
\bauthor{\bsnm{{Brosius}}, \binits{J.W.}},
\bauthor{\bsnm{{Daw}}, \binits{A.N.}}:
\byear{2015},
\batitle{{Quasi-periodic Fluctuations and Chromospheric Evaporation in a Solar
  Flare Ribbon Observed by IRIS}}.
\bjtitle{Astrophys. J.}
\bvolume{810},
\bfpage{45}.
\doiurl{10.1088/0004-637X/810/1/45}.
\end{barticle}
\endbibitem

\bibitem[\protect\citeauthoryear{{Cally} and {Hansen}}{2011}]{CalHan11aa}
\begin{barticle}
\bauthor{\bsnm{{Cally}}, \binits{P.S.}},
\bauthor{\bsnm{{Hansen}}, \binits{S.C.}}:
\byear{2011},
\batitle{{Benchmarking Fast-to-Alfv{\'e}n Mode Conversion in a Cold
  Magneto\-hydrodynamic Plasma}}.
\bjtitle{Astrophys. J.}
\bvolume{738},
\bfpage{119}.
\doiurl{10.1088/0004-637X/738/2/119}.
\end{barticle}
\endbibitem

\bibitem[\protect\citeauthoryear{{Cameron} and
  {Sch{\"u}ssler}}{2015}]{2015Sci...347.1333C}
\begin{barticle}
\bauthor{\bsnm{{Cameron}}, \binits{R.}},
\bauthor{\bsnm{{Sch{\"u}ssler}}, \binits{M.}}:
\byear{2015},
\batitle{{The crucial role of surface magnetic fields for the solar dynamo}}.
\bjtitle{Science}
\bvolume{347},
\bfpage{1333}.
\doiurl{10.1126/science.1261470}.
\end{barticle}
\endbibitem

\bibitem[\protect\citeauthoryear{{Cameron}
  \textit{et~al.}}{2014}]{2014JGRA..119..680C}
\begin{barticle}
\bauthor{\bsnm{{Cameron}}, \binits{R.H.}},
\bauthor{\bsnm{{Jiang}}, \binits{J.}},
\bauthor{\bsnm{{Sch{\"u}ssler}}, \binits{M.}},
\bauthor{\bsnm{{Gizon}}, \binits{L.}}:
\byear{2014},
\batitle{{Physical causes of solar cycle amplitude variability}}.
\bjtitle{Journal of Geophysical Research (Space Physics)}
\bvolume{119},
\bfpage{680}.
\doiurl{10.1002/2013JA019498}.
\end{barticle}
\endbibitem

\bibitem[\protect\citeauthoryear{{Chatterjee} and
  {Fan}}{2013}]{2013ApJ...778L...8C}
\begin{barticle}
\bauthor{\bsnm{{Chatterjee}}, \binits{P.}},
\bauthor{\bsnm{{Fan}}, \binits{Y.}}:
\byear{2013},
\batitle{{Simulation of Homologous and Cannibalistic Coronal Mass Ejections
  produced by the Emergence of a Twisted Flux Rope into the Solar Corona}}.
\bjtitle{Astrophys. J., Lett.}
\bvolume{778},
\bfpage{L8}.
\doiurl{10.1088/2041-8205/778/1/L8}.
\end{barticle}
\endbibitem

\bibitem[\protect\citeauthoryear{{Chen}
  \textit{et~al.}}{2013}]{2013ApJ...763L..21C}
\begin{barticle}
\bauthor{\bsnm{{Chen}}, \binits{B.}},
\bauthor{\bsnm{{Bastian}}, \binits{T.S.}},
\bauthor{\bsnm{{White}}, \binits{S.M.}},
\bauthor{\bsnm{{Gary}}, \binits{D.E.}},
\bauthor{\bsnm{{Perley}}, \binits{R.}},
\bauthor{\bsnm{{Rupen}}, \binits{M.}},
\bauthor{\bsnm{{Carlson}}, \binits{B.}}:
\byear{2013},
\batitle{{Tracing Electron Beams in the Sun's Corona with Radio Dynamic Imaging
  Spectroscopy}}.
\bjtitle{Astrophys. J., Lett.}
\bvolume{763},
\bfpage{L21}.
\doiurl{10.1088/2041-8205/763/1/L21}.
\end{barticle}
\endbibitem

\bibitem[\protect\citeauthoryear{{Chen}
  \textit{et~al.}}{2015}]{2015ApJ...808L..24C}
\begin{barticle}
\bauthor{\bsnm{{Chen}}, \binits{H.}},
\bauthor{\bsnm{{Zhang}}, \binits{J.}},
\bauthor{\bsnm{{Ma}}, \binits{S.}},
\bauthor{\bsnm{{Yang}}, \binits{S.}},
\bauthor{\bsnm{{Li}}, \binits{L.}},
\bauthor{\bsnm{{Huang}}, \binits{X.}},
\bauthor{\bsnm{{Xiao}}, \binits{J.}}:
\byear{2015},
\batitle{{Confined Flares in Solar Active Region 12192 from 2014 October 18 to
  29}}.
\bjtitle{Astrophys. J., Lett.}
\bvolume{808},
\bfpage{L24}.
\doiurl{10.1088/2041-8205/808/1/L24}.
\end{barticle}
\endbibitem

\bibitem[\protect\citeauthoryear{{Cheng}
  \textit{et~al.}}{2012a}]{2012ApJ...761...62C}
\begin{barticle}
\bauthor{\bsnm{{Cheng}}, \binits{X.}},
\bauthor{\bsnm{{Zhang}}, \binits{J.}},
\bauthor{\bsnm{{Saar}}, \binits{S.H.}},
\bauthor{\bsnm{{Ding}}, \binits{M.D.}}:
\byear{2012}a,
\batitle{{Differential Emission Measure Analysis of Multiple Structural
  Components of Coronal Mass Ejections in the Inner Corona}}.
\bjtitle{Astrophys. J.}
\bvolume{761},
\bfpage{62}.
\doiurl{10.1088/0004-637X/761/1/62}.
\end{barticle}
\endbibitem

\bibitem[\protect\citeauthoryear{{Cheng}
  \textit{et~al.}}{2012b}]{2012ApJ...745L...5C}
\begin{barticle}
\bauthor{\bsnm{{Cheng}}, \binits{X.}},
\bauthor{\bsnm{{Zhang}}, \binits{J.}},
\bauthor{\bsnm{{Olmedo}}, \binits{O.}},
\bauthor{\bsnm{{Vourlidas}}, \binits{A.}},
\bauthor{\bsnm{{Ding}}, \binits{M.D.}},
\bauthor{\bsnm{{Liu}}, \binits{Y.}}:
\byear{2012}b,
\batitle{{Investigation of the Formation and Separation of an
  Extreme-ultraviolet Wave from the Expansion of a Coronal Mass Ejection}}.
\bjtitle{Astrophys. J., Lett.}
\bvolume{745},
\bfpage{L5}.
\doiurl{10.1088/2041-8205/745/1/L5}.
\end{barticle}
\endbibitem

\bibitem[\protect\citeauthoryear{{Cheng}
  \textit{et~al.}}{2013}]{2013ApJ...769L..25C}
\begin{barticle}
\bauthor{\bsnm{{Cheng}}, \binits{X.}},
\bauthor{\bsnm{{Zhang}}, \binits{J.}},
\bauthor{\bsnm{{Ding}}, \binits{M.D.}},
\bauthor{\bsnm{{Olmedo}}, \binits{O.}},
\bauthor{\bsnm{{Sun}}, \binits{X.D.}},
\bauthor{\bsnm{{Guo}}, \binits{Y.}},
\bauthor{\bsnm{{Liu}}, \binits{Y.}}:
\byear{2013},
\batitle{{Investigating Two Successive Flux Rope Eruptions in a Solar Active
  Region}}.
\bjtitle{Astrophys. J., Lett.}
\bvolume{769},
\bfpage{L25}.
\doiurl{10.1088/2041-8205/769/2/L25}.
\end{barticle}
\endbibitem

\bibitem[\protect\citeauthoryear{{Cheung} and
  {DeRosa}}{2012}]{2012ApJ...757..147C}
\begin{barticle}
\bauthor{\bsnm{{Cheung}}, \binits{M.C.M.}},
\bauthor{\bsnm{{DeRosa}}, \binits{M.L.}}:
\byear{2012},
\batitle{{A Method for Data-driven Simulations of Evolving Solar Active
  Regions}}.
\bjtitle{Astrophys. J.}
\bvolume{757},
\bfpage{147}.
\doiurl{10.1088/0004-637X/757/2/147}.
\end{barticle}
\endbibitem

\bibitem[\protect\citeauthoryear{Cirtain \textit{et~al.}}{2013}]{Cirtain2013}
\begin{barticle}
\bauthor{\bsnm{Cirtain}, \binits{J.W.}},
\bauthor{\bsnm{Golub}, \binits{L.}},
\bauthor{\bsnm{Winebarger}, \binits{A.R.}},
\bauthor{\bsnm{De~Pontieu}, \binits{B.}},
\bauthor{\bsnm{Kobayashi}, \binits{K.}},
\bauthor{\bsnm{Moore}, \binits{R.L.}},
\bauthor{\bsnm{Walsh}, \binits{R.W.}},
\bauthor{\bsnm{Korreck}, \binits{K.E.}},
\bauthor{\bsnm{Weber}, \binits{M.}},
\bauthor{\bsnm{McCauley}, \binits{P.}},
\bauthor{\bsnm{Title}, \binits{A.}},
\bauthor{\bsnm{Kuzin}, \binits{S.}},
\bauthor{\bsnm{DeForest}, \binits{C.E.}}:
\byear{2013},
\batitle{Energy release in the solar corona from spatially resolved magnetic
  braids}.
\bjtitle{Nature}
\bvolume{493}(\bissue{7433}),
\bfpage{501}.
\bisbn{0028-0836}.
\burl{http://dx.doi.org/10.1038/nature11772}.
\end{barticle}
\endbibitem

\bibitem[\protect\citeauthoryear{{Clette}
  \textit{et~al.}}{2014}]{2014SSRv..186...35C}
\begin{barticle}
\bauthor{\bsnm{{Clette}}, \binits{F.}},
\bauthor{\bsnm{{Svalgaard}}, \binits{L.}},
\bauthor{\bsnm{{Vaquero}}, \binits{J.M.}},
\bauthor{\bsnm{{Cliver}}, \binits{E.W.}}:
\byear{2014},
\batitle{{Revisiting the Sunspot Number. A 400-Year Perspective on the Solar
  Cycle}}.
\bjtitle{Space Sci. Rev.}
\bvolume{186},
\bfpage{35}.
\doiurl{10.1007/s11214-014-0074-2}.
\end{barticle}
\endbibitem

\bibitem[\protect\citeauthoryear{{Colaninno}, {Vourlidas}, and
  {Wu}}{2013}]{2013JGRA..118.6866C}
\begin{barticle}
\bauthor{\bsnm{{Colaninno}}, \binits{R.C.}},
\bauthor{\bsnm{{Vourlidas}}, \binits{A.}},
\bauthor{\bsnm{{Wu}}, \binits{C.C.}}:
\byear{2013},
\batitle{{Quantitative comparison of methods for predicting the arrival of
  coronal mass ejections at Earth based on multiview imaging}}.
\bjtitle{Journal of Geophysical Research (Space Physics)}
\bvolume{118},
\bfpage{6866}.
\doiurl{10.1002/2013JA019205}.
\end{barticle}
\endbibitem

\bibitem[\protect\citeauthoryear{{Cossette}, {Charbonneau}, and
  {Smolarkiewicz}}{2013}]{2013ApJ...777L..29C}
\begin{barticle}
\bauthor{\bsnm{{Cossette}}, \binits{J.-F.}},
\bauthor{\bsnm{{Charbonneau}}, \binits{P.}},
\bauthor{\bsnm{{Smolarkiewicz}}, \binits{P.K.}}:
\byear{2013},
\batitle{{Cyclic Thermal Signature in a Global MHD Simulation of Solar
  Convection}}.
\bjtitle{Astrophys. J., Lett.}
\bvolume{777},
\bfpage{L29}.
\doiurl{10.1088/2041-8205/777/2/L29}.
\end{barticle}
\endbibitem

\bibitem[\protect\citeauthoryear{{Cranmer} and {van
  Ballegooijen}}{2005}]{Cravan05aa}
\begin{barticle}
\bauthor{\bsnm{{Cranmer}}, \binits{S.R.}},
\bauthor{\bsnm{{van Ballegooijen}}, \binits{A.A.}}:
\byear{2005},
\batitle{{On the Generation, Propagation, and Reflection of Alfv{\'e}n Waves
  from the Solar Photosphere to the Distant Heliosphere}}.
\bjtitle{Astrophys. J., Suppl. Ser.}
\bvolume{156},
\bfpage{265}.
\doiurl{10.1086/426507}.
\end{barticle}
\endbibitem

\bibitem[\protect\citeauthoryear{Culhane \textit{et~al.}}{2007}]{Culhane2007}
\begin{barticle}
\bauthor{\bsnm{Culhane}, \binits{L.}},
\bauthor{\bsnm{Harra}, \binits{L.K.}},
\bauthor{\bsnm{Baker}, \binits{D.}},
\bauthor{\bparticle{van} \bsnm{Driel-Gesztelyi}, \binits{L.}},
\bauthor{\bsnm{Sun}, \binits{J.}},
\bauthor{\bsnm{Doschek}, \binits{G.A.}},
\bauthor{\bsnm{Brooks}, \binits{D.H.}},
\bauthor{\bsnm{Lundquist}, \binits{L.L.}},
\bauthor{\bsnm{Kamio}, \binits{S.}},
\bauthor{\bsnm{Young}, \binits{P.R.}},
\bauthor{\bsnm{Hansteen}, \binits{V.H.}}:
\byear{2007},
\batitle{Hinode euv study of jets in the sun’s south polar corona}.
\bjtitle{Publications of the Astronomical Society of Japan}
\bvolume{59}(\bissue{sp3}),
\bfpage{S751}.
\doiurl{10.1093/pasj/59.sp3.S751}.
\burl{http://pasj.oxfordjournals.org/content/59/sp3/S751.abstract}.
\end{barticle}
\endbibitem

\bibitem[\protect\citeauthoryear{{De Moortel} and
  {Nakariakov}}{2012}]{De-Nak12aa}
\begin{barticle}
\bauthor{\bsnm{{De Moortel}}, \binits{I.}},
\bauthor{\bsnm{{Nakariakov}}, \binits{V.M.}}:
\byear{2012},
\batitle{{Magnetohydrodynamic waves and coronal seismology: an overview of
  recent results}}.
\bjtitle{Royal Society of London Philosophical Transactions Series A}
\bvolume{370},
\bfpage{3193}.
\doiurl{10.1098/rsta.2011.0640}.
\end{barticle}
\endbibitem

\bibitem[\protect\citeauthoryear{{De Moortel} and {Pascoe}}{2012}]{De-Pas12aa}
\begin{barticle}
\bauthor{\bsnm{{De Moortel}}, \binits{I.}},
\bauthor{\bsnm{{Pascoe}}, \binits{D.J.}}:
\byear{2012},
\batitle{{The Effects of Line-of-sight Integration on Multistrand Coronal Loop
  Oscillations}}.
\bjtitle{Astrophys. J.}
\bvolume{746},
\bfpage{31}.
\doiurl{10.1088/0004-637X/746/1/31}.
\end{barticle}
\endbibitem

\bibitem[\protect\citeauthoryear{{De Moortel}
  \textit{et~al.}}{2014}]{De-McIThr14aa}
\begin{barticle}
\bauthor{\bsnm{{De Moortel}}, \binits{I.}},
\bauthor{\bsnm{{McIntosh}}, \binits{S.W.}},
\bauthor{\bsnm{{Threlfall}}, \binits{J.}},
\bauthor{\bsnm{{Bethge}}, \binits{C.}},
\bauthor{\bsnm{{Liu}}, \binits{J.}}:
\byear{2014},
\batitle{{Potential Evidence for the Onset of Alfv{\'e}nic Turbulence in
  Trans-equatorial Coronal Loops}}.
\bjtitle{Astrophys. J., Lett.}
\bvolume{782},
\bfpage{L34}.
\doiurl{10.1088/2041-8205/782/2/L34}.
\end{barticle}
\endbibitem

\bibitem[\protect\citeauthoryear{De~Pontieu
  \textit{et~al.}}{2014}]{DePontieu2014}
\begin{botherref}
\oauthor{\bsnm{De~Pontieu}, \binits{B.}},
\oauthor{\bparticle{Rouppe van~der} \bsnm{Voort}, \binits{L.}},
\oauthor{\bsnm{McIntosh}, \binits{S.W.}},
\oauthor{\bsnm{Pereira}, \binits{T.M.D.}},
\oauthor{\bsnm{Carlsson}, \binits{M.}},
\oauthor{\bsnm{Hansteen}, \binits{V.}},
\oauthor{\bsnm{Skogsrud}, \binits{H.}},
\oauthor{\bsnm{Lemen}, \binits{J.}},
\oauthor{\bsnm{Title}, \binits{A.}},
\oauthor{\bsnm{Boerner}, \binits{P.}},
\oauthor{\bsnm{Hurlburt}, \binits{N.}},
\oauthor{\bsnm{Tarbell}, \binits{T.D.}},
\oauthor{\bsnm{Wuelser}, \binits{J.P.}},
\oauthor{\bsnm{De~Luca}, \binits{E.E.}},
\oauthor{\bsnm{Golub}, \binits{L.}},
\oauthor{\bsnm{McKillop}, \binits{S.}},
\oauthor{\bsnm{Reeves}, \binits{K.}},
\oauthor{\bsnm{Saar}, \binits{S.}},
\oauthor{\bsnm{Testa}, \binits{P.}},
\oauthor{\bsnm{Tian}, \binits{H.}},
\oauthor{\bsnm{Kankelborg}, \binits{C.}},
\oauthor{\bsnm{Jaeggli}, \binits{S.}},
\oauthor{\bsnm{Kleint}, \binits{L.}},
\oauthor{\bsnm{Martinez-Sykora}, \binits{J.}}:
2014,
On the prevalence of small-scale twist in the solar chromosphere and transition
  region.
\textit{Science}
\textbf{346}(6207).
\end{botherref}
\endbibitem

\bibitem[\protect\citeauthoryear{{De Pontieu}
  \textit{et~al.}}{2014a}]{De-RouMcI14aa}
\begin{barticle}
\bauthor{\bsnm{{De Pontieu}}, \binits{B.}},
\bauthor{\bsnm{{Rouppe van der Voort}}, \binits{L.}},
\bauthor{\bsnm{{McIntosh}}, \binits{S.W.}},
\bauthor{\bsnm{{Pereira}}, \binits{T.M.D.}},
\bauthor{\bsnm{{Carlsson}}, \binits{M.}},
\bauthor{\bsnm{{Hansteen}}, \binits{V.}},
\bauthor{\bsnm{{Skogsrud}}, \binits{H.}},
\bauthor{\bsnm{{Lemen}}, \binits{J.}},
\bauthor{\bsnm{{Title}}, \binits{A.}},
\bauthor{\bsnm{{Boerner}}, \binits{P.}},
\bauthor{\bsnm{{Hurlburt}}, \binits{N.}},
\bauthor{\bsnm{{Tarbell}}, \binits{T.D.}},
\bauthor{\bsnm{{Wuelser}}, \binits{J.P.}},
\bauthor{\bsnm{{De Luca}}, \binits{E.E.}},
\bauthor{\bsnm{{Golub}}, \binits{L.}},
\bauthor{\bsnm{{McKillop}}, \binits{S.}},
\bauthor{\bsnm{{Reeves}}, \binits{K.}},
\bauthor{\bsnm{{Saar}}, \binits{S.}},
\bauthor{\bsnm{{Testa}}, \binits{P.}},
\bauthor{\bsnm{{Tian}}, \binits{H.}},
\bauthor{\bsnm{{Kankelborg}}, \binits{C.}},
\bauthor{\bsnm{{Jaeggli}}, \binits{S.}},
\bauthor{\bsnm{{Kleint}}, \binits{L.}},
\bauthor{\bsnm{{Martinez-Sykora}}, \binits{J.}}:
\byear{2014}a,
\batitle{{On the prevalence of small-scale twist in the solar chromosphere and
  transition region}}.
\bjtitle{Science}
\bvolume{346},
\bfpage{1255732}.
\doiurl{10.1126/science.1255732}.
\end{barticle}
\endbibitem

\bibitem[\protect\citeauthoryear{{De Pontieu}
  \textit{et~al.}}{2014b}]{2014SoPh..289.2733D}
\begin{barticle}
\bauthor{\bsnm{{De Pontieu}}, \binits{B.}},
\bauthor{\bsnm{{Title}}, \binits{A.M.}},
\bauthor{\bsnm{{Lemen}}, \binits{J.R.}},
\bauthor{\bsnm{{Kushner}}, \binits{G.D.}},
\bauthor{\bsnm{{Akin}}, \binits{D.J.}},
\bauthor{\bsnm{{Allard}}, \binits{B.}},
\bauthor{\bsnm{{Berger}}, \binits{T.}},
\bauthor{\bsnm{{Boerner}}, \binits{P.}},
\bauthor{\bsnm{{Cheung}}, \binits{M.}},
\bauthor{\bsnm{{Chou}}, \binits{C.}},
\bauthor{\bsnm{{Drake}}, \binits{J.F.}},
\bauthor{\bsnm{{Duncan}}, \binits{D.W.}},
\bauthor{\bsnm{{Freeland}}, \binits{S.}},
\bauthor{\bsnm{{Heyman}}, \binits{G.F.}},
\bauthor{\bsnm{{Hoffman}}, \binits{C.}},
\bauthor{\bsnm{{Hurlburt}}, \binits{N.E.}},
\bauthor{\bsnm{{Lindgren}}, \binits{R.W.}},
\bauthor{\bsnm{{Mathur}}, \binits{D.}},
\bauthor{\bsnm{{Rehse}}, \binits{R.}},
\bauthor{\bsnm{{Sabolish}}, \binits{D.}},
\bauthor{\bsnm{{Seguin}}, \binits{R.}},
\bauthor{\bsnm{{Schrijver}}, \binits{C.J.}},
\bauthor{\bsnm{{Tarbell}}, \binits{T.D.}},
\bauthor{\bsnm{{W{\"u}lser}}, \binits{J.-P.}},
\bauthor{\bsnm{{Wolfson}}, \binits{C.J.}},
\bauthor{\bsnm{{Yanari}}, \binits{C.}},
\bauthor{\bsnm{{Mudge}}, \binits{J.}},
\bauthor{\bsnm{{Nguyen-Phuc}}, \binits{N.}},
\bauthor{\bsnm{{Timmons}}, \binits{R.}},
\bauthor{\bsnm{{van Bezooijen}}, \binits{R.}},
\bauthor{\bsnm{{Weingrod}}, \binits{I.}},
\bauthor{\bsnm{{Brookner}}, \binits{R.}},
\bauthor{\bsnm{{Butcher}}, \binits{G.}},
\bauthor{\bsnm{{Dougherty}}, \binits{B.}},
\bauthor{\bsnm{{Eder}}, \binits{J.}},
\bauthor{\bsnm{{Knagenhjelm}}, \binits{V.}},
\bauthor{\bsnm{{Larsen}}, \binits{S.}},
\bauthor{\bsnm{{Mansir}}, \binits{D.}},
\bauthor{\bsnm{{Phan}}, \binits{L.}},
\bauthor{\bsnm{{Boyle}}, \binits{P.}},
\bauthor{\bsnm{{Cheimets}}, \binits{P.N.}},
\bauthor{\bsnm{{DeLuca}}, \binits{E.E.}},
\bauthor{\bsnm{{Golub}}, \binits{L.}},
\bauthor{\bsnm{{Gates}}, \binits{R.}},
\bauthor{\bsnm{{Hertz}}, \binits{E.}},
\bauthor{\bsnm{{McKillop}}, \binits{S.}},
\bauthor{\bsnm{{Park}}, \binits{S.}},
\bauthor{\bsnm{{Perry}}, \binits{T.}},
\bauthor{\bsnm{{Podgorski}}, \binits{W.A.}},
\bauthor{\bsnm{{Reeves}}, \binits{K.}},
\bauthor{\bsnm{{Saar}}, \binits{S.}},
\bauthor{\bsnm{{Testa}}, \binits{P.}},
\bauthor{\bsnm{{Tian}}, \binits{H.}},
\bauthor{\bsnm{{Weber}}, \binits{M.}},
\bauthor{\bsnm{{Dunn}}, \binits{C.}},
\bauthor{\bsnm{{Eccles}}, \binits{S.}},
\bauthor{\bsnm{{Jaeggli}}, \binits{S.A.}},
\bauthor{\bsnm{{Kankelborg}}, \binits{C.C.}},
\bauthor{\bsnm{{Mashburn}}, \binits{K.}},
\bauthor{\bsnm{{Pust}}, \binits{N.}},
\bauthor{\bsnm{{Springer}}, \binits{L.}},
\bauthor{\bsnm{{Carvalho}}, \binits{R.}},
\bauthor{\bsnm{{Kleint}}, \binits{L.}},
\bauthor{\bsnm{{Marmie}}, \binits{J.}},
\bauthor{\bsnm{{Mazmanian}}, \binits{E.}},
\bauthor{\bsnm{{Pereira}}, \binits{T.M.D.}},
\bauthor{\bsnm{{Sawyer}}, \binits{S.}},
\bauthor{\bsnm{{Strong}}, \binits{J.}},
\bauthor{\bsnm{{Worden}}, \binits{S.P.}},
\bauthor{\bsnm{{Carlsson}}, \binits{M.}},
\bauthor{\bsnm{{Hansteen}}, \binits{V.H.}},
\bauthor{\bsnm{{Leenaarts}}, \binits{J.}},
\bauthor{\bsnm{{Wiesmann}}, \binits{M.}},
\bauthor{\bsnm{{Aloise}}, \binits{J.}},
\bauthor{\bsnm{{Chu}}, \binits{K.-C.}},
\bauthor{\bsnm{{Bush}}, \binits{R.I.}},
\bauthor{\bsnm{{Scherrer}}, \binits{P.H.}},
\bauthor{\bsnm{{Brekke}}, \binits{P.}},
\bauthor{\bsnm{{Martinez-Sykora}}, \binits{J.}},
\bauthor{\bsnm{{Lites}}, \binits{B.W.}},
\bauthor{\bsnm{{McIntosh}}, \binits{S.W.}},
\bauthor{\bsnm{{Uitenbroek}}, \binits{H.}},
\bauthor{\bsnm{{Okamoto}}, \binits{T.J.}},
\bauthor{\bsnm{{Gummin}}, \binits{M.A.}},
\bauthor{\bsnm{{Auker}}, \binits{G.}},
\bauthor{\bsnm{{Jerram}}, \binits{P.}},
\bauthor{\bsnm{{Pool}}, \binits{P.}},
\bauthor{\bsnm{{Waltham}}, \binits{N.}}:
\byear{2014}b,
\batitle{{The Interface Region Imaging Spectrograph (IRIS)}}.
\bjtitle{Solar Phys.}
\bvolume{289},
\bfpage{2733}.
\doiurl{10.1007/s11207-014-0485-y}.
\end{barticle}
\endbibitem

\bibitem[\protect\citeauthoryear{{DeForest}, {Howard}, and
  {McComas}}{2013}]{2013ApJ...769...43D}
\begin{barticle}
\bauthor{\bsnm{{DeForest}}, \binits{C.E.}},
\bauthor{\bsnm{{Howard}}, \binits{T.A.}},
\bauthor{\bsnm{{McComas}}, \binits{D.J.}}:
\byear{2013},
\batitle{{Tracking Coronal Features from the Low Corona to Earth: A
  Quantitative Analysis of the 2008 December 12 Coronal Mass Ejection}}.
\bjtitle{Astrophys. J.}
\bvolume{769},
\bfpage{43}.
\doiurl{10.1088/0004-637X/769/1/43}.
\end{barticle}
\endbibitem

\bibitem[\protect\citeauthoryear{{Deng}
  \textit{et~al.}}{2013}]{2013ApJ...769..112D}
\begin{barticle}
\bauthor{\bsnm{{Deng}}, \binits{N.}},
\bauthor{\bsnm{{Tritschler}}, \binits{A.}},
\bauthor{\bsnm{{Jing}}, \binits{J.}},
\bauthor{\bsnm{{Chen}}, \binits{X.}},
\bauthor{\bsnm{{Liu}}, \binits{C.}},
\bauthor{\bsnm{{Reardon}}, \binits{K.}},
\bauthor{\bsnm{{Denker}}, \binits{C.}},
\bauthor{\bsnm{{Xu}}, \binits{Y.}},
\bauthor{\bsnm{{Wang}}, \binits{H.}}:
\byear{2013},
\batitle{{High-cadence and High-resolution H{$\alpha$} Imaging Spectroscopy of
  a Circular Flare's Remote Ribbon with IBIS}}.
\bjtitle{Astrophys. J.}
\bvolume{769},
\bfpage{112}.
\doiurl{10.1088/0004-637X/769/2/112}.
\end{barticle}
\endbibitem

\bibitem[\protect\citeauthoryear{{DeRosa}
  \textit{et~al.}}{2015}]{2015arXiv150805455D}
\begin{botherref}
\oauthor{\bsnm{{DeRosa}}, \binits{M.L.}},
\oauthor{\bsnm{{Wheatland}}, \binits{M.S.}},
\oauthor{\bsnm{{Leka}}, \binits{K.D.}},
\oauthor{\bsnm{{Barnes}}, \binits{G.}},
\oauthor{\bsnm{{Amari}}, \binits{T.}},
\oauthor{\bsnm{{Canou}}, \binits{A.}},
\oauthor{\bsnm{{Gilchrist}}, \binits{S.A.}},
\oauthor{\bsnm{{Thalmann}}, \binits{J.K.}},
\oauthor{\bsnm{{Valori}}, \binits{G.}},
\oauthor{\bsnm{{Wiegelmann}}, \binits{T.}},
\oauthor{\bsnm{{Schrijver}}, \binits{C.J.}},
\oauthor{\bsnm{{Malanushenko}}, \binits{A.}},
\oauthor{\bsnm{{Sun}}, \binits{X.}},
\oauthor{\bsnm{{R{\'e}gnier}}, \binits{S.}}:
2015,
{The Influence of Spatial Resolution on Nonlinear Force-Free Modeling}.
\textit{Solar Phys.}
in press.
\end{botherref}
\endbibitem

\bibitem[\protect\citeauthoryear{DeRosa
  \textit{et~al.}}{2009}]{derosa+etal2008}
\begin{barticle}
\bauthor{\bsnm{DeRosa}, \binits{M.L.}},
\bauthor{\bsnm{Schrijver}, \binits{C.J.}},
\bauthor{\bsnm{Barnes}, \binits{G.}},
\bauthor{\bsnm{Leka}, \binits{K.D.}},
\bauthor{\bsnm{Lites}, \binits{B.W.}},
\bauthor{\bsnm{Aschwanden}, \binits{M.J.}},
\bauthor{\bsnm{Amari}, \binits{T.}},
\bauthor{\bsnm{Canou}, \binits{A.}},
\bauthor{\bsnm{McTiernan}, \binits{J.M.}},
\bauthor{\bsnm{Regnier}, \binits{S.}},
\bauthor{\bsnm{Thalmann}, \binits{J.K.}},
\bauthor{\bsnm{Valori}, \binits{G.}},
\bauthor{\bsnm{Wheatland}, \binits{M.S.}},
\bauthor{\bsnm{Wiegelmann}, \binits{T.}},
\bauthor{\bsnm{Cheung}, \binits{M.C.M.}},
\bauthor{\bsnm{Conlon}, \binits{P.A.}},
\bauthor{\bsnm{Fuhrmann}, \binits{M.}},
\bauthor{\bsnm{Inhester}, \binits{B.}},
\bauthor{\bsnm{Tadesse}, \binits{T.}}:
\byear{2009},
\batitle{{A critical assessment of the feasibility of nonlinear force-free
  field modeling of the solar corona}}.
\bjtitle{Astrophys. J.}
\bvolume{696},
\bfpage{1780}.
\end{barticle}
\endbibitem

\bibitem[\protect\citeauthoryear{{Dickson} and
  {Kontar}}{2013}]{2013SoPh..284..405D}
\begin{barticle}
\bauthor{\bsnm{{Dickson}}, \binits{E.C.M.}},
\bauthor{\bsnm{{Kontar}}, \binits{E.P.}}:
\byear{2013},
\batitle{{Measurements of Electron Anisotropy in Solar Flares Using Albedo with
  RHESSI X-Ray Data}}.
\bjtitle{Solar Phys.}
\bvolume{284},
\bfpage{405}.
\doiurl{10.1007/s11207-012-0178-3}.
\end{barticle}
\endbibitem

\bibitem[\protect\citeauthoryear{{Downs}
  \textit{et~al.}}{2012}]{2012ApJ...750..134D}
\begin{barticle}
\bauthor{\bsnm{{Downs}}, \binits{C.}},
\bauthor{\bsnm{{Roussev}}, \binits{I.I.}},
\bauthor{\bsnm{{van der Holst}}, \binits{B.}},
\bauthor{\bsnm{{Lugaz}}, \binits{N.}},
\bauthor{\bsnm{{Sokolov}}, \binits{I.V.}}:
\byear{2012},
\batitle{{Understanding SDO/AIA Observations of the 2010 June 13 EUV Wave
  Event: Direct Insight from a Global Thermodynamic MHD Simulation}}.
\bjtitle{Astrophys. J.}
\bvolume{750},
\bfpage{134}.
\end{barticle}
\endbibitem

\bibitem[\protect\citeauthoryear{{Downs}
  \textit{et~al.}}{2013}]{lovejoymhd2013}
\begin{barticle}
\bauthor{\bsnm{{Downs}}, \binits{C.}},
\bauthor{\bsnm{{Linker}}, \binits{J.A.}},
\bauthor{\bsnm{{Miki{\'c}}}, \binits{Z.}},
\bauthor{\bsnm{{Riley}}, \binits{P.}},
\bauthor{\bsnm{{Schrijver}}, \binits{C.J.}},
\bauthor{\bsnm{{Saint-Hilaire}}, \binits{P.}}:
\byear{2013},
\batitle{{Probing the Solar Magnetic Field with a Sun-Grazing Comet}}.
\bjtitle{Science}
\bvolume{340},
\bfpage{1196}.
\doiurl{10.1126/science.1236550}.
\end{barticle}
\endbibitem

\bibitem[\protect\citeauthoryear{{Emslie}
  \textit{et~al.}}{2012}]{2012ApJ...759...71E}
\begin{barticle}
\bauthor{\bsnm{{Emslie}}, \binits{A.G.}},
\bauthor{\bsnm{{Dennis}}, \binits{B.R.}},
\bauthor{\bsnm{{Shih}}, \binits{A.Y.}},
\bauthor{\bsnm{{Chamberlin}}, \binits{P.C.}},
\bauthor{\bsnm{{Mewaldt}}, \binits{R.A.}},
\bauthor{\bsnm{{Moore}}, \binits{C.S.}},
\bauthor{\bsnm{{Share}}, \binits{G.H.}},
\bauthor{\bsnm{{Vourlidas}}, \binits{A.}},
\bauthor{\bsnm{{Welsch}}, \binits{B.T.}}:
\byear{2012},
\batitle{{Global Energetics of Thirty-eight Large Solar Eruptive Events}}.
\bjtitle{Astrophys. J.}
\bvolume{759},
\bfpage{71}.
\doiurl{10.1088/0004-637X/759/1/71}.
\end{barticle}
\endbibitem

\bibitem[\protect\citeauthoryear{{Fan}}{2012}]{2012ApJ...758...60F}
\begin{barticle}
\bauthor{\bsnm{{Fan}}, \binits{Y.}}:
\byear{2012},
\batitle{{Thermal Signatures of Tether-cutting Reconnections in Pre-eruption
  Coronal Flux Ropes: Hot Central Voids in Coronal Cavities}}.
\bjtitle{Astrophys. J.}
\bvolume{758},
\bfpage{60}.
\doiurl{10.1088/0004-637X/758/1/60}.
\end{barticle}
\endbibitem

\bibitem[\protect\citeauthoryear{{Fan} and {Fang}}{2014}]{2014ApJ...789...35F}
\begin{barticle}
\bauthor{\bsnm{{Fan}}, \binits{Y.}},
\bauthor{\bsnm{{Fang}}, \binits{F.}}:
\byear{2014},
\batitle{{A Simulation of Convective Dynamo in the Solar Convective Envelope:
  Maintenance of the Solar-like Differential Rotation and Emerging Flux}}.
\bjtitle{Astrophys. J.}
\bvolume{789},
\bfpage{35}.
\doiurl{10.1088/0004-637X/789/1/35}.
\end{barticle}
\endbibitem

\bibitem[\protect\citeauthoryear{{Fisher}
  \textit{et~al.}}{2015}]{2015SpWea..13..369F}
\begin{barticle}
\bauthor{\bsnm{{Fisher}}, \binits{G.H.}},
\bauthor{\bsnm{{Abbett}}, \binits{W.P.}},
\bauthor{\bsnm{{Bercik}}, \binits{D.J.}},
\bauthor{\bsnm{{Kazachenko}}, \binits{M.D.}},
\bauthor{\bsnm{{Lynch}}, \binits{B.J.}},
\bauthor{\bsnm{{Welsch}}, \binits{B.T.}},
\bauthor{\bsnm{{Hoeksema}}, \binits{J.T.}},
\bauthor{\bsnm{{Hayashi}}, \binits{K.}},
\bauthor{\bsnm{{Liu}}, \binits{Y.}},
\bauthor{\bsnm{{Norton}}, \binits{A.A.}},
\bauthor{\bsnm{{Dalda}}, \binits{A.S.}},
\bauthor{\bsnm{{Sun}}, \binits{X.}},
\bauthor{\bsnm{{DeRosa}}, \binits{M.L.}},
\bauthor{\bsnm{{Cheung}}, \binits{M.C.M.}}:
\byear{2015},
\batitle{{The Coronal Global Evolutionary Model: Using HMI Vector Magnetogram
  and Doppler Data to Model the Buildup of Free Magnetic Energy in the Solar
  Corona}}.
\bjtitle{Space Weather}
\bvolume{13},
\bfpage{369}.
\doiurl{10.1002/2015SW001191}.
\end{barticle}
\endbibitem

\bibitem[\protect\citeauthoryear{{Giacconi}
  \textit{et~al.}}{1965}]{Giacconi1965}
\begin{barticle}
\bauthor{\bsnm{{Giacconi}}, \binits{R.}},
\bauthor{\bsnm{{Reidy}}, \binits{W.P.}},
\bauthor{\bsnm{{Zehnpfennig}}, \binits{T.}},
\bauthor{\bsnm{{Lindsay}}, \binits{J.C.}},
\bauthor{\bsnm{{Muney}}, \binits{W.S.}}:
\byear{1965},
\batitle{{Solar X-Ray Image Obtained Using Grazing-Incidence Optics.}}
\bjtitle{Astrophys. J.}
\bvolume{142},
\bfpage{1274}.
\doiurl{10.1086/148404}.
\end{barticle}
\endbibitem

\bibitem[\protect\citeauthoryear{{Gibson}}{2015}]{2015ASSL..415..323G}
\begin{bchapter}
\bauthor{\bsnm{{Gibson}}, \binits{S.}}:
\byear{2015},
\bctitle{{Coronal Cavities: Observations and Implications for the Magnetic
  Environment of Prominences}}.
In: \beditor{\bsnm{{Vial}}, \binits{J.-C.}},
\beditor{\bsnm{{Engvold}}, \binits{O.}} (eds.)
\bbtitle{Astrophysics and Space Science Library},
\bsertitle{Astrophysics and Space Science Library}
\bseriesno{415},
\bfpage{323}.
\end{bchapter}
\endbibitem

\bibitem[\protect\citeauthoryear{G{\'o}mez}{2010}]{Gomez2011}
\begin{bchapter}
\bauthor{\bsnm{G{\'o}mez}, \binits{D.O.}}:
\byear{2010},
\bctitle{Heating of coronal active regions}.
In: \bbtitle{The Physics of Sun and Star Spots},
\bsertitle{Proceedings of the International Astronomical Union}
\bseriesno{6},
\bfpage{44}.
\doiurl{10.1017/S1743921311014980}.
\end{bchapter}
\endbibitem

\bibitem[\protect\citeauthoryear{{G{\'o}mez}, {Martens}, and
  {Golub}}{1993}]{Gomez1993}
\begin{barticle}
\bauthor{\bsnm{{G{\'o}mez}}, \binits{D.O.}},
\bauthor{\bsnm{{Martens}}, \binits{P.C.H.}},
\bauthor{\bsnm{{Golub}}, \binits{L.}}:
\byear{1993},
\batitle{{Normal incidence X-ray telescope power spectra of X-ray emission from
  solar active regions. I - Observations. II - Theory}}.
\bjtitle{Astrophys. J.}
\bvolume{405},
\bfpage{767}.
\doiurl{10.1086/172405}.
\end{barticle}
\endbibitem

\bibitem[\protect\citeauthoryear{{Goode} and {Cao}}{2012}]{2012SPIE.8444E..03G}
\begin{bchapter}
\bauthor{\bsnm{{Goode}}, \binits{P.R.}},
\bauthor{\bsnm{{Cao}}, \binits{W.}}:
\byear{2012},
\bctitle{{The 1.6 m off-axis New Solar Telescope (NST) in Big Bear}}.
In: \bbtitle{Society of Photo-Optical Instrumentation Engineers (SPIE)
  Conference Series},
\bsertitle{Society of Photo-Optical Instrumentation Engineers (SPIE) Conference
  Series}
\bseriesno{8444},
\bfpage{3}.
\doiurl{10.1117/12.925494}.
\end{bchapter}
\endbibitem

\bibitem[\protect\citeauthoryear{{Graham} and
  {Cauzzi}}{2015}]{2015ApJ...807L..22G}
\begin{barticle}
\bauthor{\bsnm{{Graham}}, \binits{D.R.}},
\bauthor{\bsnm{{Cauzzi}}, \binits{G.}}:
\byear{2015},
\batitle{{Temporal Evolution of Multiple Evaporating Ribbon Sources in a Solar
  Flare}}.
\bjtitle{Astrophys. J., Lett.}
\bvolume{807},
\bfpage{L22}.
\doiurl{10.1088/2041-8205/807/2/L22}.
\end{barticle}
\endbibitem

\bibitem[\protect\citeauthoryear{{Graham}
  \textit{et~al.}}{2013}]{2013ApJ...767...83G}
\begin{barticle}
\bauthor{\bsnm{{Graham}}, \binits{D.R.}},
\bauthor{\bsnm{{Hannah}}, \binits{I.G.}},
\bauthor{\bsnm{{Fletcher}}, \binits{L.}},
\bauthor{\bsnm{{Milligan}}, \binits{R.O.}}:
\byear{2013},
\batitle{{The Emission Measure Distribution of Impulsive Phase Flare
  Footpoints}}.
\bjtitle{Astrophys. J.}
\bvolume{767},
\bfpage{83}.
\doiurl{10.1088/0004-637X/767/1/83}.
\end{barticle}
\endbibitem

\bibitem[\protect\citeauthoryear{{Hannah} and
  {Kontar}}{2012}]{2012A&A...539A.146H}
\begin{barticle}
\bauthor{\bsnm{{Hannah}}, \binits{I.G.}},
\bauthor{\bsnm{{Kontar}}, \binits{E.P.}}:
\byear{2012},
\batitle{{Differential emission measures from the regularized inversion of
  Hinode and SDO data}}.
\bjtitle{A{\&}A}
\bvolume{539},
\bfpage{A146}.
\doiurl{10.1051/0004-6361/201117576}.
\end{barticle}
\endbibitem

\bibitem[\protect\citeauthoryear{{Hannah} and
  {Kontar}}{2013}]{2013A&A...553A..10H}
\begin{barticle}
\bauthor{\bsnm{{Hannah}}, \binits{I.G.}},
\bauthor{\bsnm{{Kontar}}, \binits{E.P.}}:
\byear{2013},
\batitle{{Multi-thermal dynamics and energetics of a coronal mass ejection in
  the low solar atmosphere}}.
\bjtitle{A{\&}A}
\bvolume{553},
\bfpage{A10}.
\doiurl{10.1051/0004-6361/201219727}.
\end{barticle}
\endbibitem

\bibitem[\protect\citeauthoryear{{Hannah}, {Kontar}, and
  {Reid}}{2013}]{2013A&A...550A..51H}
\begin{barticle}
\bauthor{\bsnm{{Hannah}}, \binits{I.G.}},
\bauthor{\bsnm{{Kontar}}, \binits{E.P.}},
\bauthor{\bsnm{{Reid}}, \binits{H.A.S.}}:
\byear{2013},
\batitle{{Effect of turbulent density-fluctuations on wave-particle
  interactions and solar flare X-ray spectra}}.
\bjtitle{A{\&}A}
\bvolume{550},
\bfpage{A51}.
\doiurl{10.1051/0004-6361/201220462}.
\end{barticle}
\endbibitem

\bibitem[\protect\citeauthoryear{{Hansen} and {Cally}}{2012}]{HanCal12aa}
\begin{barticle}
\bauthor{\bsnm{{Hansen}}, \binits{S.C.}},
\bauthor{\bsnm{{Cally}}, \binits{P.S.}}:
\byear{2012},
\batitle{{Benchmarking Fast-to-Alfv\'en Mode Conversion in a Cold MHD Plasma.
  II. How to get Alfv\'en waves through the Solar Transition Region}}.
\bjtitle{Astrophys. J.}
\bvolume{751},
\bfpage{31}.
\doiurl{doi:10.1088/0004-637X/751/1/31}.
\end{barticle}
\endbibitem

\bibitem[\protect\citeauthoryear{{Hansteen}
  \textit{et~al.}}{2014}]{2014Sci...346E.315H}
\begin{barticle}
\bauthor{\bsnm{{Hansteen}}, \binits{V.}},
\bauthor{\bsnm{{De Pontieu}}, \binits{B.}},
\bauthor{\bsnm{{Carlsson}}, \binits{M.}},
\bauthor{\bsnm{{Lemen}}, \binits{J.}},
\bauthor{\bsnm{{Title}}, \binits{A.}},
\bauthor{\bsnm{{Boerner}}, \binits{P.}},
\bauthor{\bsnm{{Hurlburt}}, \binits{N.}},
\bauthor{\bsnm{{Tarbell}}, \binits{T.D.}},
\bauthor{\bsnm{{Wuelser}}, \binits{J.P.}},
\bauthor{\bsnm{{Pereira}}, \binits{T.M.D.}},
\bauthor{\bsnm{{De Luca}}, \binits{E.E.}},
\bauthor{\bsnm{{Golub}}, \binits{L.}},
\bauthor{\bsnm{{McKillop}}, \binits{S.}},
\bauthor{\bsnm{{Reeves}}, \binits{K.}},
\bauthor{\bsnm{{Saar}}, \binits{S.}},
\bauthor{\bsnm{{Testa}}, \binits{P.}},
\bauthor{\bsnm{{Tian}}, \binits{H.}},
\bauthor{\bsnm{{Kankelborg}}, \binits{C.}},
\bauthor{\bsnm{{Jaeggli}}, \binits{S.}},
\bauthor{\bsnm{{Kleint}}, \binits{L.}},
\bauthor{\bsnm{{Mart{\'{\i}}nez-Sykora}}, \binits{J.}}:
\byear{2014},
\batitle{{The unresolved fine structure resolved: IRIS observations of the
  solar transition region}}.
\bjtitle{Science}
\bvolume{346}.
\doiurl{10.1126/science.1255757}.
\end{barticle}
\endbibitem

\bibitem[\protect\citeauthoryear{{Harrison}
  \textit{et~al.}}{2012}]{2012ApJ...750...45H}
\begin{barticle}
\bauthor{\bsnm{{Harrison}}, \binits{R.A.}},
\bauthor{\bsnm{{Davies}}, \binits{J.A.}},
\bauthor{\bsnm{{M{\"o}stl}}, \binits{C.}},
\bauthor{\bsnm{{Liu}}, \binits{Y.}},
\bauthor{\bsnm{{Temmer}}, \binits{M.}},
\bauthor{\bsnm{{Bisi}}, \binits{M.M.}},
\bauthor{\bsnm{{Eastwood}}, \binits{J.P.}},
\bauthor{\bsnm{{de Koning}}, \binits{C.A.}},
\bauthor{\bsnm{{Nitta}}, \binits{N.}},
\bauthor{\bsnm{{Rollett}}, \binits{T.}},
\bauthor{\bsnm{{Farrugia}}, \binits{C.J.}},
\bauthor{\bsnm{{Forsyth}}, \binits{R.J.}},
\bauthor{\bsnm{{Jackson}}, \binits{B.V.}},
\bauthor{\bsnm{{Jensen}}, \binits{E.A.}},
\bauthor{\bsnm{{Kilpua}}, \binits{E.K.J.}},
\bauthor{\bsnm{{Odstrcil}}, \binits{D.}},
\bauthor{\bsnm{{Webb}}, \binits{D.F.}}:
\byear{2012},
\batitle{{An Analysis of the Origin and Propagation of the Multiple Coronal
  Mass Ejections of 2010 August 1}}.
\bjtitle{Astrophys. J.}
\bvolume{750},
\bfpage{45}.
\end{barticle}
\endbibitem

\bibitem[\protect\citeauthoryear{{Hayashi}
  \textit{et~al.}}{2015}]{2015SoPh..290.1507H}
\begin{barticle}
\bauthor{\bsnm{{Hayashi}}, \binits{K.}},
\bauthor{\bsnm{{Hoeksema}}, \binits{J.T.}},
\bauthor{\bsnm{{Liu}}, \binits{Y.}},
\bauthor{\bsnm{{Bobra}}, \binits{M.G.}},
\bauthor{\bsnm{{Sun}}, \binits{X.D.}},
\bauthor{\bsnm{{Norton}}, \binits{A.A.}}:
\byear{2015},
\batitle{{The Helioseismic and Magnetic Imager (HMI) Vector Magnetic Field
  Pipeline: Magnetohydrodynamics Simulation Module for the Global Solar
  Corona}}.
\bjtitle{Solar Phys.}
\bvolume{290},
\bfpage{1507}.
\doiurl{10.1007/s11207-015-0686-z}.
\end{barticle}
\endbibitem

\bibitem[\protect\citeauthoryear{{Hazra}, {Karak}, and
  {Choudhuri}}{2014}]{2014ApJ...782...93H}
\begin{barticle}
\bauthor{\bsnm{{Hazra}}, \binits{G.}},
\bauthor{\bsnm{{Karak}}, \binits{B.B.}},
\bauthor{\bsnm{{Choudhuri}}, \binits{A.R.}}:
\byear{2014},
\batitle{{Is a Deep One-cell Meridional Circulation Essential for the Flux
  Transport Solar Dynamo?}}
\bjtitle{Astrophys. J.}
\bvolume{782},
\bfpage{93}.
\doiurl{10.1088/0004-637X/782/2/93}.
\end{barticle}
\endbibitem

\bibitem[\protect\citeauthoryear{{Heinzel} and
  {Kleint}}{2014}]{2014ApJ...794L..23H}
\begin{barticle}
\bauthor{\bsnm{{Heinzel}}, \binits{P.}},
\bauthor{\bsnm{{Kleint}}, \binits{L.}}:
\byear{2014},
\batitle{{Hydrogen Balmer Continuum in Solar Flares Detected by the Interface
  Region Imaging Spectrograph (IRIS)}}.
\bjtitle{Astrophys. J., Lett.}
\bvolume{794},
\bfpage{L23}.
\doiurl{10.1088/2041-8205/794/2/L23}.
\end{barticle}
\endbibitem

\bibitem[\protect\citeauthoryear{{Howard} and
  {DeForest}}{2012}]{2012ApJ...746...64H}
\begin{barticle}
\bauthor{\bsnm{{Howard}}, \binits{T.A.}},
\bauthor{\bsnm{{DeForest}}, \binits{C.E.}}:
\byear{2012},
\batitle{{Inner Heliospheric Flux Rope Evolution via Imaging of Coronal Mass
  Ejections}}.
\bjtitle{Astrophys. J.}
\bvolume{746},
\bfpage{64}.
\doiurl{10.1088/0004-637X/746/1/64}.
\end{barticle}
\endbibitem

\bibitem[\protect\citeauthoryear{{Jiang}, {Cameron}, and
  {Sch{\"u}ssler}}{2015}]{2015ApJ...808L..28J}
\begin{barticle}
\bauthor{\bsnm{{Jiang}}, \binits{J.}},
\bauthor{\bsnm{{Cameron}}, \binits{R.H.}},
\bauthor{\bsnm{{Sch{\"u}ssler}}, \binits{M.}}:
\byear{2015},
\batitle{{The Cause of the Weak Solar Cycle 24}}.
\bjtitle{Astrophys. J., Lett.}
\bvolume{808},
\bfpage{L28}.
\doiurl{10.1088/2041-8205/808/1/L28}.
\end{barticle}
\endbibitem

\bibitem[\protect\citeauthoryear{{Kaiser}
  \textit{et~al.}}{2008}]{2008SSRv..136....5K}
\begin{barticle}
\bauthor{\bsnm{{Kaiser}}, \binits{M.L.}},
\bauthor{\bsnm{{Kucera}}, \binits{T.A.}},
\bauthor{\bsnm{{Davila}}, \binits{J.M.}},
\bauthor{\bsnm{{St.~Cyr}}, \binits{O.C.}},
\bauthor{\bsnm{{Guhathakurta}}, \binits{M.}},
\bauthor{\bsnm{{Christian}}, \binits{E.}}:
\byear{2008},
\batitle{{The STEREO Mission: An Introduction}}.
\bjtitle{Space Science Reviews}
\bvolume{136},
\bfpage{5}.
\doiurl{10.1007/s11214-007-9277-0}.
\end{barticle}
\endbibitem

\bibitem[\protect\citeauthoryear{{K{\"a}pyl{\"a}}, {Mantere}, and
  {Brandenburg}}{2012}]{2012ApJ...755L..22K}
\begin{barticle}
\bauthor{\bsnm{{K{\"a}pyl{\"a}}}, \binits{P.J.}},
\bauthor{\bsnm{{Mantere}}, \binits{M.J.}},
\bauthor{\bsnm{{Brandenburg}}, \binits{A.}}:
\byear{2012},
\batitle{{Cyclic Magnetic Activity due to Turbulent Convection in Spherical
  Wedge Geometry}}.
\bjtitle{Astrophys. J., Lett.}
\bvolume{755},
\bfpage{L22}.
\doiurl{10.1088/2041-8205/755/1/L22}.
\end{barticle}
\endbibitem

\bibitem[\protect\citeauthoryear{{Karak}
  \textit{et~al.}}{2014}]{2014SSRv..186..561K}
\begin{barticle}
\bauthor{\bsnm{{Karak}}, \binits{B.B.}},
\bauthor{\bsnm{{Jiang}}, \binits{J.}},
\bauthor{\bsnm{{Miesch}}, \binits{M.S.}},
\bauthor{\bsnm{{Charbonneau}}, \binits{P.}},
\bauthor{\bsnm{{Choudhuri}}, \binits{A.R.}}:
\byear{2014},
\batitle{{Flux Transport Dynamos: From Kinematics to Dynamics}}.
\bjtitle{Space Sci. Rev.}
\bvolume{186},
\bfpage{561}.
\doiurl{10.1007/s11214-014-0099-6}.
\end{barticle}
\endbibitem

\bibitem[\protect\citeauthoryear{{Karpen}, {Antiochos}, and
  {DeVore}}{2012}]{2012ApJ...760...81K}
\begin{barticle}
\bauthor{\bsnm{{Karpen}}, \binits{J.T.}},
\bauthor{\bsnm{{Antiochos}}, \binits{S.K.}},
\bauthor{\bsnm{{DeVore}}, \binits{C.R.}}:
\byear{2012},
\batitle{{The Mechanisms for the Onset and Explosive Eruption of Coronal Mass
  Ejections and Eruptive Flares}}.
\bjtitle{Astrophys. J.}
\bvolume{760},
\bfpage{81}.
\doiurl{10.1088/0004-637X/760/1/81}.
\end{barticle}
\endbibitem

\bibitem[\protect\citeauthoryear{{Kennedy}
  \textit{et~al.}}{2015}]{2015A&A...578A..72K}
\begin{barticle}
\bauthor{\bsnm{{Kennedy}}, \binits{M.B.}},
\bauthor{\bsnm{{Milligan}}, \binits{R.O.}},
\bauthor{\bsnm{{Allred}}, \binits{J.C.}},
\bauthor{\bsnm{{Mathioudakis}}, \binits{M.}},
\bauthor{\bsnm{{Keenan}}, \binits{F.P.}}:
\byear{2015},
\batitle{{Radiative hydrodynamic modelling and observations of the X-class
  solar flare on 2011 March 9}}.
\bjtitle{A{\&}A}
\bvolume{578},
\bfpage{A72}.
\doiurl{10.1051/0004-6361/201425144}.
\end{barticle}
\endbibitem

\bibitem[\protect\citeauthoryear{{Kerr} and
  {Fletcher}}{2014}]{2014ApJ...783...98K}
\begin{barticle}
\bauthor{\bsnm{{Kerr}}, \binits{G.S.}},
\bauthor{\bsnm{{Fletcher}}, \binits{L.}}:
\byear{2014},
\batitle{{Physical Properties of White-light Sources in the 2011 February 15
  Solar Flare}}.
\bjtitle{Astrophys. J.}
\bvolume{783},
\bfpage{98}.
\doiurl{10.1088/0004-637X/783/2/98}.
\end{barticle}
\endbibitem

\bibitem[\protect\citeauthoryear{{Kienreich}
  \textit{et~al.}}{2013}]{2013SoPh..286..201K}
\begin{barticle}
\bauthor{\bsnm{{Kienreich}}, \binits{I.W.}},
\bauthor{\bsnm{{Muhr}}, \binits{N.}},
\bauthor{\bsnm{{Veronig}}, \binits{A.M.}},
\bauthor{\bsnm{{Berghmans}}, \binits{D.}},
\bauthor{\bsnm{{De Groof}}, \binits{A.}},
\bauthor{\bsnm{{Temmer}}, \binits{M.}},
\bauthor{\bsnm{{Vr{\v s}nak}}, \binits{B.}},
\bauthor{\bsnm{{Seaton}}, \binits{D.B.}}:
\byear{2013},
\batitle{{Solar TErrestrial Relations Observatory-A (STEREO-A) and PRoject for
  On-Board Autonomy 2 (PROBA2) Quadrature Observations of Reflections of Three
  EUV Waves from a Coronal Hole}}.
\bjtitle{Solar Phys.}
\bvolume{286},
\bfpage{201}.
\doiurl{10.1007/s11207-012-0023-8}.
\end{barticle}
\endbibitem

\bibitem[\protect\citeauthoryear{{Kliem}
  \textit{et~al.}}{2014}]{2014ApJ...789...46K}
\begin{barticle}
\bauthor{\bsnm{{Kliem}}, \binits{B.}},
\bauthor{\bsnm{{Lin}}, \binits{J.}},
\bauthor{\bsnm{{Forbes}}, \binits{T.G.}},
\bauthor{\bsnm{{Priest}}, \binits{E.R.}},
\bauthor{\bsnm{{T{\"o}r{\"o}k}}, \binits{T.}}:
\byear{2014},
\batitle{{Catastrophe versus Instability for the Eruption of a Toroidal Solar
  Magnetic Flux Rope}}.
\bjtitle{Astrophys. J.}
\bvolume{789},
\bfpage{46}.
\doiurl{10.1088/0004-637X/789/1/46}.
\end{barticle}
\endbibitem

\bibitem[\protect\citeauthoryear{Kosugi \textit{et~al.}}{2007}]{kosugi+etl2007}
\begin{barticle}
\bauthor{\bsnm{Kosugi}, \binits{T.}},
\bauthor{\bsnm{Matsuzaki}, \binits{K.}},
\bauthor{\bsnm{Sakao}, \binits{T.}},
\bauthor{\bsnm{Shimizu}, \binits{T.}},
\bauthor{\bsnm{Sone}, \binits{Y.}},
\bauthor{\bsnm{Tachikawa}, \binits{S.}},
\bauthor{\bsnm{Hashimoto}, \binits{T.}},
\bauthor{\bsnm{Minesugi}, \binits{K.}},
\bauthor{\bsnm{Ohnishi}, \binits{A.}},
\bauthor{\bsnm{Yamada}, \binits{T.}},
\bauthor{\bsnm{Tsuneta}, \binits{S.}},
\bauthor{\bsnm{Hara}, \binits{H.}},
\bauthor{\bsnm{Ichimoto}, \binits{K.}},
\bauthor{\bsnm{Suematsu}, \binits{Y.}},
\bauthor{\bsnm{Shimojo}, \binits{M.}},
\bauthor{\bsnm{Watanabe}, \binits{T.}},
\bauthor{\bsnm{Shimada}, \binits{S.}},
\bauthor{\bsnm{Davis}, \binits{J.M.}},
\bauthor{\bsnm{Hill}, \binits{L.D.}},
\bauthor{\bsnm{Owens}, \binits{J.K.}},
\bauthor{\bsnm{Title}, \binits{A.M.}},
\bauthor{\bsnm{Culhane}, \binits{J.L.}},
\bauthor{\bsnm{Harra}, \binits{L.K.}},
\bauthor{\bsnm{Doschek}, \binits{G.A.}},
\bauthor{\bsnm{Golub}, \binits{L.}}:
\byear{2007},
\batitle{{The Hinode (Solar-B) mission: an overview}}.
\bjtitle{Solar Phys.}
\bvolume{243},
\bfpage{1}.
\end{barticle}
\endbibitem

\bibitem[\protect\citeauthoryear{{Krucker} and
  {Battaglia}}{2014}]{2014ApJ...780..107K}
\begin{barticle}
\bauthor{\bsnm{{Krucker}}, \binits{S.}},
\bauthor{\bsnm{{Battaglia}}, \binits{M.}}:
\byear{2014},
\batitle{{Particle Densities within the Acceleration Region of a Solar Flare}}.
\bjtitle{Astrophys. J.}
\bvolume{780},
\bfpage{107}.
\doiurl{10.1088/0004-637X/780/1/107}.
\end{barticle}
\endbibitem

\bibitem[\protect\citeauthoryear{{Krucker}
  \textit{et~al.}}{2015}]{2015ApJ...802...19K}
\begin{barticle}
\bauthor{\bsnm{{Krucker}}, \binits{S.}},
\bauthor{\bsnm{{Saint-Hilaire}}, \binits{P.}},
\bauthor{\bsnm{{Hudson}}, \binits{H.S.}},
\bauthor{\bsnm{{Haberreiter}}, \binits{M.}},
\bauthor{\bsnm{{Martinez-Oliveros}}, \binits{J.C.}},
\bauthor{\bsnm{{Fivian}}, \binits{M.D.}},
\bauthor{\bsnm{{Hurford}}, \binits{G.}},
\bauthor{\bsnm{{Kleint}}, \binits{L.}},
\bauthor{\bsnm{{Battaglia}}, \binits{M.}},
\bauthor{\bsnm{{Kuhar}}, \binits{M.}},
\bauthor{\bsnm{{Arnold}}, \binits{N.G.}}:
\byear{2015},
\batitle{{Co-Spatial White Light and Hard X-Ray Flare Footpoints Seen Above the
  Solar Limb}}.
\bjtitle{Astrophys. J.}
\bvolume{802},
\bfpage{19}.
\doiurl{10.1088/0004-637X/802/1/19}.
\end{barticle}
\endbibitem

\bibitem[\protect\citeauthoryear{{Kumar}
  \textit{et~al.}}{2012}]{2012ApJ...746...67K}
\begin{barticle}
\bauthor{\bsnm{{Kumar}}, \binits{P.}},
\bauthor{\bsnm{{Cho}}, \binits{K.-S.}},
\bauthor{\bsnm{{Bong}}, \binits{S.-C.}},
\bauthor{\bsnm{{Park}}, \binits{S.-H.}},
\bauthor{\bsnm{{Kim}}, \binits{Y.H.}}:
\byear{2012},
\batitle{{Initiation of Coronal Mass Ejection and Associated Flare Caused by
  Helical Kink Instability Observed by SDO/AIA}}.
\bjtitle{Astrophys. J.}
\bvolume{746},
\bfpage{67}.
\doiurl{10.1088/0004-637X/746/1/67}.
\end{barticle}
\endbibitem

\bibitem[\protect\citeauthoryear{{Lee}
  \textit{et~al.}}{2013}]{2013SoPh..285..349L}
\begin{barticle}
\bauthor{\bsnm{{Lee}}, \binits{C.O.}},
\bauthor{\bsnm{{Arge}}, \binits{C.N.}},
\bauthor{\bsnm{{Odstr{\v c}il}}, \binits{D.}},
\bauthor{\bsnm{{Millward}}, \binits{G.}},
\bauthor{\bsnm{{Pizzo}}, \binits{V.}},
\bauthor{\bsnm{{Quinn}}, \binits{J.M.}},
\bauthor{\bsnm{{Henney}}, \binits{C.J.}}:
\byear{2013},
\batitle{{Ensemble Modeling of CME Propagation}}.
\bjtitle{Solar Phys.}
\bvolume{285},
\bfpage{349}.
\doiurl{10.1007/s11207-012-9980-1}.
\end{barticle}
\endbibitem

\bibitem[\protect\citeauthoryear{Leibacher
  \textit{et~al.}}{2010}]{2010SoPh..263....1.}
\begin{barticle}
\bauthor{\bsnm{Leibacher}, \binits{J.}},
\bauthor{\bsnm{Sakurai}, \binits{T.}},
\bauthor{\bsnm{Schrijver}, \binits{C.J.}},
\bauthor{\bsnm{{van Driel-Gesztelyi}}, \binits{L.}}:
\byear{2010},
\batitle{{Solar Observation Target Identification Convention for use in Solar
  Physics}}.
\bjtitle{Solar Phys.}
\bvolume{263},
\bfpage{1}.
\end{barticle}
\endbibitem

\bibitem[\protect\citeauthoryear{{Lemen} \textit{et~al.}}{2012}]{aiainstrument}
\begin{barticle}
\bauthor{\bsnm{{Lemen}}, \binits{J.R.}},
\bauthor{\bsnm{{Title}}, \binits{A.M.}},
\bauthor{\bsnm{{Akin}}, \binits{D.J.}},
\bauthor{\bsnm{{Boerner}}, \binits{P.F.}},
\bauthor{\bsnm{{Chou}}, \binits{C.}},
\bauthor{\bsnm{{Drake}}, \binits{J.F.}},
\bauthor{\bsnm{{Duncan}}, \binits{D.W.}},
\bauthor{\bsnm{{Edwards}}, \binits{C.G.}},
\bauthor{\bsnm{{Friedlaender}}, \binits{F.M.}},
\bauthor{\bsnm{{Heyman}}, \binits{G.F.}},
\bauthor{\bsnm{{Hurlburt}}, \binits{N.E.}},
\bauthor{\bsnm{{Katz}}, \binits{N.L.}},
\bauthor{\bsnm{{Kushner}}, \binits{G.D.}},
\bauthor{\bsnm{{Levay}}, \binits{M.}},
\bauthor{\bsnm{{Lindgren}}, \binits{R.W.}},
\bauthor{\bsnm{{Mathur}}, \binits{D.P.}},
\bauthor{\bsnm{{McFeaters}}, \binits{E.L.}},
\bauthor{\bsnm{{Mitchell}}, \binits{S.}},
\bauthor{\bsnm{{Rehse}}, \binits{R.A.}},
\bauthor{\bsnm{{Schrijver}}, \binits{C.J.}},
\bauthor{\bsnm{{Springer}}, \binits{L.A.}},
\bauthor{\bsnm{{Stern}}, \binits{R.A.}},
\bauthor{\bsnm{{Tarbell}}, \binits{T.D.}},
\bauthor{\bsnm{{Wuelser}}, \binits{J.-P.}},
\bauthor{\bsnm{{Wolfson}}, \binits{C.J.}},
\bauthor{\bsnm{{Yanari}}, \binits{C.}},
\bauthor{\bsnm{{Bookbinder}}, \binits{J.A.}},
\bauthor{\bsnm{{Cheimets}}, \binits{P.N.}},
\bauthor{\bsnm{{Caldwell}}, \binits{D.}},
\bauthor{\bsnm{{Deluca}}, \binits{E.E.}},
\bauthor{\bsnm{{Gates}}, \binits{R.}},
\bauthor{\bsnm{{Golub}}, \binits{L.}},
\bauthor{\bsnm{{Park}}, \binits{S.}},
\bauthor{\bsnm{{Podgorski}}, \binits{W.A.}},
\bauthor{\bsnm{{Bush}}, \binits{R.I.}},
\bauthor{\bsnm{{Scherrer}}, \binits{P.H.}},
\bauthor{\bsnm{{Gummin}}, \binits{M.A.}},
\bauthor{\bsnm{{Smith}}, \binits{P.}},
\bauthor{\bsnm{{Auker}}, \binits{G.}},
\bauthor{\bsnm{{Jerram}}, \binits{P.}},
\bauthor{\bsnm{{Pool}}, \binits{P.}},
\bauthor{\bsnm{{Soufli}}, \binits{R.}},
\bauthor{\bsnm{{Windt}}, \binits{D.L.}},
\bauthor{\bsnm{{Beardsley}}, \binits{S.}},
\bauthor{\bsnm{{Clapp}}, \binits{M.}},
\bauthor{\bsnm{{Lang}}, \binits{J.}},
\bauthor{\bsnm{{Waltham}}, \binits{N.}}:
\byear{2012},
\batitle{{The Atmospheric Imaging Assembly (AIA) on the Solar Dynamics
  Observatory (SDO)}}.
\bjtitle{Solar Phys.}
\bvolume{275},
\bfpage{17}.
\end{barticle}
\endbibitem

\bibitem[\protect\citeauthoryear{{Li}
  \textit{et~al.}}{2012}]{2012ApJ...746...13L}
\begin{barticle}
\bauthor{\bsnm{{Li}}, \binits{T.}},
\bauthor{\bsnm{{Zhang}}, \binits{J.}},
\bauthor{\bsnm{{Yang}}, \binits{S.}},
\bauthor{\bsnm{{Liu}}, \binits{W.}}:
\byear{2012},
\batitle{{SDO/AIA Observations of Secondary Waves Generated by Interaction of
  the 2011 June 7 Global EUV Wave with Solar Coronal Structures}}.
\bjtitle{Astrophys. J.}
\bvolume{746},
\bfpage{13}.
\doiurl{10.1088/0004-637X/746/1/13}.
\end{barticle}
\endbibitem

\bibitem[\protect\citeauthoryear{{Lin}
  \textit{et~al.}}{2002}]{2002SoPh..210....3L}
\begin{barticle}
\bauthor{\bsnm{{Lin}}, \binits{R.P.}},
\bauthor{\bsnm{{Dennis}}, \binits{B.R.}},
\bauthor{\bsnm{{Hurford}}, \binits{G.J.}},
\bauthor{\bsnm{{Smith}}, \binits{D.M.}},
\bauthor{\bsnm{{Zehnder}}, \binits{A.}},
\bauthor{\bsnm{{Harvey}}, \binits{P.R.}},
\bauthor{\bsnm{{Curtis}}, \binits{D.W.}},
\bauthor{\bsnm{{Pankow}}, \binits{D.}},
\bauthor{\bsnm{{Turin}}, \binits{P.}},
\bauthor{\bsnm{{Bester}}, \binits{M.}},
\bauthor{\bsnm{{Csillaghy}}, \binits{A.}},
\bauthor{\bsnm{{Lewis}}, \binits{M.}},
\bauthor{\bsnm{{Madden}}, \binits{N.}},
\bauthor{\bsnm{{van Beek}}, \binits{H.F.}},
\bauthor{\bsnm{{Appleby}}, \binits{M.}},
\bauthor{\bsnm{{Raudorf}}, \binits{T.}},
\bauthor{\bsnm{{McTiernan}}, \binits{J.}},
\bauthor{\bsnm{{Ramaty}}, \binits{R.}},
\bauthor{\bsnm{{Schmahl}}, \binits{E.}},
\bauthor{\bsnm{{Schwartz}}, \binits{R.}},
\bauthor{\bsnm{{Krucker}}, \binits{S.}},
\bauthor{\bsnm{{Abiad}}, \binits{R.}},
\bauthor{\bsnm{{Quinn}}, \binits{T.}},
\bauthor{\bsnm{{Berg}}, \binits{P.}},
\bauthor{\bsnm{{Hashii}}, \binits{M.}},
\bauthor{\bsnm{{Sterling}}, \binits{R.}},
\bauthor{\bsnm{{Jackson}}, \binits{R.}},
\bauthor{\bsnm{{Pratt}}, \binits{R.}},
\bauthor{\bsnm{{Campbell}}, \binits{R.D.}},
\bauthor{\bsnm{{Malone}}, \binits{D.}},
\bauthor{\bsnm{{Landis}}, \binits{D.}},
\bauthor{\bsnm{{Barrington-Leigh}}, \binits{C.P.}},
\bauthor{\bsnm{{Slassi-Sennou}}, \binits{S.}},
\bauthor{\bsnm{{Cork}}, \binits{C.}},
\bauthor{\bsnm{{Clark}}, \binits{D.}},
\bauthor{\bsnm{{Amato}}, \binits{D.}},
\bauthor{\bsnm{{Orwig}}, \binits{L.}},
\bauthor{\bsnm{{Boyle}}, \binits{R.}},
\bauthor{\bsnm{{Banks}}, \binits{I.S.}},
\bauthor{\bsnm{{Shirey}}, \binits{K.}},
\bauthor{\bsnm{{Tolbert}}, \binits{A.K.}},
\bauthor{\bsnm{{Zarro}}, \binits{D.}},
\bauthor{\bsnm{{Snow}}, \binits{F.}},
\bauthor{\bsnm{{Thomsen}}, \binits{K.}},
\bauthor{\bsnm{{Henneck}}, \binits{R.}},
\bauthor{\bsnm{{McHedlishvili}}, \binits{A.}},
\bauthor{\bsnm{{Ming}}, \binits{P.}},
\bauthor{\bsnm{{Fivian}}, \binits{M.}},
\bauthor{\bsnm{{Jordan}}, \binits{J.}},
\bauthor{\bsnm{{Wanner}}, \binits{R.}},
\bauthor{\bsnm{{Crubb}}, \binits{J.}},
\bauthor{\bsnm{{Preble}}, \binits{J.}},
\bauthor{\bsnm{{Matranga}}, \binits{M.}},
\bauthor{\bsnm{{Benz}}, \binits{A.}},
\bauthor{\bsnm{{Hudson}}, \binits{H.}},
\bauthor{\bsnm{{Canfield}}, \binits{R.C.}},
\bauthor{\bsnm{{Holman}}, \binits{G.D.}},
\bauthor{\bsnm{{Crannell}}, \binits{C.}},
\bauthor{\bsnm{{Kosugi}}, \binits{T.}},
\bauthor{\bsnm{{Emslie}}, \binits{A.G.}},
\bauthor{\bsnm{{Vilmer}}, \binits{N.}},
\bauthor{\bsnm{{Brown}}, \binits{J.C.}},
\bauthor{\bsnm{{Johns-Krull}}, \binits{C.}},
\bauthor{\bsnm{{Aschwanden}}, \binits{M.}},
\bauthor{\bsnm{{Metcalf}}, \binits{T.}},
\bauthor{\bsnm{{Conway}}, \binits{A.}}:
\byear{2002},
\batitle{{The Reuven Ramaty High-Energy Solar Spectroscopic Imager (RHESSI)}}.
\bjtitle{Solar Phys.}
\bvolume{210},
\bfpage{3}.
\doiurl{10.1023/A:1022428818870}.
\end{barticle}
\endbibitem

\bibitem[\protect\citeauthoryear{{Liu} \textit{et~al.}}{2014a}]{LiuMcIDe-14aa}
\begin{barticle}
\bauthor{\bsnm{{Liu}}, \binits{J.}},
\bauthor{\bsnm{{McIntosh}}, \binits{S.W.}},
\bauthor{\bsnm{{De Moortel}}, \binits{I.}},
\bauthor{\bsnm{{Threlfall}}, \binits{J.}},
\bauthor{\bsnm{{Bethge}}, \binits{C.}}:
\byear{2014}a,
\batitle{{Statistical Evidence for the Existence of Alfv{\'e}nic Turbulence in
  Solar Coronal Loops}}.
\bjtitle{Astrophys. J.}
\bvolume{797},
\bfpage{7}.
\doiurl{10.1088/0004-637X/797/1/7}.
\end{barticle}
\endbibitem

\bibitem[\protect\citeauthoryear{{Liu} and {Ofman}}{2014}]{2014SoPh..289.3233L}
\begin{barticle}
\bauthor{\bsnm{{Liu}}, \binits{W.}},
\bauthor{\bsnm{{Ofman}}, \binits{L.}}:
\byear{2014},
\batitle{{Advances in Observing Various Coronal EUV Waves in the SDO Era and
  Their Seismological Applications (Invited Review)}}.
\bjtitle{Solar Phys.}
\bvolume{289},
\bfpage{3233}.
\end{barticle}
\endbibitem

\bibitem[\protect\citeauthoryear{{Liu}
  \textit{et~al.}}{2012a}]{2012ApJ...753...52L}
\begin{barticle}
\bauthor{\bsnm{{Liu}}, \binits{W.}},
\bauthor{\bsnm{{Ofman}}, \binits{L.}},
\bauthor{\bsnm{{Nitta}}, \binits{N.V.}},
\bauthor{\bsnm{{Aschwanden}}, \binits{M.J.}},
\bauthor{\bsnm{{Schrijver}}, \binits{C.J.}},
\bauthor{\bsnm{{Title}}, \binits{A.M.}},
\bauthor{\bsnm{{Tarbell}}, \binits{T.D.}}:
\byear{2012}a,
\batitle{{Quasi-periodic Fast-mode Wave Trains within a Global EUV Wave and
  Sequential Transverse Oscillations Detected by SDO/AIA}}.
\bjtitle{Astrophys. J.}
\bvolume{753},
\bfpage{52}.
\end{barticle}
\endbibitem

\bibitem[\protect\citeauthoryear{{Liu}
  \textit{et~al.}}{2012b}]{2012ApJ...746L..15L}
\begin{barticle}
\bauthor{\bsnm{{Liu}}, \binits{Y.D.}},
\bauthor{\bsnm{{Luhmann}}, \binits{J.G.}},
\bauthor{\bsnm{{M{\"o}stl}}, \binits{C.}},
\bauthor{\bsnm{{Martinez-Oliveros}}, \binits{J.C.}},
\bauthor{\bsnm{{Bale}}, \binits{S.D.}},
\bauthor{\bsnm{{Lin}}, \binits{R.P.}},
\bauthor{\bsnm{{Harrison}}, \binits{R.A.}},
\bauthor{\bsnm{{Temmer}}, \binits{M.}},
\bauthor{\bsnm{{Webb}}, \binits{D.F.}},
\bauthor{\bsnm{{Odstrcil}}, \binits{D.}}:
\byear{2012}b,
\batitle{{Interactions between Coronal Mass Ejections Viewed in Coordinated
  Imaging and in situ Observations}}.
\bjtitle{Astrophys. J., Lett.}
\bvolume{746},
\bfpage{L15}.
\doiurl{10.1088/2041-8205/746/2/L15}.
\end{barticle}
\endbibitem

\bibitem[\protect\citeauthoryear{{Liu}
  \textit{et~al.}}{2013}]{2013ApJ...769...45L}
\begin{barticle}
\bauthor{\bsnm{{Liu}}, \binits{Y.D.}},
\bauthor{\bsnm{{Luhmann}}, \binits{J.G.}},
\bauthor{\bsnm{{Lugaz}}, \binits{N.}},
\bauthor{\bsnm{{M{\"o}stl}}, \binits{C.}},
\bauthor{\bsnm{{Davies}}, \binits{J.A.}},
\bauthor{\bsnm{{Bale}}, \binits{S.D.}},
\bauthor{\bsnm{{Lin}}, \binits{R.P.}}:
\byear{2013},
\batitle{{On Sun-to-Earth Propagation of Coronal Mass Ejections}}.
\bjtitle{Astrophys. J.}
\bvolume{769},
\bfpage{45}.
\doiurl{10.1088/0004-637X/769/1/45}.
\end{barticle}
\endbibitem

\bibitem[\protect\citeauthoryear{{Liu}
  \textit{et~al.}}{2014b}]{2014NatCo...5E3481L}
\begin{barticle}
\bauthor{\bsnm{{Liu}}, \binits{Y.D.}},
\bauthor{\bsnm{{Luhmann}}, \binits{J.G.}},
\bauthor{\bsnm{{Kajdi{\v c}}}, \binits{P.}},
\bauthor{\bsnm{{Kilpua}}, \binits{E.K.J.}},
\bauthor{\bsnm{{Lugaz}}, \binits{N.}},
\bauthor{\bsnm{{Nitta}}, \binits{N.V.}},
\bauthor{\bsnm{{M{\"o}stl}}, \binits{C.}},
\bauthor{\bsnm{{Lavraud}}, \binits{B.}},
\bauthor{\bsnm{{Bale}}, \binits{S.D.}},
\bauthor{\bsnm{{Farrugia}}, \binits{C.J.}},
\bauthor{\bsnm{{Galvin}}, \binits{A.B.}}:
\byear{2014}b,
\batitle{{Observations of an extreme storm in interplanetary space caused by
  successive coronal mass ejections}}.
\bjtitle{Nature Communications}
\bvolume{5},
\bfpage{3481}.
\doiurl{10.1038/ncomms4481}.
\end{barticle}
\endbibitem

\bibitem[\protect\citeauthoryear{{Liu}
  \textit{et~al.}}{2014c}]{2014RAA....14..705L}
\begin{barticle}
\bauthor{\bsnm{{Liu}}, \binits{Z.}},
\bauthor{\bsnm{{Xu}}, \binits{J.}},
\bauthor{\bsnm{{Gu}}, \binits{B.-Z.}},
\bauthor{\bsnm{{Wang}}, \binits{S.}},
\bauthor{\bsnm{{You}}, \binits{J.-Q.}},
\bauthor{\bsnm{{Shen}}, \binits{L.-X.}},
\bauthor{\bsnm{{Lu}}, \binits{R.-W.}},
\bauthor{\bsnm{{Jin}}, \binits{Z.-Y.}},
\bauthor{\bsnm{{Chen}}, \binits{L.-F.}},
\bauthor{\bsnm{{Lou}}, \binits{K.}},
\bauthor{\bsnm{{Li}}, \binits{Z.}},
\bauthor{\bsnm{{Liu}}, \binits{G.-Q.}},
\bauthor{\bsnm{{Xu}}, \binits{Z.}},
\bauthor{\bsnm{{Rao}}, \binits{C.-H.}},
\bauthor{\bsnm{{Hu}}, \binits{Q.-Q.}},
\bauthor{\bsnm{{Li}}, \binits{R.-F.}},
\bauthor{\bsnm{{Fu}}, \binits{H.-W.}},
\bauthor{\bsnm{{Wang}}, \binits{F.}},
\bauthor{\bsnm{{Bao}}, \binits{M.-X.}},
\bauthor{\bsnm{{Wu}}, \binits{M.-C.}},
\bauthor{\bsnm{{Zhang}}, \binits{B.-R.}}:
\byear{2014}c,
\batitle{{New vacuum solar telescope and observations with high resolution}}.
\bjtitle{Research in Astronomy and Astrophysics}
\bvolume{14},
\bfpage{705}.
\doiurl{10.1088/1674-4527/14/6/009}.
\end{barticle}
\endbibitem

\bibitem[\protect\citeauthoryear{{Long}
  \textit{et~al.}}{2015}]{2015ApJ...799..224L}
\begin{barticle}
\bauthor{\bsnm{{Long}}, \binits{D.M.}},
\bauthor{\bsnm{{Baker}}, \binits{D.}},
\bauthor{\bsnm{{Williams}}, \binits{D.R.}},
\bauthor{\bsnm{{Carley}}, \binits{E.P.}},
\bauthor{\bsnm{{Gallagher}}, \binits{P.T.}},
\bauthor{\bsnm{{Zucca}}, \binits{P.}}:
\byear{2015},
\batitle{{The Energetics of a Global Shock Wave in the Low Solar Corona}}.
\bjtitle{Astrophys. J.}
\bvolume{799},
\bfpage{224}.
\doiurl{10.1088/0004-637X/799/2/224}.
\end{barticle}
\endbibitem

\bibitem[\protect\citeauthoryear{{Lugaz}
  \textit{et~al.}}{2012}]{2012ApJ...759...68L}
\begin{barticle}
\bauthor{\bsnm{{Lugaz}}, \binits{N.}},
\bauthor{\bsnm{{Farrugia}}, \binits{C.J.}},
\bauthor{\bsnm{{Davies}}, \binits{J.A.}},
\bauthor{\bsnm{{M{\"o}stl}}, \binits{C.}},
\bauthor{\bsnm{{Davis}}, \binits{C.J.}},
\bauthor{\bsnm{{Roussev}}, \binits{I.I.}},
\bauthor{\bsnm{{Temmer}}, \binits{M.}}:
\byear{2012},
\batitle{{The Deflection of the Two Interacting Coronal Mass Ejections of 2010
  May 23-24 as Revealed by Combined in Situ Measurements and Heliospheric
  Imaging}}.
\bjtitle{Astrophys. J.}
\bvolume{759},
\bfpage{68}.
\doiurl{10.1088/0004-637X/759/1/68}.
\end{barticle}
\endbibitem

\bibitem[\protect\citeauthoryear{{Lynch} and
  {Edmondson}}{2013}]{2013ApJ...764...87L}
\begin{barticle}
\bauthor{\bsnm{{Lynch}}, \binits{B.J.}},
\bauthor{\bsnm{{Edmondson}}, \binits{J.K.}}:
\byear{2013},
\batitle{{Sympathetic Magnetic Breakout Coronal Mass Ejections from
  Pseudostreamers}}.
\bjtitle{Astrophys. J.}
\bvolume{764},
\bfpage{87}.
\end{barticle}
\endbibitem

\bibitem[\protect\citeauthoryear{{Malanushenko}
  \textit{et~al.}}{2014}]{2014ApJ...783..102M}
\begin{barticle}
\bauthor{\bsnm{{Malanushenko}}, \binits{A.}},
\bauthor{\bsnm{{Schrijver}}, \binits{C.J.}},
\bauthor{\bsnm{{DeRosa}}, \binits{M.L.}},
\bauthor{\bsnm{{Wheatland}}, \binits{M.S.}}:
\byear{2014},
\batitle{{Using Coronal Loops to Reconstruct the Magnetic Field of an Active
  Region before and after a Major Flare}}.
\bjtitle{Astrophys. J.}
\bvolume{783},
\bfpage{102}.
\doiurl{10.1088/0004-637X/783/2/102}.
\end{barticle}
\endbibitem

\bibitem[\protect\citeauthoryear{{Mann}, {Vocks}, and
  {Breitling}}{2011}]{2011pre7.conf..507M}
\begin{botherref}
\oauthor{\bsnm{{Mann}}, \binits{G.}},
\oauthor{\bsnm{{Vocks}}, \binits{C.}},
\oauthor{\bsnm{{Breitling}}, \binits{F.}}:
2011,
{Solar Observations with LOFAR}.
\textit{Planetary, Solar and Heliospheric Radio Emissions (PRE VII)},
507.
\end{botherref}
\endbibitem

\bibitem[\protect\citeauthoryear{{Mart{\'{\i}}nez Oliveros}
  \textit{et~al.}}{2012}]{2012ApJ...753L..26M}
\begin{barticle}
\bauthor{\bsnm{{Mart{\'{\i}}nez Oliveros}}, \binits{J.-C.}},
\bauthor{\bsnm{{Hudson}}, \binits{H.S.}},
\bauthor{\bsnm{{Hurford}}, \binits{G.J.}},
\bauthor{\bsnm{{Krucker}}, \binits{S.}},
\bauthor{\bsnm{{Lin}}, \binits{R.P.}},
\bauthor{\bsnm{{Lindsey}}, \binits{C.}},
\bauthor{\bsnm{{Couvidat}}, \binits{S.}},
\bauthor{\bsnm{{Schou}}, \binits{J.}},
\bauthor{\bsnm{{Thompson}}, \binits{W.T.}}:
\byear{2012},
\batitle{{The Height of a White-light Flare and Its Hard X-Ray Sources}}.
\bjtitle{Astrophys. J., Lett.}
\bvolume{753},
\bfpage{L26}.
\doiurl{10.1088/2041-8205/753/2/L26}.
\end{barticle}
\endbibitem

\bibitem[\protect\citeauthoryear{{Masada}, {Yamada}, and
  {Kageyama}}{2013}]{2013ApJ...778...11M}
\begin{barticle}
\bauthor{\bsnm{{Masada}}, \binits{Y.}},
\bauthor{\bsnm{{Yamada}}, \binits{K.}},
\bauthor{\bsnm{{Kageyama}}, \binits{A.}}:
\byear{2013},
\batitle{{Effects of Penetrative Convection on Solar Dynamo}}.
\bjtitle{Astrophys. J.}
\bvolume{778},
\bfpage{11}.
\doiurl{10.1088/0004-637X/778/1/11}.
\end{barticle}
\endbibitem

\bibitem[\protect\citeauthoryear{{Mathioudakis}, {Jess}, and
  {Erd{\'e}lyi}}{2013}]{MatJesErd13aa}
\begin{barticle}
\bauthor{\bsnm{{Mathioudakis}}, \binits{M.}},
\bauthor{\bsnm{{Jess}}, \binits{D.B.}},
\bauthor{\bsnm{{Erd{\'e}lyi}}, \binits{R.}}:
\byear{2013},
\batitle{{Alfv{\'e}n Waves in the Solar Atmosphere. From Theory to
  Observations}}.
\bjtitle{Space Sci. Rev.}
\bvolume{175},
\bfpage{1}.
\doiurl{10.1007/s11214-012-9944-7}.
\end{barticle}
\endbibitem

\bibitem[\protect\citeauthoryear{{McComas}
  \textit{et~al.}}{2013}]{2013ApJ...779....2M}
\begin{barticle}
\bauthor{\bsnm{{McComas}}, \binits{D.J.}},
\bauthor{\bsnm{{Angold}}, \binits{N.}},
\bauthor{\bsnm{{Elliott}}, \binits{H.A.}},
\bauthor{\bsnm{{Livadiotis}}, \binits{G.}},
\bauthor{\bsnm{{Schwadron}}, \binits{N.A.}},
\bauthor{\bsnm{{Skoug}}, \binits{R.M.}},
\bauthor{\bsnm{{Smith}}, \binits{C.W.}}:
\byear{2013},
\batitle{{Weakest Solar Wind of the Space Age and the Current ''Mini'' Solar
  Maximum}}.
\bjtitle{Astrophys. J.}
\bvolume{779},
\bfpage{2}.
\doiurl{10.1088/0004-637X/779/1/2}.
\end{barticle}
\endbibitem

\bibitem[\protect\citeauthoryear{{McIntosh} and {De
  Pontieu}}{2012}]{McIDe-12aa}
\begin{barticle}
\bauthor{\bsnm{{McIntosh}}, \binits{S.W.}},
\bauthor{\bsnm{{De Pontieu}}, \binits{B.}}:
\byear{2012},
\batitle{{Estimating the ''Dark'' Energy Content of the Solar Corona}}.
\bjtitle{Astrophys. J.}
\bvolume{761},
\bfpage{138}.
\doiurl{10.1088/0004-637X/761/2/138}.
\end{barticle}
\endbibitem

\bibitem[\protect\citeauthoryear{{McIntosh}
  \textit{et~al.}}{2011}]{McIde-Car11aa}
\begin{barticle}
\bauthor{\bsnm{{McIntosh}}, \binits{S.W.}},
\bauthor{\bsnm{{de Pontieu}}, \binits{B.}},
\bauthor{\bsnm{{Carlsson}}, \binits{M.}},
\bauthor{\bsnm{{Hansteen}}, \binits{V.}},
\bauthor{\bsnm{{Boerner}}, \binits{P.}},
\bauthor{\bsnm{{Goossens}}, \binits{M.}}:
\byear{2011},
\batitle{{Alfv{\'e}nic waves with sufficient energy to power the quiet solar
  corona and fast solar wind}}.
\bjtitle{Nature}
\bvolume{475},
\bfpage{477}.
\doiurl{10.1038/nature10235}.
\end{barticle}
\endbibitem

\bibitem[\protect\citeauthoryear{{Melrose} and
  {Wheatland}}{2014}]{2014SoPh..289..881M}
\begin{barticle}
\bauthor{\bsnm{{Melrose}}, \binits{D.B.}},
\bauthor{\bsnm{{Wheatland}}, \binits{M.S.}}:
\byear{2014},
\batitle{{Bulk Energization of Electrons in Solar Flares by Alfv{\'e}n Waves}}.
\bjtitle{Solar Phys.}
\bvolume{289},
\bfpage{881}.
\doiurl{10.1007/s11207-013-0376-7}.
\end{barticle}
\endbibitem

\bibitem[\protect\citeauthoryear{{Miyake}
  \textit{et~al.}}{2012}]{2012Natur.486..240M}
\begin{barticle}
\bauthor{\bsnm{{Miyake}}, \binits{F.}},
\bauthor{\bsnm{{Nagaya}}, \binits{K.}},
\bauthor{\bsnm{{Masuda}}, \binits{K.}},
\bauthor{\bsnm{{Nakamura}}, \binits{T.}}:
\byear{2012},
\batitle{{A signature of cosmic-ray increase in AD 774-775 from tree rings in
  Japan}}.
\bjtitle{Nature}
\bvolume{486},
\bfpage{240}.
\doiurl{10.1038/nature11123}.
\end{barticle}
\endbibitem

\bibitem[\protect\citeauthoryear{{Morton}, {Tomczyk}, and
  {Pinto}}{2015}]{MorTomPin15aa}
\begin{barticle}
\bauthor{\bsnm{{Morton}}, \binits{R.J.}},
\bauthor{\bsnm{{Tomczyk}}, \binits{S.}},
\bauthor{\bsnm{{Pinto}}, \binits{R.}}:
\byear{2015},
\batitle{{Investigating Alfv{\'e}nic wave propagation in coronal open-field
  regions}}.
\bjtitle{Nature Communications}
\bvolume{6},
\bfpage{7813}.
\doiurl{10.1038/ncomms8813}.
\end{barticle}
\endbibitem

\bibitem[\protect\citeauthoryear{{M{\"o}stl} and
  {Davies}}{2013}]{2013SoPh..285..411M}
\begin{barticle}
\bauthor{\bsnm{{M{\"o}stl}}, \binits{C.}},
\bauthor{\bsnm{{Davies}}, \binits{J.A.}}:
\byear{2013},
\batitle{{Speeds and Arrival Times of Solar Transients Approximated by
  Self-similar Expanding Circular Fronts}}.
\bjtitle{Solar Phys.}
\bvolume{285},
\bfpage{411}.
\doiurl{10.1007/s11207-012-9978-8}.
\end{barticle}
\endbibitem

\bibitem[\protect\citeauthoryear{{M{\"o}stl}
  \textit{et~al.}}{2012}]{2012ApJ...758...10M}
\begin{barticle}
\bauthor{\bsnm{{M{\"o}stl}}, \binits{C.}},
\bauthor{\bsnm{{Farrugia}}, \binits{C.J.}},
\bauthor{\bsnm{{Kilpua}}, \binits{E.K.J.}},
\bauthor{\bsnm{{Jian}}, \binits{L.K.}},
\bauthor{\bsnm{{Liu}}, \binits{Y.}},
\bauthor{\bsnm{{Eastwood}}, \binits{J.P.}},
\bauthor{\bsnm{{Harrison}}, \binits{R.A.}},
\bauthor{\bsnm{{Webb}}, \binits{D.F.}},
\bauthor{\bsnm{{Temmer}}, \binits{M.}},
\bauthor{\bsnm{{Odstrcil}}, \binits{D.}},
\bauthor{\bsnm{{Davies}}, \binits{J.A.}},
\bauthor{\bsnm{{Rollett}}, \binits{T.}},
\bauthor{\bsnm{{Luhmann}}, \binits{J.G.}},
\bauthor{\bsnm{{Nitta}}, \binits{N.}},
\bauthor{\bsnm{{Mulligan}}, \binits{T.}},
\bauthor{\bsnm{{Jensen}}, \binits{E.A.}},
\bauthor{\bsnm{{Forsyth}}, \binits{R.}},
\bauthor{\bsnm{{Lavraud}}, \binits{B.}},
\bauthor{\bsnm{{de Koning}}, \binits{C.A.}},
\bauthor{\bsnm{{Veronig}}, \binits{A.M.}},
\bauthor{\bsnm{{Galvin}}, \binits{A.B.}},
\bauthor{\bsnm{{Zhang}}, \binits{T.L.}},
\bauthor{\bsnm{{Anderson}}, \binits{B.J.}}:
\byear{2012},
\batitle{{Multi-point Shock and Flux Rope Analysis of Multiple Interplanetary
  Coronal Mass Ejections around 2010 August 1 in the Inner Heliosphere}}.
\bjtitle{Astrophys. J.}
\bvolume{758},
\bfpage{10}.
\doiurl{10.1088/0004-637X/758/1/10}.
\end{barticle}
\endbibitem

\bibitem[\protect\citeauthoryear{{M{\"o}stl}
  \textit{et~al.}}{2014}]{2014ApJ...787..119M}
\begin{barticle}
\bauthor{\bsnm{{M{\"o}stl}}, \binits{C.}},
\bauthor{\bsnm{{Amla}}, \binits{K.}},
\bauthor{\bsnm{{Hall}}, \binits{J.R.}},
\bauthor{\bsnm{{Liewer}}, \binits{P.C.}},
\bauthor{\bsnm{{De Jong}}, \binits{E.M.}},
\bauthor{\bsnm{{Colaninno}}, \binits{R.C.}},
\bauthor{\bsnm{{Veronig}}, \binits{A.M.}},
\bauthor{\bsnm{{Rollett}}, \binits{T.}},
\bauthor{\bsnm{{Temmer}}, \binits{M.}},
\bauthor{\bsnm{{Peinhart}}, \binits{V.}},
\bauthor{\bsnm{{Davies}}, \binits{J.A.}},
\bauthor{\bsnm{{Lugaz}}, \binits{N.}},
\bauthor{\bsnm{{Liu}}, \binits{Y.D.}},
\bauthor{\bsnm{{Farrugia}}, \binits{C.J.}},
\bauthor{\bsnm{{Luhmann}}, \binits{J.G.}},
\bauthor{\bsnm{{Vr{\v s}nak}}, \binits{B.}},
\bauthor{\bsnm{{Harrison}}, \binits{R.A.}},
\bauthor{\bsnm{{Galvin}}, \binits{A.B.}}:
\byear{2014},
\batitle{{Connecting Speeds, Directions and Arrival Times of 22 Coronal Mass
  Ejections from the Sun to 1 AU}}.
\bjtitle{Astrophys. J.}
\bvolume{787},
\bfpage{119}.
\doiurl{10.1088/0004-637X/787/2/119}.
\end{barticle}
\endbibitem

\bibitem[\protect\citeauthoryear{{Mu{\~n}oz-Jaramillo}
  \textit{et~al.}}{2012}]{2012ApJ...753..146M}
\begin{barticle}
\bauthor{\bsnm{{Mu{\~n}oz-Jaramillo}}, \binits{A.}},
\bauthor{\bsnm{{Sheeley}}, \binits{N.R.}},
\bauthor{\bsnm{{Zhang}}, \binits{J.}},
\bauthor{\bsnm{{DeLuca}}, \binits{E.E.}}:
\byear{2012},
\batitle{{Calibrating 100 Years of Polar Faculae Measurements: Implications for
  the Evolution of the Heliospheric Magnetic Field}}.
\bjtitle{Astrophys. J.}
\bvolume{753},
\bfpage{146}.
\doiurl{10.1088/0004-637X/753/2/146}.
\end{barticle}
\endbibitem

\bibitem[\protect\citeauthoryear{{Muhr}
  \textit{et~al.}}{2014}]{2014SoPh..289.4563M}
\begin{barticle}
\bauthor{\bsnm{{Muhr}}, \binits{N.}},
\bauthor{\bsnm{{Veronig}}, \binits{A.M.}},
\bauthor{\bsnm{{Kienreich}}, \binits{I.W.}},
\bauthor{\bsnm{{Vr{\v s}nak}}, \binits{B.}},
\bauthor{\bsnm{{Temmer}}, \binits{M.}},
\bauthor{\bsnm{{Bein}}, \binits{B.M.}}:
\byear{2014},
\batitle{{Statistical Analysis of Large-Scale EUV Waves Observed by
  STEREO/EUVI}}.
\bjtitle{Solar Phys.}
\bvolume{289},
\bfpage{4563}.
\doiurl{10.1007/s11207-014-0594-7}.
\end{barticle}
\endbibitem

\bibitem[\protect\citeauthoryear{Mumford
  \textit{et~al.}}{2013}]{mumford-proc-scipy-2013}
\begin{bchapter}
\bauthor{\bsnm{Mumford}, \binits{S.}},
\bauthor{\bsnm{P\'erez-Su\'arez}, \binits{D.}},
\bauthor{\bsnm{Christe}, \binits{S.}},
\bauthor{\bsnm{Mayer}, \binits{F.}},
\bauthor{\bsnm{Hewett}, \binits{R.J.}}:
\byear{2013},
\bctitle{Sunpy: Python for solar physicists}.
In: \beditor{\bsnm{{van der Walt}}, \binits{S.}},
\beditor{\bsnm{Millman}, \binits{J.}},
\beditor{\bsnm{Huff}, \binits{K.}} (eds.)
\bbtitle{Proceedings of the 12th Python in Science Conference},
\bfpage{74 }.
\end{bchapter}
\endbibitem

\bibitem[\protect\citeauthoryear{Nakariakov and Verwichte}{2005}]{NakVer05aa}
\begin{botherref}
\oauthor{\bsnm{Nakariakov}, \binits{V.M.}},
\oauthor{\bsnm{Verwichte}, \binits{E.}}:
2005,
Coronal waves and oscillations.
\textit{Living Rev. Solar Phys.}
\textbf{2}(3).
\url{http://www.livingreviews.org/lrsp-2005-3}.
\end{botherref}
\endbibitem

\bibitem[\protect\citeauthoryear{{Nelson}
  \textit{et~al.}}{2013}]{2013ApJ...762...73N}
\begin{barticle}
\bauthor{\bsnm{{Nelson}}, \binits{N.J.}},
\bauthor{\bsnm{{Brown}}, \binits{B.P.}},
\bauthor{\bsnm{{Brun}}, \binits{A.S.}},
\bauthor{\bsnm{{Miesch}}, \binits{M.S.}},
\bauthor{\bsnm{{Toomre}}, \binits{J.}}:
\byear{2013},
\batitle{{Magnetic Wreaths and Cycles in Convective Dynamos}}.
\bjtitle{Astrophys. J.}
\bvolume{762},
\bfpage{73}.
\doiurl{10.1088/0004-637X/762/2/73}.
\end{barticle}
\endbibitem

\bibitem[\protect\citeauthoryear{{Nelson}
  \textit{et~al.}}{2014}]{2014SoPh..289..441N}
\begin{barticle}
\bauthor{\bsnm{{Nelson}}, \binits{N.J.}},
\bauthor{\bsnm{{Brown}}, \binits{B.P.}},
\bauthor{\bsnm{{Brun}}, \binits{A.}},
\bauthor{\bsnm{{Miesch}}, \binits{M.S.}},
\bauthor{\bsnm{{Toomre}}, \binits{J.}}:
\byear{2014},
\batitle{{Buoyant Magnetic Loops Generated by Global Convective Dynamo
  Action}}.
\bjtitle{Solar Phys.}
\bvolume{289},
\bfpage{441}.
\doiurl{10.1007/s11207-012-0221-4}.
\end{barticle}
\endbibitem

\bibitem[\protect\citeauthoryear{{Neuh{\"a}user} and
  {Neuh{\"a}user}}{2015}]{2015AN....336..225N}
\begin{barticle}
\bauthor{\bsnm{{Neuh{\"a}user}}, \binits{R.}},
\bauthor{\bsnm{{Neuh{\"a}user}}, \binits{D.L.}}:
\byear{2015},
\batitle{{Solar activity around AD 775 from aurorae and radiocarbon}}.
\bjtitle{Astronomische Nachrichten}
\bvolume{336},
\bfpage{225}.
\doiurl{10.1002/asna.201412160}.
\end{barticle}
\endbibitem

\bibitem[\protect\citeauthoryear{{Nitta}
  \textit{et~al.}}{2013}]{2013ApJ...776...58N}
\begin{barticle}
\bauthor{\bsnm{{Nitta}}, \binits{N.V.}},
\bauthor{\bsnm{{Schrijver}}, \binits{C.J.}},
\bauthor{\bsnm{{Title}}, \binits{A.M.}},
\bauthor{\bsnm{{Liu}}, \binits{W.}}:
\byear{2013},
\batitle{{Large-scale Coronal Propagating Fronts in Solar Eruptions as Observed
  by the Atmospheric Imaging Assembly on Board the Solar Dynamics Observatory -
  an Ensemble Study}}.
\bjtitle{Astrophys. J.}
\bvolume{776},
\bfpage{58}.
\end{barticle}
\endbibitem

\bibitem[\protect\citeauthoryear{{Nogami}
  \textit{et~al.}}{2014}]{2014PASJ...66L...4N}
\begin{barticle}
\bauthor{\bsnm{{Nogami}}, \binits{D.}},
\bauthor{\bsnm{{Notsu}}, \binits{Y.}},
\bauthor{\bsnm{{Honda}}, \binits{S.}},
\bauthor{\bsnm{{Maehara}}, \binits{H.}},
\bauthor{\bsnm{{Notsu}}, \binits{S.}},
\bauthor{\bsnm{{Shibayama}}, \binits{T.}},
\bauthor{\bsnm{{Shibata}}, \binits{K.}}:
\byear{2014},
\batitle{{Two sun-like superflare stars rotating as slow as the Sun*}}.
\bjtitle{PASJ}
\bvolume{66},
\bfpage{L4}.
\doiurl{10.1093/pasj/psu012}.
\end{barticle}
\endbibitem

\bibitem[\protect\citeauthoryear{{Oberoi}
  \textit{et~al.}}{2011}]{2011ApJ...728L..27O}
\begin{barticle}
\bauthor{\bsnm{{Oberoi}}, \binits{D.}},
\bauthor{\bsnm{{Matthews}}, \binits{L.D.}},
\bauthor{\bsnm{{Cairns}}, \binits{I.H.}},
\bauthor{\bsnm{{Emrich}}, \binits{D.}},
\bauthor{\bsnm{{Lobzin}}, \binits{V.}},
\bauthor{\bsnm{{Lonsdale}}, \binits{C.J.}},
\bauthor{\bsnm{{Morgan}}, \binits{E.H.}},
\bauthor{\bsnm{{Prabu}}, \binits{T.}},
\bauthor{\bsnm{{Vedantham}}, \binits{H.}},
\bauthor{\bsnm{{Wayth}}, \binits{R.B.}},
\bauthor{\bsnm{{Williams}}, \binits{A.}},
\bauthor{\bsnm{{Williams}}, \binits{C.}},
\bauthor{\bsnm{{White}}, \binits{S.M.}},
\bauthor{\bsnm{{Allen}}, \binits{G.}},
\bauthor{\bsnm{{Arcus}}, \binits{W.}},
\bauthor{\bsnm{{Barnes}}, \binits{D.}},
\bauthor{\bsnm{{Benkevitch}}, \binits{L.}},
\bauthor{\bsnm{{Bernardi}}, \binits{G.}},
\bauthor{\bsnm{{Bowman}}, \binits{J.D.}},
\bauthor{\bsnm{{Briggs}}, \binits{F.H.}},
\bauthor{\bsnm{{Bunton}}, \binits{J.D.}},
\bauthor{\bsnm{{Burns}}, \binits{S.}},
\bauthor{\bsnm{{Cappallo}}, \binits{R.C.}},
\bauthor{\bsnm{{Clark}}, \binits{M.A.}},
\bauthor{\bsnm{{Corey}}, \binits{B.E.}},
\bauthor{\bsnm{{Dawson}}, \binits{M.}},
\bauthor{\bsnm{{DeBoer}}, \binits{D.}},
\bauthor{\bsnm{{De Gans}}, \binits{A.}},
\bauthor{\bsnm{{deSouza}}, \binits{L.}},
\bauthor{\bsnm{{Derome}}, \binits{M.}},
\bauthor{\bsnm{{Edgar}}, \binits{R.G.}},
\bauthor{\bsnm{{Elton}}, \binits{T.}},
\bauthor{\bsnm{{Goeke}}, \binits{R.}},
\bauthor{\bsnm{{Gopalakrishna}}, \binits{M.R.}},
\bauthor{\bsnm{{Greenhill}}, \binits{L.J.}},
\bauthor{\bsnm{{Hazelton}}, \binits{B.}},
\bauthor{\bsnm{{Herne}}, \binits{D.}},
\bauthor{\bsnm{{Hewitt}}, \binits{J.N.}},
\bauthor{\bsnm{{Kamini}}, \binits{P.A.}},
\bauthor{\bsnm{{Kaplan}}, \binits{D.L.}},
\bauthor{\bsnm{{Kasper}}, \binits{J.C.}},
\bauthor{\bsnm{{Kennedy}}, \binits{R.}},
\bauthor{\bsnm{{Kincaid}}, \binits{B.B.}},
\bauthor{\bsnm{{Kocz}}, \binits{J.}},
\bauthor{\bsnm{{Koeing}}, \binits{R.}},
\bauthor{\bsnm{{Kowald}}, \binits{E.}},
\bauthor{\bsnm{{Lynch}}, \binits{M.J.}},
\bauthor{\bsnm{{Madhavi}}, \binits{S.}},
\bauthor{\bsnm{{McWhirter}}, \binits{S.R.}},
\bauthor{\bsnm{{Mitchell}}, \binits{D.A.}},
\bauthor{\bsnm{{Morales}}, \binits{M.F.}},
\bauthor{\bsnm{{Ng}}, \binits{A.}},
\bauthor{\bsnm{{Ord}}, \binits{S.M.}},
\bauthor{\bsnm{{Pathikulangara}}, \binits{J.}},
\bauthor{\bsnm{{Rogers}}, \binits{A.E.E.}},
\bauthor{\bsnm{{Roshi}}, \binits{A.}},
\bauthor{\bsnm{{Salah}}, \binits{J.E.}},
\bauthor{\bsnm{{Sault}}, \binits{R.J.}},
\bauthor{\bsnm{{Schinckel}}, \binits{A.}},
\bauthor{\bsnm{{Udaya Shankar}}, \binits{N.}},
\bauthor{\bsnm{{Srivani}}, \binits{K.S.}},
\bauthor{\bsnm{{Stevens}}, \binits{J.}},
\bauthor{\bsnm{{Subrahmanyan}}, \binits{R.}},
\bauthor{\bsnm{{Thakkar}}, \binits{D.}},
\bauthor{\bsnm{{Tingay}}, \binits{S.J.}},
\bauthor{\bsnm{{Tuthill}}, \binits{J.}},
\bauthor{\bsnm{{Vaccarella}}, \binits{A.}},
\bauthor{\bsnm{{Waterson}}, \binits{M.}},
\bauthor{\bsnm{{Webster}}, \binits{R.L.}},
\bauthor{\bsnm{{Whitney}}, \binits{A.R.}}:
\byear{2011},
\batitle{{First Spectroscopic Imaging Observations of the Sun at Low Radio
  Frequencies with the Murchison Widefield Array Prototype}}.
\bjtitle{Astrophys. J., Lett.}
\bvolume{728},
\bfpage{L27}.
\doiurl{10.1088/2041-8205/728/2/L27}.
\end{barticle}
\endbibitem

\bibitem[\protect\citeauthoryear{{Ofman}}{2010}]{2010LRSP....7....4O}
\begin{barticle}
\bauthor{\bsnm{{Ofman}}, \binits{L.}}:
\byear{2010},
\batitle{{Wave Modeling of the Solar Wind}}.
\bjtitle{Living Reviews in Solar Physics}
\bvolume{7},
\bfpage{4}.
\doiurl{10.12942/lrsp-2010-4}.
\end{barticle}
\endbibitem

\bibitem[\protect\citeauthoryear{{Oka}
  \textit{et~al.}}{2015}]{2015ApJ...799..129O}
\begin{barticle}
\bauthor{\bsnm{{Oka}}, \binits{M.}},
\bauthor{\bsnm{{Krucker}}, \binits{S.}},
\bauthor{\bsnm{{Hudson}}, \binits{H.S.}},
\bauthor{\bsnm{{Saint-Hilaire}}, \binits{P.}}:
\byear{2015},
\batitle{{Electron Energy Partition in the Above-the-looptop Solar Hard X-Ray
  Sources}}.
\bjtitle{Astrophys. J.}
\bvolume{799},
\bfpage{129}.
\doiurl{10.1088/0004-637X/799/2/129}.
\end{barticle}
\endbibitem

\bibitem[\protect\citeauthoryear{{Olmedo}
  \textit{et~al.}}{2012}]{2012ApJ...756..143O}
\begin{barticle}
\bauthor{\bsnm{{Olmedo}}, \binits{O.}},
\bauthor{\bsnm{{Vourlidas}}, \binits{A.}},
\bauthor{\bsnm{{Zhang}}, \binits{J.}},
\bauthor{\bsnm{{Cheng}}, \binits{X.}}:
\byear{2012},
\batitle{{Secondary Waves and/or the ''Reflection'' from and ''Transmission''
  through a Coronal Hole of an Extreme Ultraviolet Wave Associated with the
  2011 February 15 X2.2 Flare Observed with SDO/AIA and STEREO/EUVI}}.
\bjtitle{Astrophys. J.}
\bvolume{756},
\bfpage{143}.
\doiurl{10.1088/0004-637X/756/2/143}.
\end{barticle}
\endbibitem

\bibitem[\protect\citeauthoryear{{Otsuji}, {Sakurai}, and
  {Kuzanyan}}{2015}]{2015PASJ...67....6O}
\begin{barticle}
\bauthor{\bsnm{{Otsuji}}, \binits{K.}},
\bauthor{\bsnm{{Sakurai}}, \binits{T.}},
\bauthor{\bsnm{{Kuzanyan}}, \binits{K.}}:
\byear{2015},
\batitle{{A statistical analysis of current helicity and twist in solar active
  regions over the phases of the solar cycle using the spectro-polarimeter data
  of Hinode}}.
\bjtitle{PASJ}
\bvolume{67},
\bfpage{6}.
\doiurl{10.1093/pasj/psu130}.
\end{barticle}
\endbibitem

\bibitem[\protect\citeauthoryear{{Panasenco}
  \textit{et~al.}}{2013}]{2013SoPh..287..391P}
\begin{barticle}
\bauthor{\bsnm{{Panasenco}}, \binits{O.}},
\bauthor{\bsnm{{Martin}}, \binits{S.F.}},
\bauthor{\bsnm{{Velli}}, \binits{M.}},
\bauthor{\bsnm{{Vourlidas}}, \binits{A.}}:
\byear{2013},
\batitle{{Origins of Rolling, Twisting, and Non-radial Propagation of Eruptive
  Solar Events}}.
\bjtitle{Solar Phys.}
\bvolume{287},
\bfpage{391}.
\doiurl{10.1007/s11207-012-0194-3}.
\end{barticle}
\endbibitem

\bibitem[\protect\citeauthoryear{{Passos} and
  {Charbonneau}}{2014}]{2014A&A...568A.113P}
\begin{barticle}
\bauthor{\bsnm{{Passos}}, \binits{D.}},
\bauthor{\bsnm{{Charbonneau}}, \binits{P.}}:
\byear{2014},
\batitle{{Characteristics of magnetic solar-like cycles in a 3D MHD simulation
  of solar convection}}.
\bjtitle{A{\&}A}
\bvolume{568},
\bfpage{A113}.
\doiurl{10.1051/0004-6361/201423700}.
\end{barticle}
\endbibitem

\bibitem[\protect\citeauthoryear{{Patsourakos} and
  {Vourlidas}}{2012}]{2012SoPh..281..187P}
\begin{barticle}
\bauthor{\bsnm{{Patsourakos}}, \binits{S.}},
\bauthor{\bsnm{{Vourlidas}}, \binits{A.}}:
\byear{2012},
\batitle{{On the Nature and Genesis of EUV Waves: A Synthesis of Observations
  from SOHO, STEREO, SDO, and Hinode (Invited Review)}}.
\bjtitle{Solar Phys.}
\bvolume{281},
\bfpage{187}.
\end{barticle}
\endbibitem

\bibitem[\protect\citeauthoryear{{Patsourakos}, {Vourlidas}, and
  {Stenborg}}{2013}]{2013ApJ...764..125P}
\begin{barticle}
\bauthor{\bsnm{{Patsourakos}}, \binits{S.}},
\bauthor{\bsnm{{Vourlidas}}, \binits{A.}},
\bauthor{\bsnm{{Stenborg}}, \binits{G.}}:
\byear{2013},
\batitle{{Direct Evidence for a Fast Coronal Mass Ejection Driven by the Prior
  Formation and Subsequent Destabilization of a Magnetic Flux Rope}}.
\bjtitle{Astrophys. J.}
\bvolume{764},
\bfpage{125}.
\doiurl{10.1088/0004-637X/764/2/125}.
\end{barticle}
\endbibitem

\bibitem[\protect\citeauthoryear{Pavai \textit{et~al.}}{2015}]{pavai_etal_2015}
\begin{botherref}
\oauthor{\bsnm{Pavai}, \binits{V.S.}},
\oauthor{\bsnm{Arlt}, \binits{R.}},
\oauthor{\bsnm{{Dasi-Espuig}}, \binits{M.}},
\oauthor{\bsnm{Krivova}, \binits{N.}},
\oauthor{\bsnm{Solanki}, \binits{S.}}:
2015,.
\textit{A{\&}A}.
\end{botherref}
\endbibitem

\bibitem[\protect\citeauthoryear{{Pesnell}, {Thompson}, and
  {Chamberlin}}{2012}]{2012SoPh..275....3P}
\begin{barticle}
\bauthor{\bsnm{{Pesnell}}, \binits{W.D.}},
\bauthor{\bsnm{{Thompson}}, \binits{B.J.}},
\bauthor{\bsnm{{Chamberlin}}, \binits{P.C.}}:
\byear{2012},
\batitle{{The Solar Dynamics Observatory (SDO)}}.
\bjtitle{Solar Phys.}
\bvolume{275},
\bfpage{3}.
\end{barticle}
\endbibitem

\bibitem[\protect\citeauthoryear{{Petrie}}{2012}]{2012ApJ...759...50P}
\begin{barticle}
\bauthor{\bsnm{{Petrie}}, \binits{G.J.D.}}:
\byear{2012},
\batitle{{The Abrupt Changes in the Photospheric Magnetic and Lorentz Force
  Vectors during Six Major Neutral-line Flares}}.
\bjtitle{Astrophys. J.}
\bvolume{759},
\bfpage{50}.
\doiurl{10.1088/0004-637X/759/1/50}.
\end{barticle}
\endbibitem

\bibitem[\protect\citeauthoryear{{Petrie}}{2013}]{2013SoPh..287..415P}
\begin{barticle}
\bauthor{\bsnm{{Petrie}}, \binits{G.J.D.}}:
\byear{2013},
\batitle{{A Spatio-temporal Description of the Abrupt Changes in the
  Photospheric Magnetic and Lorentz-Force Vectors During the 15 February 2011
  X2.2 Flare}}.
\bjtitle{Solar Phys.}
\bvolume{287},
\bfpage{415}.
\doiurl{10.1007/s11207-012-0071-0}.
\end{barticle}
\endbibitem

\bibitem[\protect\citeauthoryear{{Population reference bureau}}{2013}]{prb2013}
\begin{bbook}
\bauthor{\bsnm{{Population reference bureau}}}:
\byear{2013},
\bbtitle{{2013 World population data sheet}},
\bpublisher{PRB},
\blocation{Washington, DC}.
\end{bbook}
\endbibitem

\bibitem[\protect\citeauthoryear{{Potgieter}}{2013}]{2013LRSP...10....3P}
\begin{barticle}
\bauthor{\bsnm{{Potgieter}}, \binits{M.}}:
\byear{2013},
\batitle{{Solar Modulation of Cosmic Rays}}.
\bjtitle{Living Reviews in Solar Physics}
\bvolume{10},
\bfpage{3}.
\doiurl{10.12942/lrsp-2013-3; accessed 2015/09/02}.
\end{barticle}
\endbibitem

\bibitem[\protect\citeauthoryear{{Raymond}
  \textit{et~al.}}{2014}]{2014ApJ...788..152R}
\begin{barticle}
\bauthor{\bsnm{{Raymond}}, \binits{J.C.}},
\bauthor{\bsnm{{McCauley}}, \binits{P.I.}},
\bauthor{\bsnm{{Cranmer}}, \binits{S.R.}},
\bauthor{\bsnm{{Downs}}, \binits{C.}}:
\byear{2014},
\batitle{{The Solar Corona as Probed by Comet Lovejoy (C/2011 W3)}}.
\bjtitle{Astrophys. J.}
\bvolume{788},
\bfpage{152}.
\doiurl{10.1088/0004-637X/788/2/152}.
\end{barticle}
\endbibitem

\bibitem[\protect\citeauthoryear{{Reale}
  \textit{et~al.}}{2013}]{2013Sci...341..251R}
\begin{barticle}
\bauthor{\bsnm{{Reale}}, \binits{F.}},
\bauthor{\bsnm{{Orlando}}, \binits{S.}},
\bauthor{\bsnm{{Testa}}, \binits{P.}},
\bauthor{\bsnm{{Peres}}, \binits{G.}},
\bauthor{\bsnm{{Landi}}, \binits{E.}},
\bauthor{\bsnm{{Schrijver}}, \binits{C.J.}}:
\byear{2013},
\batitle{{Bright Hot Impacts by Erupted Fragments Falling Back on the Sun: A
  Template for Stellar Accretion}}.
\bjtitle{Science}
\bvolume{341},
\bfpage{251}.
\doiurl{10.1126/science.1235692}.
\end{barticle}
\endbibitem

\bibitem[\protect\citeauthoryear{{Rempel}}{2014}]{2014ApJ...789..132R}
\begin{barticle}
\bauthor{\bsnm{{Rempel}}, \binits{M.}}:
\byear{2014},
\batitle{{Numerical Simulations of Quiet Sun Magnetism: On the Contribution
  from a Small-scale Dynamo}}.
\bjtitle{Astrophys. J.}
\bvolume{789},
\bfpage{132}.
\doiurl{10.1088/0004-637X/789/2/132}.
\end{barticle}
\endbibitem

\bibitem[\protect\citeauthoryear{{Rempel} and
  {Cheung}}{2014}]{2014ApJ...785...90R}
\begin{barticle}
\bauthor{\bsnm{{Rempel}}, \binits{M.}},
\bauthor{\bsnm{{Cheung}}, \binits{M.C.M.}}:
\byear{2014},
\batitle{{Numerical Simulations of Active Region Scale Flux Emergence: From
  Spot Formation to Decay}}.
\bjtitle{Astrophys. J.}
\bvolume{785},
\bfpage{90}.
\doiurl{10.1088/0004-637X/785/2/90}.
\end{barticle}
\endbibitem

\bibitem[\protect\citeauthoryear{{Rollett}
  \textit{et~al.}}{2012}]{2012SoPh..276..293R}
\begin{barticle}
\bauthor{\bsnm{{Rollett}}, \binits{T.}},
\bauthor{\bsnm{{M{\"o}stl}}, \binits{C.}},
\bauthor{\bsnm{{Temmer}}, \binits{M.}},
\bauthor{\bsnm{{Veronig}}, \binits{A.M.}},
\bauthor{\bsnm{{Farrugia}}, \binits{C.J.}},
\bauthor{\bsnm{{Biernat}}, \binits{H.K.}}:
\byear{2012},
\batitle{{Constraining the Kinematics of Coronal Mass Ejections in the Inner
  Heliosphere with In-Situ Signatures}}.
\bjtitle{Solar Phys.}
\bvolume{276},
\bfpage{293}.
\doiurl{10.1007/s11207-011-9897-0}.
\end{barticle}
\endbibitem

\bibitem[\protect\citeauthoryear{{Rubio da Costa}
  \textit{et~al.}}{2015}]{2015ApJ...804...56R}
\begin{barticle}
\bauthor{\bsnm{{Rubio da Costa}}, \binits{F.}},
\bauthor{\bsnm{{Kleint}}, \binits{L.}},
\bauthor{\bsnm{{Petrosian}}, \binits{V.}},
\bauthor{\bsnm{{Sainz Dalda}}, \binits{A.}},
\bauthor{\bsnm{{Liu}}, \binits{W.}}:
\byear{2015},
\batitle{{Solar Flare Chromospheric Line Emission: Comparison Between IBIS
  High-resolution Observations and Radiative Hydrodynamic Simulations}}.
\bjtitle{Astrophys. J.}
\bvolume{804},
\bfpage{56}.
\doiurl{10.1088/0004-637X/804/1/56}.
\end{barticle}
\endbibitem

\bibitem[\protect\citeauthoryear{{Russell} and
  {Fletcher}}{2013}]{2013ApJ...765...81R}
\begin{barticle}
\bauthor{\bsnm{{Russell}}, \binits{A.J.B.}},
\bauthor{\bsnm{{Fletcher}}, \binits{L.}}:
\byear{2013},
\batitle{{Propagation of Alfv{\'e}nic Waves from Corona to Chromosphere and
  Consequences for Solar Flares}}.
\bjtitle{Astrophys. J.}
\bvolume{765},
\bfpage{81}.
\doiurl{10.1088/0004-637X/765/2/81}.
\end{barticle}
\endbibitem

\bibitem[\protect\citeauthoryear{{Schad}, {Timmer}, and
  {Roth}}{2013}]{2013ApJ...778L..38S}
\begin{barticle}
\bauthor{\bsnm{{Schad}}, \binits{A.}},
\bauthor{\bsnm{{Timmer}}, \binits{J.}},
\bauthor{\bsnm{{Roth}}, \binits{M.}}:
\byear{2013},
\batitle{{Global Helioseismic Evidence for a Deeply Penetrating Solar
  Meridional Flow Consisting of Multiple Flow Cells}}.
\bjtitle{Astrophys. J., Lett.}
\bvolume{778},
\bfpage{L38}.
\doiurl{10.1088/2041-8205/778/2/L38}.
\end{barticle}
\endbibitem

\bibitem[\protect\citeauthoryear{{Scherrer}
  \textit{et~al.}}{2012}]{2012SoPh..275..207S}
\begin{barticle}
\bauthor{\bsnm{{Scherrer}}, \binits{P.H.}},
\bauthor{\bsnm{{Schou}}, \binits{J.}},
\bauthor{\bsnm{{Bush}}, \binits{R.I.}},
\bauthor{\bsnm{{Kosovichev}}, \binits{A.G.}},
\bauthor{\bsnm{{Bogart}}, \binits{R.S.}},
\bauthor{\bsnm{{Hoeksema}}, \binits{J.T.}},
\bauthor{\bsnm{{Liu}}, \binits{Y.}},
\bauthor{\bsnm{{Duvall}}, \binits{T.L.}},
\bauthor{\bsnm{{Zhao}}, \binits{J.}},
\bauthor{\bsnm{{Title}}, \binits{A.M.}},
\bauthor{\bsnm{{Schrijver}}, \binits{C.J.}},
\bauthor{\bsnm{{Tarbell}}, \binits{T.D.}},
\bauthor{\bsnm{{Tomczyk}}, \binits{S.}}:
\byear{2012},
\batitle{{The Helioseismic and Magnetic Imager (HMI) Investigation for the
  Solar Dynamics Observatory (SDO)}}.
\bjtitle{Solar Phys.}
\bvolume{275},
\bfpage{207}.
\doiurl{10.1007/s11207-011-9834-2}.
\end{barticle}
\endbibitem

\bibitem[\protect\citeauthoryear{Schmelz \textit{et~al.}}{2014}]{Schmelz2014}
\begin{barticle}
\bauthor{\bsnm{Schmelz}, \binits{J.T.}},
\bauthor{\bsnm{Pathak}, \binits{S.}},
\bauthor{\bsnm{Brooks}, \binits{D.H.}},
\bauthor{\bsnm{Christian}, \binits{G.M.}},
\bauthor{\bsnm{Dhaliwal}, \binits{R.S.}}:
\byear{2014},
\batitle{Hot topic, warm loops, cooling plasma? multithermal analysis of active
  region loops}.
\bjtitle{The Astrophysical Journal}
\bvolume{795}(\bissue{2}),
\bfpage{171}.
\burl{http://stacks.iop.org/0004-637X/795/i=2/a=171}.
\end{barticle}
\endbibitem

\bibitem[\protect\citeauthoryear{{Schmidt}
  \textit{et~al.}}{2012}]{2012AN....333..796S}
\begin{barticle}
\bauthor{\bsnm{{Schmidt}}, \binits{W.}},
\bauthor{\bsnm{{von der L{\"u}he}}, \binits{O.}},
\bauthor{\bsnm{{Volkmer}}, \binits{R.}},
\bauthor{\bsnm{{Denker}}, \binits{C.}},
\bauthor{\bsnm{{Solanki}}, \binits{S.K.}},
\bauthor{\bsnm{{Balthasar}}, \binits{H.}},
\bauthor{\bsnm{{Bello Gonzalez}}, \binits{N.}},
\bauthor{\bsnm{{Berkefeld}}, \binits{T.}},
\bauthor{\bsnm{{Collados}}, \binits{M.}},
\bauthor{\bsnm{{Fischer}}, \binits{A.}},
\bauthor{\bsnm{{Halbgewachs}}, \binits{C.}},
\bauthor{\bsnm{{Heidecke}}, \binits{F.}},
\bauthor{\bsnm{{Hofmann}}, \binits{A.}},
\bauthor{\bsnm{{Kneer}}, \binits{F.}},
\bauthor{\bsnm{{Lagg}}, \binits{A.}},
\bauthor{\bsnm{{Nicklas}}, \binits{H.}},
\bauthor{\bsnm{{Popow}}, \binits{E.}},
\bauthor{\bsnm{{Puschmann}}, \binits{K.G.}},
\bauthor{\bsnm{{Schmidt}}, \binits{D.}},
\bauthor{\bsnm{{Sigwarth}}, \binits{M.}},
\bauthor{\bsnm{{Sobotka}}, \binits{M.}},
\bauthor{\bsnm{{Soltau}}, \binits{D.}},
\bauthor{\bsnm{{Staude}}, \binits{J.}},
\bauthor{\bsnm{{Strassmeier}}, \binits{K.G.}},
\bauthor{\bsnm{{Waldmann }}, \binits{T.A.}}:
\byear{2012},
\batitle{{The 1.5 meter solar telescope GREGOR}}.
\bjtitle{Astronomische Nachrichten}
\bvolume{333},
\bfpage{796}.
\doiurl{10.1002/asna.201211725}.
\end{barticle}
\endbibitem

\bibitem[\protect\citeauthoryear{{Schmieder}, {Aulanier}, and {Vr{\v
  s}nak}}{2015}]{2015SoPh..tmp...64S}
\begin{botherref}
\oauthor{\bsnm{{Schmieder}}, \binits{B.}},
\oauthor{\bsnm{{Aulanier}}, \binits{G.}},
\oauthor{\bsnm{{Vr{\v s}nak}}, \binits{B.}}:
2015,
{Flare-CME Models: An Observational Perspective (Invited Review)}.
\textit{Solar Phys.}
in press.
\doiurl{10.1007/s11207-015-0712-1}.
\end{botherref}
\endbibitem

\bibitem[\protect\citeauthoryear{{Schrijver} and
  {Beer}}{2014}]{2014EOSTr..95Q.201S}
\begin{barticle}
\bauthor{\bsnm{{Schrijver}}, \binits{C.J.}},
\bauthor{\bsnm{{Beer}}, \binits{J.}}:
\byear{2014},
\batitle{{Space Weather From Explosions on the Sun: How Bad Could It Be?}}
\bjtitle{EOS Transactions}
\bvolume{95},
\bfpage{201}.
\doiurl{10.1002/2014EO240001}.
\end{barticle}
\endbibitem

\bibitem[\protect\citeauthoryear{{Sharykin} and
  {Kosovichev}}{2014}]{2014ApJ...788L..18S}
\begin{barticle}
\bauthor{\bsnm{{Sharykin}}, \binits{I.N.}},
\bauthor{\bsnm{{Kosovichev}}, \binits{A.G.}}:
\byear{2014},
\batitle{{Fine Structure of Flare Ribbons and Evolution of Electric Currents}}.
\bjtitle{Astrophys. J., Lett.}
\bvolume{788},
\bfpage{L18}.
\doiurl{10.1088/2041-8205/788/1/L18}.
\end{barticle}
\endbibitem

\bibitem[\protect\citeauthoryear{{Shen} and {Liu}}{2012}]{2012ApJ...754....7S}
\begin{barticle}
\bauthor{\bsnm{{Shen}}, \binits{Y.}},
\bauthor{\bsnm{{Liu}}, \binits{Y.}}:
\byear{2012},
\batitle{{Evidence for the Wave Nature of an Extreme Ultraviolet Wave Observed
  by the Atmospheric Imaging Assembly on Board the Solar Dynamics
  Observatory}}.
\bjtitle{Astrophys. J.}
\bvolume{754},
\bfpage{7}.
\doiurl{10.1088/0004-637X/754/1/7}.
\end{barticle}
\endbibitem

\bibitem[\protect\citeauthoryear{{Solanki}, {Krivova}, and
  {Haigh}}{2013}]{2013ARA&A..51..311S}
\begin{barticle}
\bauthor{\bsnm{{Solanki}}, \binits{S.K.}},
\bauthor{\bsnm{{Krivova}}, \binits{N.A.}},
\bauthor{\bsnm{{Haigh}}, \binits{J.D.}}:
\byear{2013},
\batitle{{Solar Irradiance Variability and Climate}}.
\bjtitle{ARA{\&}A}
\bvolume{51},
\bfpage{311}.
\doiurl{10.1146/annurev-astro-082812-141007}.
\end{barticle}
\endbibitem

\bibitem[\protect\citeauthoryear{{Steinhilber}
  \textit{et~al.}}{2012}]{2012PNAS..109.5967S}
\begin{barticle}
\bauthor{\bsnm{{Steinhilber}}, \binits{F.}},
\bauthor{\bsnm{{Abreu}}, \binits{J.A.}},
\bauthor{\bsnm{{Beer}}, \binits{J.}},
\bauthor{\bsnm{{Brunner}}, \binits{I.}},
\bauthor{\bsnm{{Christl}}, \binits{M.}},
\bauthor{\bsnm{{Fischer}}, \binits{H.}},
\bauthor{\bsnm{{Heikkila}}, \binits{U.}},
\bauthor{\bsnm{{Kubik}}, \binits{P.W.}},
\bauthor{\bsnm{{Mann}}, \binits{M.}},
\bauthor{\bsnm{{McCracken}}, \binits{K.G.}},
\bauthor{\bsnm{{Miller}}, \binits{H.}},
\bauthor{\bsnm{{Miyahara}}, \binits{H.}},
\bauthor{\bsnm{{Oerter}}, \binits{H.}},
\bauthor{\bsnm{{Wilhelms}}, \binits{F.}}:
\byear{2012},
\batitle{{9,400 years of cosmic radiation and solar activity from ice cores and
  tree rings}}.
\bjtitle{Proceedings of the National Academy of Science}
\bvolume{109},
\bfpage{5967}.
\doiurl{10.1073/pnas.1118965109}.
\end{barticle}
\endbibitem

\bibitem[\protect\citeauthoryear{{Su}
  \textit{et~al.}}{2014}]{2014ApJ...788..150S}
\begin{barticle}
\bauthor{\bsnm{{Su}}, \binits{J.T.}},
\bauthor{\bsnm{{Jing}}, \binits{J.}},
\bauthor{\bsnm{{Wang}}, \binits{S.}},
\bauthor{\bsnm{{Wiegelmann}}, \binits{T.}},
\bauthor{\bsnm{{Wang}}, \binits{H.M.}}:
\byear{2014},
\batitle{{Statistical Study of Free Magnetic Energy and Flare Productivity of
  Solar Active Regions}}.
\bjtitle{Astrophys. J.}
\bvolume{788},
\bfpage{150}.
\doiurl{10.1088/0004-637X/788/2/150}.
\end{barticle}
\endbibitem

\bibitem[\protect\citeauthoryear{{Su}
  \textit{et~al.}}{2013}]{2013NatPh...9..489S}
\begin{barticle}
\bauthor{\bsnm{{Su}}, \binits{Y.}},
\bauthor{\bsnm{{Veronig}}, \binits{A.M.}},
\bauthor{\bsnm{{Holman}}, \binits{G.D.}},
\bauthor{\bsnm{{Dennis}}, \binits{B.R.}},
\bauthor{\bsnm{{Wang}}, \binits{T.}},
\bauthor{\bsnm{{Temmer}}, \binits{M.}},
\bauthor{\bsnm{{Gan}}, \binits{W.}}:
\byear{2013},
\batitle{{Imaging coronal magnetic-field reconnection in a solar flare}}.
\bjtitle{Nature Physics}
\bvolume{9},
\bfpage{489}.
\doiurl{10.1038/nphys2675}.
\end{barticle}
\endbibitem

\bibitem[\protect\citeauthoryear{{Sun}
  \textit{et~al.}}{2012}]{2012ApJ...748...77S}
\begin{barticle}
\bauthor{\bsnm{{Sun}}, \binits{X.}},
\bauthor{\bsnm{{Hoeksema}}, \binits{J.T.}},
\bauthor{\bsnm{{Liu}}, \binits{Y.}},
\bauthor{\bsnm{{Wiegelmann}}, \binits{T.}},
\bauthor{\bsnm{{Hayashi}}, \binits{K.}},
\bauthor{\bsnm{{Chen}}, \binits{Q.}},
\bauthor{\bsnm{{Thalmann}}, \binits{J.}}:
\byear{2012},
\batitle{{Evolution of Magnetic Field and Energy in a Major Eruptive Active
  Region Based on SDO/HMI Observation}}.
\bjtitle{Astrophys. J.}
\bvolume{748},
\bfpage{77}.
\doiurl{10.1088/0004-637X/748/2/77}.
\end{barticle}
\endbibitem

\bibitem[\protect\citeauthoryear{{Sun}
  \textit{et~al.}}{2015}]{2015ApJ...804L..28S}
\begin{barticle}
\bauthor{\bsnm{{Sun}}, \binits{X.}},
\bauthor{\bsnm{{Bobra}}, \binits{M.G.}},
\bauthor{\bsnm{{Hoeksema}}, \binits{J.T.}},
\bauthor{\bsnm{{Liu}}, \binits{Y.}},
\bauthor{\bsnm{{Li}}, \binits{Y.}},
\bauthor{\bsnm{{Shen}}, \binits{C.}},
\bauthor{\bsnm{{Couvidat}}, \binits{S.}},
\bauthor{\bsnm{{Norton}}, \binits{A.A.}},
\bauthor{\bsnm{{Fisher}}, \binits{G.H.}}:
\byear{2015},
\batitle{{Why Is the Great Solar Active Region 12192 Flare-rich but CME-poor?}}
\bjtitle{Astrophys. J., Lett.}
\bvolume{804},
\bfpage{L28}.
\doiurl{10.1088/2041-8205/804/2/L28}.
\end{barticle}
\endbibitem

\bibitem[\protect\citeauthoryear{{Tadesse}, {Wiegelmann}, and
  {MacNeice}}{2015}]{2015SoPh..290.1159T}
\begin{barticle}
\bauthor{\bsnm{{Tadesse}}, \binits{T.}},
\bauthor{\bsnm{{Wiegelmann}}, \binits{T.}},
\bauthor{\bsnm{{MacNeice}}, \binits{P.J.}}:
\byear{2015},
\batitle{{Effect of the Size of the Computational Domain on Spherical Nonlinear
  Force-Free Modeling of a Coronal Magnetic Field Using SDO/HMI Data}}.
\bjtitle{Solar Phys.}
\bvolume{290},
\bfpage{1159}.
\doiurl{10.1007/s11207-015-0664-5}.
\end{barticle}
\endbibitem

\bibitem[\protect\citeauthoryear{{Tan}}{2014}]{2014ApJ...795..140T}
\begin{barticle}
\bauthor{\bsnm{{Tan}}, \binits{B.}}:
\byear{2014},
\batitle{{Coronal Heating Driven by a Magnetic Gradient Pumping Mechanism in
  Solar Plasmas}}.
\bjtitle{Astrophys. J.}
\bvolume{795},
\bfpage{140}.
\doiurl{10.1088/0004-637X/795/2/140}.
\end{barticle}
\endbibitem

\bibitem[\protect\citeauthoryear{{Temmer}
  \textit{et~al.}}{2012}]{2012ApJ...749...57T}
\begin{barticle}
\bauthor{\bsnm{{Temmer}}, \binits{M.}},
\bauthor{\bsnm{{Vr{\v s}nak}}, \binits{B.}},
\bauthor{\bsnm{{Rollett}}, \binits{T.}},
\bauthor{\bsnm{{Bein}}, \binits{B.}},
\bauthor{\bsnm{{de Koning}}, \binits{C.A.}},
\bauthor{\bsnm{{Liu}}, \binits{Y.}},
\bauthor{\bsnm{{Bosman}}, \binits{E.}},
\bauthor{\bsnm{{Davies}}, \binits{J.A.}},
\bauthor{\bsnm{{M{\"o}stl}}, \binits{C.}},
\bauthor{\bsnm{{{\v Z}ic}}, \binits{T.}},
\bauthor{\bsnm{{Veronig}}, \binits{A.M.}},
\bauthor{\bsnm{{Bothmer}}, \binits{V.}},
\bauthor{\bsnm{{Harrison}}, \binits{R.}},
\bauthor{\bsnm{{Nitta}}, \binits{N.}},
\bauthor{\bsnm{{Bisi}}, \binits{M.}},
\bauthor{\bsnm{{Flor}}, \binits{O.}},
\bauthor{\bsnm{{Eastwood}}, \binits{J.}},
\bauthor{\bsnm{{Odstrcil}}, \binits{D.}},
\bauthor{\bsnm{{Forsyth}}, \binits{R.}}:
\byear{2012},
\batitle{{Characteristics of Kinematics of a Coronal Mass Ejection during the
  2010 August 1 CME-CME Interaction Event}}.
\bjtitle{Astrophys. J.}
\bvolume{749},
\bfpage{57}.
\doiurl{10.1088/0004-637X/749/1/57}.
\end{barticle}
\endbibitem

\bibitem[\protect\citeauthoryear{{Temmer}
  \textit{et~al.}}{2014}]{2014ApJ...785...85T}
\begin{barticle}
\bauthor{\bsnm{{Temmer}}, \binits{M.}},
\bauthor{\bsnm{{Veronig}}, \binits{A.M.}},
\bauthor{\bsnm{{Peinhart}}, \binits{V.}},
\bauthor{\bsnm{{Vr{\v s}nak}}, \binits{B.}}:
\byear{2014},
\batitle{{Asymmetry in the CME-CME Interaction Process for the Events from 2011
  February 14-15}}.
\bjtitle{Astrophys. J.}
\bvolume{785},
\bfpage{85}.
\doiurl{10.1088/0004-637X/785/2/85}.
\end{barticle}
\endbibitem

\bibitem[\protect\citeauthoryear{Testa \textit{et~al.}}{2014}]{Testa2014}
\begin{barticle}
\bauthor{\bsnm{Testa}, \binits{P.}},
\bauthor{\bsnm{De~Pontieu}, \binits{B.}},
\bauthor{\bsnm{Allred}, \binits{J.}},
\bauthor{\bsnm{Carlsson}, \binits{M.}},
\bauthor{\bsnm{Reale}, \binits{F.}},
\bauthor{\bsnm{Daw}, \binits{A.}},
\bauthor{\bsnm{Hansteen}, \binits{V.}},
\bauthor{\bsnm{Martinez-Sykora}, \binits{J.}},
\bauthor{\bsnm{Liu}, \binits{W.}},
\bauthor{\bsnm{DeLuca}, \binits{E.E.}},
\bauthor{\bsnm{Golub}, \binits{L.}},
\bauthor{\bsnm{McKillop}, \binits{S.}},
\bauthor{\bsnm{Reeves}, \binits{K.}},
\bauthor{\bsnm{Saar}, \binits{S.}},
\bauthor{\bsnm{Tian}, \binits{H.}},
\bauthor{\bsnm{Lemen}, \binits{J.}},
\bauthor{\bsnm{Title}, \binits{A.}},
\bauthor{\bsnm{Boerner}, \binits{P.}},
\bauthor{\bsnm{Hurlburt}, \binits{N.}},
\bauthor{\bsnm{Tarbell}, \binits{T.D.}},
\bauthor{\bsnm{Wuelser}, \binits{J.P.}},
\bauthor{\bsnm{Kleint}, \binits{L.}},
\bauthor{\bsnm{Kankelborg}, \binits{C.}},
\bauthor{\bsnm{Jaeggli}, \binits{S.}}:
\byear{2014},
\batitle{Evidence of nonthermal particles in coronal loops heated impulsively
  by nanoflares}.
\bjtitle{Science}
\bvolume{346}(\bissue{6207}).
\doiurl{10.1126/science.1255724}.
\burl{http://www.sciencemag.org/content/346/6207/1255724.abstract}.
\end{barticle}
\endbibitem

\bibitem[\protect\citeauthoryear{{Thalmann}
  \textit{et~al.}}{2015}]{2015ApJ...801L..23T}
\begin{barticle}
\bauthor{\bsnm{{Thalmann}}, \binits{J.K.}},
\bauthor{\bsnm{{Su}}, \binits{Y.}},
\bauthor{\bsnm{{Temmer}}, \binits{M.}},
\bauthor{\bsnm{{Veronig}}, \binits{A.M.}}:
\byear{2015},
\batitle{{The Confined X-class Flares of Solar Active Region 2192}}.
\bjtitle{Astrophys. J., Lett.}
\bvolume{801},
\bfpage{L23}.
\doiurl{10.1088/2041-8205/801/2/L23}.
\end{barticle}
\endbibitem

\bibitem[\protect\citeauthoryear{{Tian}
  \textit{et~al.}}{2014}]{2014ApJ...797L..14T}
\begin{barticle}
\bauthor{\bsnm{{Tian}}, \binits{H.}},
\bauthor{\bsnm{{Li}}, \binits{G.}},
\bauthor{\bsnm{{Reeves}}, \binits{K.K.}},
\bauthor{\bsnm{{Raymond}}, \binits{J.C.}},
\bauthor{\bsnm{{Guo}}, \binits{F.}},
\bauthor{\bsnm{{Liu}}, \binits{W.}},
\bauthor{\bsnm{{Chen}}, \binits{B.}},
\bauthor{\bsnm{{Murphy}}, \binits{N.A.}}:
\byear{2014},
\batitle{{Imaging and Spectroscopic Observations of Magnetic Reconnection and
  Chromospheric Evaporation in a Solar Flare}}.
\bjtitle{Astrophys. J., Lett.}
\bvolume{797},
\bfpage{L14}.
\doiurl{10.1088/2041-8205/797/2/L14}.
\end{barticle}
\endbibitem

\bibitem[\protect\citeauthoryear{{Tiwari}
  \textit{et~al.}}{2015}]{2015GeoRL..42.5702T}
\begin{barticle}
\bauthor{\bsnm{{Tiwari}}, \binits{S.K.}},
\bauthor{\bsnm{{Falconer}}, \binits{D.A.}},
\bauthor{\bsnm{{Moore}}, \binits{R.L.}},
\bauthor{\bsnm{{Venkatakrishnan}}, \binits{P.}},
\bauthor{\bsnm{{Winebarger}}, \binits{A.R.}},
\bauthor{\bsnm{{Khazanov}}, \binits{I.G.}}:
\byear{2015},
\batitle{{Near-Sun speed of CMEs and the magnetic nonpotentiality of their
  source active regions}}.
\bjtitle{Geophys. Res. Lett.}
\bvolume{42},
\bfpage{5702}.
\doiurl{10.1002/2015GL064865}.
\end{barticle}
\endbibitem

\bibitem[\protect\citeauthoryear{{Tomczyk} and {McIntosh}}{2009}]{TomMcI09aa}
\begin{barticle}
\bauthor{\bsnm{{Tomczyk}}, \binits{S.}},
\bauthor{\bsnm{{McIntosh}}, \binits{S.W.}}:
\byear{2009},
\batitle{{Time-Distance Seismology of the Solar Corona with CoMP}}.
\bjtitle{Astrophys. J.}
\bvolume{697},
\bfpage{1384}.
\doiurl{10.1088/0004-637X/697/2/1384}.
\end{barticle}
\endbibitem

\bibitem[\protect\citeauthoryear{{Tomczyk}
  \textit{et~al.}}{2007}]{TomMcIKei07aa}
\begin{barticle}
\bauthor{\bsnm{{Tomczyk}}, \binits{S.}},
\bauthor{\bsnm{{McIntosh}}, \binits{S.W.}},
\bauthor{\bsnm{{Keil}}, \binits{S.L.}},
\bauthor{\bsnm{{Judge}}, \binits{P.G.}},
\bauthor{\bsnm{{Schad}}, \binits{T.}},
\bauthor{\bsnm{{Seeley}}, \binits{D.H.}},
\bauthor{\bsnm{{Edmondson}}, \binits{J.}}:
\byear{2007},
\batitle{{Alfv{\'e}n Waves in the Solar Corona}}.
\bjtitle{Science}
\bvolume{317},
\bfpage{1192}.
\doiurl{10.1126/science.1143304}.
\end{barticle}
\endbibitem

\bibitem[\protect\citeauthoryear{{Tomczyk}
  \textit{et~al.}}{2008}]{2008SoPh..247..411T}
\begin{barticle}
\bauthor{\bsnm{{Tomczyk}}, \binits{S.}},
\bauthor{\bsnm{{Card}}, \binits{G.L.}},
\bauthor{\bsnm{{Darnell}}, \binits{T.}},
\bauthor{\bsnm{{Elmore}}, \binits{D.F.}},
\bauthor{\bsnm{{Lull}}, \binits{R.}},
\bauthor{\bsnm{{Nelson}}, \binits{P.G.}},
\bauthor{\bsnm{{Streander}}, \binits{K.V.}},
\bauthor{\bsnm{{Burkepile}}, \binits{J.}},
\bauthor{\bsnm{{Casini}}, \binits{R.}},
\bauthor{\bsnm{{Judge}}, \binits{P.G.}}:
\byear{2008},
\batitle{{An Instrument to Measure Coronal Emission Line Polarization}}.
\bjtitle{Solar Phys.}
\bvolume{247},
\bfpage{411}.
\doiurl{10.1007/s11207-007-9103-6}.
\end{barticle}
\endbibitem

\bibitem[\protect\citeauthoryear{{T{\"o}r{\"o}k}
  \textit{et~al.}}{2013}]{2013SoPh..286..453T}
\begin{barticle}
\bauthor{\bsnm{{T{\"o}r{\"o}k}}, \binits{T.}},
\bauthor{\bsnm{{Temmer}}, \binits{M.}},
\bauthor{\bsnm{{Valori}}, \binits{G.}},
\bauthor{\bsnm{{Veronig}}, \binits{A.M.}},
\bauthor{\bsnm{{van Driel-Gesztelyi}}, \binits{L.}},
\bauthor{\bsnm{{Vr{\v s}nak}}, \binits{B.}}:
\byear{2013},
\batitle{{Initiation of Coronal Mass Ejections by Sunspot Rotation}}.
\bjtitle{Solar Phys.}
\bvolume{286},
\bfpage{453}.
\doiurl{10.1007/s11207-013-0269-9}.
\end{barticle}
\endbibitem

\bibitem[\protect\citeauthoryear{{Upton} and
  {Hathaway}}{2014}]{2014ApJ...792..142U}
\begin{barticle}
\bauthor{\bsnm{{Upton}}, \binits{L.}},
\bauthor{\bsnm{{Hathaway}}, \binits{D.H.}}:
\byear{2014},
\batitle{{Effects of Meridional Flow Variations on Solar Cycles 23 and 24}}.
\bjtitle{Astrophys. J.}
\bvolume{792},
\bfpage{142}.
\doiurl{10.1088/0004-637X/792/2/142}.
\end{barticle}
\endbibitem

\bibitem[\protect\citeauthoryear{{Usoskin}}{2008}]{usoskin_2008}
\begin{barticle}
\bauthor{\bsnm{{Usoskin}}, \binits{I.G.}}:
\byear{2008},
\batitle{{A History of Solar Activity over Millennia}}.
\bjtitle{Living Reviews in Solar Physics}
\bvolume{5},
\bfpage{3}.
\end{barticle}
\endbibitem

\bibitem[\protect\citeauthoryear{{Usoskin}
  \textit{et~al.}}{2013}]{2013A&A...552L...3U}
\begin{barticle}
\bauthor{\bsnm{{Usoskin}}, \binits{I.G.}},
\bauthor{\bsnm{{Kromer}}, \binits{B.}},
\bauthor{\bsnm{{Ludlow}}, \binits{F.}},
\bauthor{\bsnm{{Beer}}, \binits{J.}},
\bauthor{\bsnm{{Friedrich}}, \binits{M.}},
\bauthor{\bsnm{{Kovaltsov}}, \binits{G.A.}},
\bauthor{\bsnm{{Solanki}}, \binits{S.K.}},
\bauthor{\bsnm{{Wacker}}, \binits{L.}}:
\byear{2013},
\batitle{{The AD775 cosmic event revisited: the Sun is to blame}}.
\bjtitle{A{\&}A}
\bvolume{552},
\bfpage{L3}.
\doiurl{10.1051/0004-6361/201321080}.
\end{barticle}
\endbibitem

\bibitem[\protect\citeauthoryear{{van Ballegooijen}, {Asgari-Targhi}, and
  {Berger}}{2014}]{2014ApJ...787...87V}
\begin{barticle}
\bauthor{\bsnm{{van Ballegooijen}}, \binits{A.A.}},
\bauthor{\bsnm{{Asgari-Targhi}}, \binits{M.}},
\bauthor{\bsnm{{Berger}}, \binits{M.A.}}:
\byear{2014},
\batitle{{On the Relationship Between Photospheric Footpoint Motions and
  Coronal Heating in Solar Active Regions}}.
\bjtitle{Astrophys. J.}
\bvolume{787},
\bfpage{87}.
\doiurl{10.1088/0004-637X/787/1/87}.
\end{barticle}
\endbibitem

\bibitem[\protect\citeauthoryear{{van Ballegooijen}
  \textit{et~al.}}{2011}]{vanAsgCra11aa}
\begin{barticle}
\bauthor{\bsnm{{van Ballegooijen}}, \binits{A.A.}},
\bauthor{\bsnm{{Asgari-Targhi}}, \binits{M.}},
\bauthor{\bsnm{{Cranmer}}, \binits{S.R.}},
\bauthor{\bsnm{{DeLuca}}, \binits{E.E.}}:
\byear{2011},
\batitle{{Heating of the Solar Chromosphere and Corona by Alfv{\'e}n Wave
  Turbulence}}.
\bjtitle{Astrophys. J.}
\bvolume{736},
\bfpage{3}.
\doiurl{10.1088/0004-637X/736/1/3}.
\end{barticle}
\endbibitem

\bibitem[\protect\citeauthoryear{{Van Doorsselaere}, {Nakariakov}, and
  {Verwichte}}{2008}]{VanNakVer08aa}
\begin{barticle}
\bauthor{\bsnm{{Van Doorsselaere}}, \binits{T.}},
\bauthor{\bsnm{{Nakariakov}}, \binits{V.M.}},
\bauthor{\bsnm{{Verwichte}}, \binits{E.}}:
\byear{2008},
\batitle{{Detection of Waves in the Solar Corona: Kink or Alfv{\'e}n?}}
\bjtitle{Astrophys. J., Lett.}
\bvolume{676},
\bfpage{L73}.
\doiurl{10.1086/587029}.
\end{barticle}
\endbibitem

\bibitem[\protect\citeauthoryear{{van Noort} and {Rouppe van der
  Voort}}{2008}]{2008A&A...489..429V}
\begin{barticle}
\bauthor{\bsnm{{van Noort}}, \binits{M.J.}},
\bauthor{\bsnm{{Rouppe van der Voort}}, \binits{L.H.M.}}:
\byear{2008},
\batitle{{Stokes imaging polarimetry using image restoration at the Swedish 1-m
  solar telescope}}.
\bjtitle{A{\&}A}
\bvolume{489},
\bfpage{429}.
\doiurl{10.1051/0004-6361:200809959}.
\end{barticle}
\endbibitem

\bibitem[\protect\citeauthoryear{{Varady}
  \textit{et~al.}}{2014}]{2014A&A...563A..51V}
\begin{barticle}
\bauthor{\bsnm{{Varady}}, \binits{M.}},
\bauthor{\bsnm{{Karlick{\'y}}}, \binits{M.}},
\bauthor{\bsnm{{Moravec}}, \binits{Z.}},
\bauthor{\bsnm{{Ka{\v s}parov{\'a}}}, \binits{J.}}:
\byear{2014},
\batitle{{Modifications of thick-target model: re-acceleration of electron
  beams by static and stochastic electric fields}}.
\bjtitle{A{\&}A}
\bvolume{563},
\bfpage{A51}.
\doiurl{10.1051/0004-6361/201322391}.
\end{barticle}
\endbibitem

\bibitem[\protect\citeauthoryear{{Vemareddy}, {Ambastha}, and
  {Maurya}}{2012}]{2012ApJ...761...60V}
\begin{barticle}
\bauthor{\bsnm{{Vemareddy}}, \binits{P.}},
\bauthor{\bsnm{{Ambastha}}, \binits{A.}},
\bauthor{\bsnm{{Maurya}}, \binits{R.A.}}:
\byear{2012},
\batitle{{On the Role of Rotating Sunspots in the Activity of Solar Active
  Region NOAA 11158}}.
\bjtitle{Astrophys. J.}
\bvolume{761},
\bfpage{60}.
\doiurl{10.1088/0004-637X/761/1/60}.
\end{barticle}
\endbibitem

\bibitem[\protect\citeauthoryear{{Verwichte}
  \textit{et~al.}}{2013}]{VerVanFou13aa}
\begin{barticle}
\bauthor{\bsnm{{Verwichte}}, \binits{E.}},
\bauthor{\bsnm{{Van Doorsselaere}}, \binits{T.}},
\bauthor{\bsnm{{Foullon}}, \binits{C.}},
\bauthor{\bsnm{{White}}, \binits{R.S.}}:
\byear{2013},
\batitle{{Coronal Alfv{\'e}n Speed Determination: Consistency between
  Seismology Using AIA/SDO Transverse Loop Oscillations and Magnetic
  Extrapolation}}.
\bjtitle{Astrophys. J.}
\bvolume{767},
\bfpage{16}.
\doiurl{10.1088/0004-637X/767/1/16}.
\end{barticle}
\endbibitem

\bibitem[\protect\citeauthoryear{{Vourlidas}
  \textit{et~al.}}{2013}]{2013SoPh..284..179V}
\begin{barticle}
\bauthor{\bsnm{{Vourlidas}}, \binits{A.}},
\bauthor{\bsnm{{Lynch}}, \binits{B.J.}},
\bauthor{\bsnm{{Howard}}, \binits{R.A.}},
\bauthor{\bsnm{{Li}}, \binits{Y.}}:
\byear{2013},
\batitle{{How Many CMEs Have Flux Ropes? Deciphering the Signatures of Shocks,
  Flux Ropes, and Prominences in Coronagraph Observations of CMEs}}.
\bjtitle{Solar Phys.}
\bvolume{284},
\bfpage{179}.
\doiurl{10.1007/s11207-012-0084-8}.
\end{barticle}
\endbibitem

\bibitem[\protect\citeauthoryear{{Vr{\v s}nak}
  \textit{et~al.}}{2013}]{2013SoPh..285..295V}
\begin{barticle}
\bauthor{\bsnm{{Vr{\v s}nak}}, \binits{B.}},
\bauthor{\bsnm{{{\v Z}ic}}, \binits{T.}},
\bauthor{\bsnm{{Vrbanec}}, \binits{D.}},
\bauthor{\bsnm{{Temmer}}, \binits{M.}},
\bauthor{\bsnm{{Rollett}}, \binits{T.}},
\bauthor{\bsnm{{M{\"o}stl}}, \binits{C.}},
\bauthor{\bsnm{{Veronig}}, \binits{A.}},
\bauthor{\bsnm{{{\v C}alogovi{\'c}}}, \binits{J.}},
\bauthor{\bsnm{{Dumbovi{\'c}}}, \binits{M.}},
\bauthor{\bsnm{{Luli{\'c}}}, \binits{S.}},
\bauthor{\bsnm{{Moon}}, \binits{Y.-J.}},
\bauthor{\bsnm{{Shanmugaraju}}, \binits{A.}}:
\byear{2013},
\batitle{{Propagation of Interplanetary Coronal Mass Ejections: The Drag-Based
  Model}}.
\bjtitle{Solar Phys.}
\bvolume{285},
\bfpage{295}.
\doiurl{10.1007/s11207-012-0035-4}.
\end{barticle}
\endbibitem

\bibitem[\protect\citeauthoryear{{Vr{\v s}nak}
  \textit{et~al.}}{2014}]{2014ApJS..213...21V}
\begin{botherref}
\oauthor{\bsnm{{Vr{\v s}nak}}, \binits{B.}},
\oauthor{\bsnm{{Temmer}}, \binits{M.}},
\oauthor{\bsnm{{{\v Z}ic}}, \binits{T.}},
\oauthor{\bsnm{{Taktakishvili}}, \binits{A.}},
\oauthor{\bsnm{{Dumbovi{\'c}}}, \binits{M.}},
\oauthor{\bsnm{{M{\"o}stl}}, \binits{C.}},
\oauthor{\bsnm{{Veronig}}, \binits{A.M.}},
\oauthor{\bsnm{{Mays}}, \binits{M.L.}},
\oauthor{\bsnm{{Odstr{\v c}il}}, \binits{D.}}:
2014,
{Heliospheric Propagation of Coronal Mass Ejections: Comparison of Numerical
  WSA-ENLIL+Cone Model and Analytical Drag-based Model}.
\textbf{213},
21.
\doiurl{10.1088/0067-0049/213/2/21}.
\end{botherref}
\endbibitem

\bibitem[\protect\citeauthoryear{{Wang}, {Liu}, and
  {Wang}}{2012}]{2012ApJ...757L...5W}
\begin{barticle}
\bauthor{\bsnm{{Wang}}, \binits{S.}},
\bauthor{\bsnm{{Liu}}, \binits{C.}},
\bauthor{\bsnm{{Wang}}, \binits{H.}}:
\byear{2012},
\batitle{{The Relationship between the Sudden Change of the Lorentz Force and
  the Magnitude of Associated Flares}}.
\bjtitle{Astrophys. J., Lett.}
\bvolume{757},
\bfpage{L5}.
\doiurl{10.1088/2041-8205/757/1/L5}.
\end{barticle}
\endbibitem

\bibitem[\protect\citeauthoryear{{Wang}
  \textit{et~al.}}{2012}]{2012ApJ...745L..17W}
\begin{barticle}
\bauthor{\bsnm{{Wang}}, \binits{S.}},
\bauthor{\bsnm{{Liu}}, \binits{C.}},
\bauthor{\bsnm{{Liu}}, \binits{R.}},
\bauthor{\bsnm{{Deng}}, \binits{N.}},
\bauthor{\bsnm{{Liu}}, \binits{Y.}},
\bauthor{\bsnm{{Wang}}, \binits{H.}}:
\byear{2012},
\batitle{{Response of the Photospheric Magnetic Field to the X2.2 Flare on 2011
  February 15}}.
\bjtitle{Astrophys. J., Lett.}
\bvolume{745},
\bfpage{L17}.
\doiurl{10.1088/2041-8205/745/2/L17}.
\end{barticle}
\endbibitem

\bibitem[\protect\citeauthoryear{{Wang}
  \textit{et~al.}}{2013}]{2013PASJ...65S..18W}
\begin{barticle}
\bauthor{\bsnm{{Wang}}, \binits{W.}},
\bauthor{\bsnm{{Yan}}, \binits{Y.}},
\bauthor{\bsnm{{Liu}}, \binits{D.}},
\bauthor{\bsnm{{Chen}}, \binits{Z.}},
\bauthor{\bsnm{{Su}}, \binits{C.}},
\bauthor{\bsnm{{Liu}}, \binits{F.}},
\bauthor{\bsnm{{Geng}}, \binits{L.}},
\bauthor{\bsnm{{Chen}}, \binits{L.}},
\bauthor{\bsnm{{Du}}, \binits{J.}}:
\byear{2013},
\batitle{{Calibration and Data Processing for a Chinese Spectral
  Radioheliograph in the Decimeter Wave Range}}.
\bjtitle{PASJ}
\bvolume{65},
\bfpage{18}.
\doiurl{10.1093/pasj/65.sp1.S18}.
\end{barticle}
\endbibitem

\bibitem[\protect\citeauthoryear{{Warren}}{2014}]{2014ApJ...786L...2W}
\begin{barticle}
\bauthor{\bsnm{{Warren}}, \binits{H.P.}}:
\byear{2014},
\batitle{{Measurements of Absolute Abundances in Solar Flares}}.
\bjtitle{Astrophys. J., Lett.}
\bvolume{786},
\bfpage{L2}.
\doiurl{10.1088/2041-8205/786/1/L2}.
\end{barticle}
\endbibitem

\bibitem[\protect\citeauthoryear{Warren, Winebarger, and
  Brooks}{2012}]{Warren2012}
\begin{barticle}
\bauthor{\bsnm{Warren}, \binits{H.P.}},
\bauthor{\bsnm{Winebarger}, \binits{A.R.}},
\bauthor{\bsnm{Brooks}, \binits{D.H.}}:
\byear{2012},
\batitle{A systematic survey of high-temperature emission in solar active
  regions}.
\bjtitle{The Astrophysical Journal}
\bvolume{759}(\bissue{2}),
\bfpage{141}.
\bisbn{0004-637X}.
\burl{http://stacks.iop.org/0004-637X/759/i=2/a=141}.
\end{barticle}
\endbibitem

\bibitem[\protect\citeauthoryear{{Watanabe}
  \textit{et~al.}}{2013}]{2013ApJ...776..123W}
\begin{barticle}
\bauthor{\bsnm{{Watanabe}}, \binits{K.}},
\bauthor{\bsnm{{Shimizu}}, \binits{T.}},
\bauthor{\bsnm{{Masuda}}, \binits{S.}},
\bauthor{\bsnm{{Ichimoto}}, \binits{K.}},
\bauthor{\bsnm{{Ohno}}, \binits{M.}}:
\byear{2013},
\batitle{{Emission Height and Temperature Distribution of White-light Emission
  Observed by Hinode/SOT from the 2012 January 27 X-class Solar Flare}}.
\bjtitle{Astrophys. J.}
\bvolume{776},
\bfpage{123}.
\doiurl{10.1088/0004-637X/776/2/123}.
\end{barticle}
\endbibitem

\bibitem[\protect\citeauthoryear{{Webb} and
  {Howard}}{2012}]{2012LRSP....9....3W}
\begin{barticle}
\bauthor{\bsnm{{Webb}}, \binits{D.F.}},
\bauthor{\bsnm{{Howard}}, \binits{T.A.}}:
\byear{2012},
\batitle{{Coronal Mass Ejections: Observations}}.
\bjtitle{Living Reviews in Solar Physics}
\bvolume{9},
\bfpage{3}.
\doiurl{10.12942/lrsp-2012-3}.
\end{barticle}
\endbibitem

\bibitem[\protect\citeauthoryear{{Wedemeyer-B{\"o}hm}
  \textit{et~al.}}{2012}]{2012Natur.486..505W}
\begin{barticle}
\bauthor{\bsnm{{Wedemeyer-B{\"o}hm}}, \binits{S.}},
\bauthor{\bsnm{{Scullion}}, \binits{E.}},
\bauthor{\bsnm{{Steiner}}, \binits{O.}},
\bauthor{\bsnm{{Rouppe van der Voort}}, \binits{L.}},
\bauthor{\bsnm{{de La Cruz Rodriguez}}, \binits{J.}},
\bauthor{\bsnm{{Fedun}}, \binits{V.}},
\bauthor{\bsnm{{Erd{\'e}lyi}}, \binits{R.}}:
\byear{2012},
\batitle{{Magnetic tornadoes as energy channels into the solar corona}}.
\bjtitle{Nature}
\bvolume{486},
\bfpage{505}.
\doiurl{10.1038/nature11202}.
\end{barticle}
\endbibitem

\bibitem[\protect\citeauthoryear{Welsch}{2015}]{Welsch2015}
\begin{barticle}
\bauthor{\bsnm{Welsch}, \binits{B.T.}}:
\byear{2015},
\batitle{The photospheric poynting flux and coronal heating}.
\bjtitle{Publications of the Astronomical Society of Japan}
\bvolume{67}(\bissue{2}).
\doiurl{10.1093/pasj/psu151}.
\burl{http://pasj.oxfordjournals.org/content/67/2/18.abstract}.
\end{barticle}
\endbibitem

\bibitem[\protect\citeauthoryear{{White} and {Verwichte}}{2012}]{WhiVer12aa}
\begin{barticle}
\bauthor{\bsnm{{White}}, \binits{R.S.}},
\bauthor{\bsnm{{Verwichte}}, \binits{E.}}:
\byear{2012},
\batitle{{Transverse coronal loop oscillations seen in unprecedented detail by
  AIA/SDO}}.
\bjtitle{A{\&}A}
\bvolume{537},
\bfpage{A49}.
\doiurl{10.1051/0004-6361/201118093}.
\end{barticle}
\endbibitem

\bibitem[\protect\citeauthoryear{{Wiegelmann}, {Thalmann}, and
  {Solanki}}{2014}]{2014A&ARv..22...78W}
\begin{barticle}
\bauthor{\bsnm{{Wiegelmann}}, \binits{T.}},
\bauthor{\bsnm{{Thalmann}}, \binits{J.K.}},
\bauthor{\bsnm{{Solanki}}, \binits{S.K.}}:
\byear{2014},
\batitle{{The magnetic field in the solar atmosphere}}.
\bjtitle{Astron. and Astrophys. Rev.}
\bvolume{22},
\bfpage{78}.
\doiurl{10.1007/s00159-014-0078-7}.
\end{barticle}
\endbibitem

\bibitem[\protect\citeauthoryear{{Winebarger}
  \textit{et~al.}}{2014}]{2014ApJ...787L..10W}
\begin{barticle}
\bauthor{\bsnm{{Winebarger}}, \binits{A.R.}},
\bauthor{\bsnm{{Cirtain}}, \binits{J.}},
\bauthor{\bsnm{{Golub}}, \binits{L.}},
\bauthor{\bsnm{{DeLuca}}, \binits{E.}},
\bauthor{\bsnm{{Savage}}, \binits{S.}},
\bauthor{\bsnm{{Alexander}}, \binits{C.}},
\bauthor{\bsnm{{Schuler}}, \binits{T.}}:
\byear{2014},
\batitle{{Discovery of Finely Structured Dynamic Solar Corona Observed in the
  Hi-C Telescope}}.
\bjtitle{Astrophys. J., Lett.}
\bvolume{787},
\bfpage{L10}.
\doiurl{10.1088/2041-8205/787/1/L10}.
\end{barticle}
\endbibitem

\bibitem[\protect\citeauthoryear{{Withbroe} and {Noyes}}{1977}]{Withbroe1977}
\begin{barticle}
\bauthor{\bsnm{{Withbroe}}, \binits{G.L.}},
\bauthor{\bsnm{{Noyes}}, \binits{R.W.}}:
\byear{1977},
\batitle{{Mass and energy flow in the solar chromosphere and corona}}.
\bjtitle{ARA{\&}A}
\bvolume{15},
\bfpage{363}.
\doiurl{10.1146/annurev.aa.15.090177.002051}.
\end{barticle}
\endbibitem

\bibitem[\protect\citeauthoryear{{Woods}
  \textit{et~al.}}{2012}]{2012SoPh..275..115W}
\begin{barticle}
\bauthor{\bsnm{{Woods}}, \binits{T.N.}},
\bauthor{\bsnm{{Eparvier}}, \binits{F.G.}},
\bauthor{\bsnm{{Hock}}, \binits{R.}},
\bauthor{\bsnm{{Jones}}, \binits{A.R.}},
\bauthor{\bsnm{{Woodraska}}, \binits{D.}},
\bauthor{\bsnm{{Judge}}, \binits{D.}},
\bauthor{\bsnm{{Didkovsky}}, \binits{L.}},
\bauthor{\bsnm{{Lean}}, \binits{J.}},
\bauthor{\bsnm{{Mariska}}, \binits{J.}},
\bauthor{\bsnm{{Warren}}, \binits{H.}},
\bauthor{\bsnm{{McMullin}}, \binits{D.}},
\bauthor{\bsnm{{Chamberlin}}, \binits{P.}},
\bauthor{\bsnm{{Berthiaume}}, \binits{G.}},
\bauthor{\bsnm{{Bailey}}, \binits{S.}},
\bauthor{\bsnm{{Fuller-Rowell}}, \binits{T.}},
\bauthor{\bsnm{{Sojka}}, \binits{J.}},
\bauthor{\bsnm{{Tobiska}}, \binits{W.K.}},
\bauthor{\bsnm{{Viereck}}, \binits{R.}}:
\byear{2012},
\batitle{{Extreme Ultraviolet Variability Experiment (EVE) on the Solar
  Dynamics Observatory (SDO): Overview of Science Objectives, Instrument
  Design, Data Products, and Model Developments}}.
\bjtitle{Solar Phys.}
\bvolume{275},
\bfpage{115}.
\doiurl{10.1007/s11207-009-9487-6}.
\end{barticle}
\endbibitem

\bibitem[\protect\citeauthoryear{{Yan}
  \textit{et~al.}}{2009}]{2009EM&P..104...97Y}
\begin{barticle}
\bauthor{\bsnm{{Yan}}, \binits{Y.}},
\bauthor{\bsnm{{Zhang}}, \binits{J.}},
\bauthor{\bsnm{{Wang}}, \binits{W.}},
\bauthor{\bsnm{{Liu}}, \binits{F.}},
\bauthor{\bsnm{{Chen}}, \binits{Z.}},
\bauthor{\bsnm{{Ji}}, \binits{G.}}:
\byear{2009},
\batitle{{The Chinese Spectral Radioheliograph - CSRH}}.
\bjtitle{Earth Moon and Planets}
\bvolume{104},
\bfpage{97}.
\doiurl{10.1007/s11038-008-9254-y}.
\end{barticle}
\endbibitem

\bibitem[\protect\citeauthoryear{{Yeates}}{2014}]{2014SoPh..289..631Y}
\begin{barticle}
\bauthor{\bsnm{{Yeates}}, \binits{A.R.}}:
\byear{2014},
\batitle{{Coronal Magnetic Field Evolution from 1996 to 2012: Continuous
  Non-potential Simulations}}.
\bjtitle{Solar Phys.}
\bvolume{289},
\bfpage{631}.
\doiurl{10.1007/s11207-013-0301-0}.
\end{barticle}
\endbibitem

\bibitem[\protect\citeauthoryear{{Young}
  \textit{et~al.}}{2013}]{2013ApJ...766..127Y}
\begin{barticle}
\bauthor{\bsnm{{Young}}, \binits{P.R.}},
\bauthor{\bsnm{{Doschek}}, \binits{G.A.}},
\bauthor{\bsnm{{Warren}}, \binits{H.P.}},
\bauthor{\bsnm{{Hara}}, \binits{H.}}:
\byear{2013},
\batitle{{Properties of a Solar Flare Kernel Observed by Hinode and SDO}}.
\bjtitle{Astrophys. J.}
\bvolume{766},
\bfpage{127}.
\doiurl{10.1088/0004-637X/766/2/127}.
\end{barticle}
\endbibitem

\bibitem[\protect\citeauthoryear{{Zhang} and {Liu}}{2011}]{2011ApJ...741L...7Z}
\begin{barticle}
\bauthor{\bsnm{{Zhang}}, \binits{J.}},
\bauthor{\bsnm{{Liu}}, \binits{Y.}}:
\byear{2011},
\batitle{{Ubiquitous Rotating Network Magnetic Fields and Extreme-ultraviolet
  Cyclones in the Quiet Sun}}.
\bjtitle{Astrophys. J., Lett.}
\bvolume{741},
\bfpage{L7}.
\doiurl{10.1088/2041-8205/741/1/L7}.
\end{barticle}
\endbibitem

\bibitem[\protect\citeauthoryear{{Zhao}
  \textit{et~al.}}{2013}]{2013ApJ...774L..29Z}
\begin{barticle}
\bauthor{\bsnm{{Zhao}}, \binits{J.}},
\bauthor{\bsnm{{Bogart}}, \binits{R.S.}},
\bauthor{\bsnm{{Kosovichev}}, \binits{A.G.}},
\bauthor{\bsnm{{Duvall}}, \binits{T.L.} \bsuffix{Jr.}},
\bauthor{\bsnm{{Hartlep}}, \binits{T.}}:
\byear{2013},
\batitle{{Detection of Equatorward Meridional Flow and Evidence of Double-cell
  Meridional Circulation inside the Sun}}.
\bjtitle{Astrophys. J., Lett.}
\bvolume{774},
\bfpage{L29}.
\doiurl{10.1088/2041-8205/774/2/L29}.
\end{barticle}
\endbibitem

\bibitem[\protect\citeauthoryear{{Zharkov}
  \textit{et~al.}}{2013}]{2013SoPh..284..315Z}
\begin{barticle}
\bauthor{\bsnm{{Zharkov}}, \binits{S.}},
\bauthor{\bsnm{{Green}}, \binits{L.M.}},
\bauthor{\bsnm{{Matthews}}, \binits{S.A.}},
\bauthor{\bsnm{{Zharkova}}, \binits{V.V.}}:
\byear{2013},
\batitle{{Properties of the 15 February 2011 Flare Seismic Sources}}.
\bjtitle{Solar Phys.}
\bvolume{284},
\bfpage{315}.
\doiurl{10.1007/s11207-012-0169-4}.
\end{barticle}
\endbibitem

\bibitem[\protect\citeauthoryear{{Zuccarello}
  \textit{et~al.}}{2012}]{2012ApJ...744...66Z}
\begin{barticle}
\bauthor{\bsnm{{Zuccarello}}, \binits{F.P.}},
\bauthor{\bsnm{{Bemporad}}, \binits{A.}},
\bauthor{\bsnm{{Jacobs}}, \binits{C.}},
\bauthor{\bsnm{{Mierla}}, \binits{M.}},
\bauthor{\bsnm{{Poedts}}, \binits{S.}},
\bauthor{\bsnm{{Zuccarello}}, \binits{F.}}:
\byear{2012},
\batitle{{The Role of Streamers in the Deflection of Coronal Mass Ejections:
  Comparison between STEREO Three-dimensional Reconstructions and Numerical
  Simulations}}.
\bjtitle{Astrophys. J.}
\bvolume{744},
\bfpage{66}.
\doiurl{10.1088/0004-637X/744/1/66}.
\end{barticle}
\endbibitem

\end{thebibliography}

\end{document}